\documentclass[twocolumn]{aastex62}

\usepackage{amssymb,amsmath}

\usepackage{natbib}
\usepackage{hyperref}
\usepackage{array}
\usepackage{afterpage}
\usepackage{graphicx}

\newcommand{\Teff}{T_{\mathrm{eff}}}
\newcommand{\logg}{\log{g}}
\newcommand{\vsini}{v \sin{i}}
\newcommand{\logRHK}{\log{R'_{\mathrm{HK}}}}
\newcommand{\SHK}{$S_{\mathrm{HK}}$}

\newcolumntype{C}[1]{>{\centering\let\newline\\\arraybackslash\hspace{0pt}}m{#1}}

\begin{document}

\title{Astrophysical Insights into Radial Velocity Jitter from an Analysis of 600 Planet-search Stars}

\author{Jacob K. Luhn}
\affiliation{Department of Astronomy, The Pennsylvania State University,  525 Davey Lab, University Park, PA 16802, USA}
\affiliation{Center for Exoplanets and Habitable Worlds, 525 Davey Lab, The Pennsylvania State University, University Park, PA, 16802, USA}
\affiliation{NSF Graduate Research Fellow}

\author{Jason T. Wright}
\affiliation{Department of Astronomy, The Pennsylvania State University,  525 Davey Lab, University Park, PA 16802, USA}
\affiliation{Center for Exoplanets and Habitable Worlds, 525 Davey Lab, The Pennsylvania State University, University Park, PA, 16802, USA}

\author{Andrew W. Howard}
\affiliation{Department of Astronomy, California Institute of Technology, Pasadena, CA, USA}

\author{Howard Isaacson}
\affiliation{Astronomy Department, University of California, Berkeley, CA, USA}


\email{jluhn@psu.edu}

\begin{abstract}
Radial velocity (RV) detection of planets is hampered by astrophysical processes on the surfaces of stars that induce a stochastic signal, or ``jitter'', which can drown out or even mimic planetary signals. Here, we empirically and carefully measure the RV jitter of more than 600 stars from the California Planet Search (CPS) sample on a star-by-star basis. As part of this process we explore the activity-RV correlation of stellar cycles and include appendices listing every ostensibly companion-induced signal we removed and every activity cycle we noted. We then use precise stellar properties from \citet{Brewer2017} to separate the sample into bins of stellar mass and examine trends with activity and with evolutionary state. We find RV jitter tracks stellar evolution and that in general, stars evolve through different stages of RV jitter: the jitter in younger stars is driven by magnetic activity, while the jitter in older stars is convectively-driven and dominated by granulation and oscillations. We identify the ``jitter minimum" -- where activity-driven and convectively-driven jitter have similar amplitudes -- for stars between 0.7 M$_{\odot}$ and 1.7 M$_{\odot}$ and find that more massive stars reach this jitter minimum later in their lifetime, in the subgiant or even giant phases. Finally, we comment on how these results can inform future radial velocity efforts, from prioritization of follow-up targets from transit surveys like \emph{TESS} to target selection of future RV surveys.  
\end{abstract}

\section{Introduction}\label{sec:introduction}
Since the first discoveries of planets orbiting other stars over a quarter century ago \citep{Campbell1988,Latham1989,wolszczan1992,Mayor1995}, more than 4000 exoplanets have been confirmed, with a multitude of planet candidates waiting to be confirmed\footnote{\url{http://exoplanetarchive.ipac.caltech.edu}}. In the early days of planet detections, the radial velocity method was the primary method of discovery. With the advent of the \emph{Kepler} mission \citep{Borucki2010}, the field exploded and the transit method has become the dominant discovery method. 

In spite of this, the importance of radial velocity measurements has not diminished. Rather, radial velocities (RVs) have become increasingly important because of their role in transit follow-up. In addition to the confirmation of planet detections via rejection of the overwhelmingly large amount of false positives, by combining the radius from transit detections with the mass from confirmed RV detections we can begin to make inferences about the bulk composition of exoplanets \citep{Weiss2014,Dressing2015}. However, radial velocity resources are already struggling to keep up with transit discoveries. Despite the recent retirement of the \emph{Kepler} spacecraft, additional planets are still being discovered with data from both the original \emph{Kepler} mission as well as the extended \emph{K2} mission \citep{Howell2014}. Further, with \emph{TESS} currently performing its 2 year primary mission \citep{Ricker2014} and having recently been approved for an extended mission through 2023, the number of transiting planets can only be expected to continue to grow faster than RV teams can reasonably follow them up. 

It is therefore critical that we understand the astrophysical drivers of stellar RV variability, or ``jitter", so as to better understand which types of stars present poor cases for RV follow-up due to the increased stochastic stellar signals. In the era of next-generation extremely precise spectrographs (HPF, EXPRES, ESPRESSO, NEID), that can achieve sub-m/s instrumental precision, the largest hurdle to finding Earth-like planets that remains is the intrinsic stellar jitter. With a stronger astrophysical understanding of RV jitter, we can avoid wasting time and resources on targets whose RV variations are not likely to permit efficient Doppler work. Intrinsic stellar jitter represents a fundamental limit for finding the smallest planets, those that are of most interest to the exoplanet community, and by knowing what physical characteristics are driving stellar RV jitter, we can find ways to overcome it, be it observationally, computationally, or statistically.

Radial velocity surveys have largely avoided stars that show signs of high levels of chromospheric activity, e.g. the Eta-Earth Survey \citep{Howard2009}. For sun-like dwarf stars, measurements of chromospheric emission in the Calcium H \& K lines, such as $\logRHK$ and \SHK{} \citep{Noyes1984,Duncan1991}, have been shown to correlate with intrinsic stellar RV variations \citep{Campbell1988,Saar1998,Santos2000,Wright2005}. In these cases, the magnetic activity in the star that manifests as chromospheric emission serves as a proxy for the presence of spots or faculae on the stellar photosphere, which suppress or enhance the convective blueshift, and also introduce rotationally modulated inhomogeneities. As a star ages on the main sequence, it loses angular momentum via magnetic winds and spins down. The result is a decrease in magnetic dynamo, a decrease in magnetically-powered features on the surface of the star, and therefore a lowered chromospheric emission \citep{Wilson1963,Kraft1967,Skumanich1972}. For stars whose RV jitter is dominated by magnetic activity, it is clear that older, ``quieter" stars are the most amenable to RV observations.

However, work by \citet{Bastien2014} has shown that even among ``quiet" stars, RV jitter can be quite large. In fact, the RV jitter in their sample showed a clear increase with decreasing $\logg$. That is, as a star evolves off the main sequence into the subgiant regime, its jitter increases substantially. This dependence on evolutionary state among subgiants was seen as early as \citet{Wright2005} and \citet{Dumusque2011a} and was even predicted both in \citet{Kjeldsen1995}, who used an analytic formula to describe the RV jitter due to p-mode oscillations of evolved stars, and again in \citet{Kjeldsen2011}, who provided a scaling for RV jitter due to granulation.\footnote{Further evidence of an evolutionary dependence on intrinsic RV variations was seen among \emph{giant} stars by \citet{Hatzes1998} and with a much larger sample in \citet{Hekker2008}.} Indeed, the relation seen in \citet{Bastien2014} used a photometric measurement that probes granulation power, suggesting that for these stars the primary driver of RV variations was granulation. The increase in granulation-induced RV variation with evolution is explained by the increase in size of a granular region as the surface gravity, and therefore surface pressure, decreases as the star expands \citep{Schwarzschild1975}. As a result, the total number of granular regions across the face of the star decreases dramatically and so the degree to which the radial velocities from regions of rising and sinking gas balance out over the disk-integrated face of the star decreases because each individual granular region is more strongly weighted \citep{Trampedach2013}. Similarly, p-mode oscillation amplitudes increase as a star evolves and so oscillation-induced RV variation should also increase with evolution.

As we show below, over the course of a stellar lifetime we therefore have periods where the RV jitter is dominated by different phenomena: the activity-dominated and the convection-dominated regimes. Convection is at least partially responsible for RV jitter even in this so-called ``activity-dominated" regime since convection is a requirement for generating a magnetic dynamo, which is ultimately responsible for the RV variations \citep{Parker1955,Haywood2016}. However, as both the  granulation and oscillation components of RV jitter increase with evolutionary state, we treat these as different phases of the ``convection-dominated" regime and compare it to the ``activity-dominated regime", where RV jitter decreases with evolution. It is reasonable then to expect that there is a ``sweet spot" in a star's evolution where the combination of these two contributions are minimized. There is as of yet no published study which marries these two regimes and empirically investigates the evolutionary dependence of RV jitter. The goal of this work is to meticulously determine the astrophysical jitter due to stellar phenomena of more than 600 California Planet Search (CPS) \citep{Howard2010} stars for which we have reliably-derived stellar parameters. 

Previous empirical investigations of RV jitter using the CPS sample \citep[e.g]{Wright2005,Isaacson2010} have taken different approaches toward measuring RV jitter. For instance, \citet{Wright2005} accounted for long term linear trends present in the radial velocities by measuring the jitter about a linear fit for every star, but first removed stars from the sample that had known companions or showed evidence of companions. \citet{Isaacson2010} removed neither planets nor long term linear trends and instead noted that they were most interested in the floor of RV jitter as many stars would have jitter that included unsubtracted components. In this work, our sample is small enough to treat each star on a case-by-case basis to ensure that the RV jitter adequately reflects the intrinsic stellar RV jitter, yet still large enough to observe bulk trends across a wide range of stellar parameters. We are therefore able to analyze each star individually to account for cases of companions, long term linear trends, and other effects that can typically inflate the reported RV jitter. Additionally, we have the benefit of many more years of observations, which is crucial for subtracting companions and long term linear trends, allowing for more accurate measurements of RV jitter. 

In \autoref{sec:observations} we describe the sample selection and radial velocity observations made by Keck-HIRES. We also describe the stellar properties used in this analysis, specifically focusing on the surface gravities and the activity metrics we use in our analysis. In \autoref{sec:jitter_calc}, we describe our calculation of RV jitter, including detailed steps on our careful vetting process and theoretical scaling of the convective components of RV jitter. \autoref{sec:results} highlights the main results of our empirical jitter calculations, investigating relations between activity, evolution, and mass. \autoref{sec:discussion} contains a discussion of our results and places them in the context of RV surveys. We summarize our main results and conclusions in \autoref{sec:summary}. 

\section{Observations}\label{sec:observations}
\subsection{Sample Selection and Stellar Properties}
\begin{figure}
\includegraphics[width=\columnwidth]{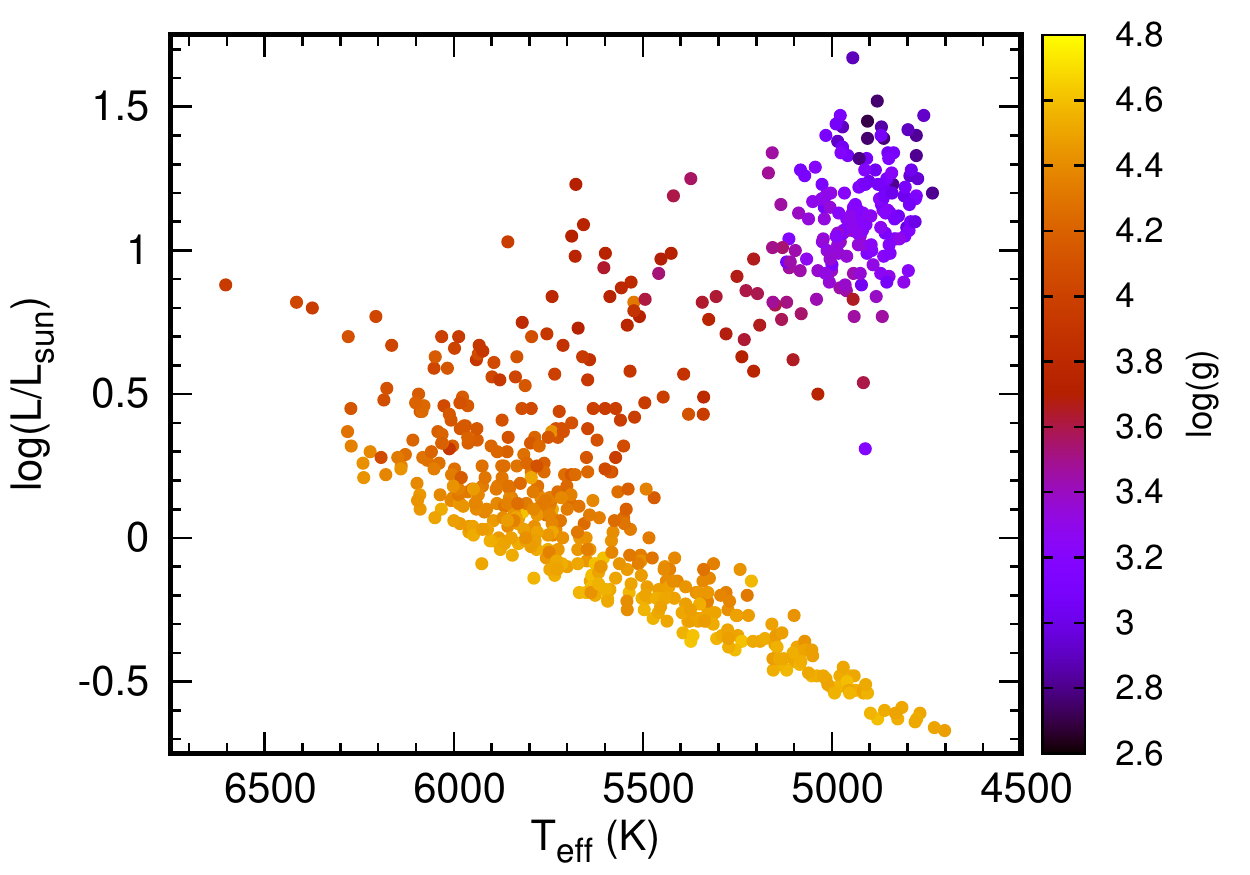}
\caption{HR diagram showing our sample. Points are colored by $\logg$ for reference.  Note the thinning of stars in our sample between $\logg \approx$ 3.5 and 3.8, which likely comes from the short evolutionary time stars spend in this regime. However, it could also arise from the selection biases of the CPS and ``retired" A-star surveys.}
\label{fig:HR_sample}
\end{figure}

The California Planet Search (CPS) has monitored the radial velocities of more than 2500 stars for as many as 20 years with typical precision of $\sim1$~m/s, providing both a long time baseline to analyze stellar jitter and high instrumental precision. Our sample is composed of stars observed as part of the CPS with stellar parameters from \citet{Brewer2016} (erratum \citep{Brewer2017} cited as B17 hereafter).  From this sample of over 1600 stars, we have identified those stars with masses above 0.7~M$_{\odot}$. This lower mass limit is mainly a result of the B17 sample, which has a lower temperature limit of $\sim4700~$K. To ensure robust mass measurements, we choose only those stars for which the spectroscopically-derived surface gravities agree with the isochrone-derived surface gravities -- both given in B17 -- to within 5\% (J. M. Brewer, private communication). We further narrow this sample down by requiring more than 10 observations from Keck/HIRES to ensure enough measurements to measure RV variability. The 10 observation requirement was chosen in order to properly disentangle center-of-mass motions due to orbital companions from the intrinsic stellar variability (see \autoref{sec:nobs} for more details on this requirement). Our final sample is made up of 617 stars. An HR diagram of our sample is shown in \autoref{fig:HR_sample}. We point out a paucity of stars between  $\logg \approx$ 3.5 and 3.8, which has two possible causes. First, an observational bias: the original CPS sample targeted mostly main sequence stars before a large number of the subgiants and giants were added to the sample as part of the ``retired" A-star survey \citep{Johnson2006}. The region of few stars is therefore likely near the boundaries of these two samples: post-main sequence stars that are at the edges of the ``subgiant" region. We note that since the ``Retired" A-star survey specifically targeted intermediate mass stars, we expect our sample of evolved stars to be biased toward those masses ($\geq 1.1$~M$_{\odot}$).

The second, and likely more important, effect is astrophysical: stars simply spend a very small portion of their lives in this region of the HR diagram, zipping through it in relatively no time at all compared to their main sequence or even giant branch lifetimes \citep{Kippenhahn2012}. For more massive stars (several solar masses), this corresponds to the classical Hertzsprung Gap, where there is a very low chance of observing stars. Since our sample is lower mass, these stars lie in the bottom edge of the Hertzsprung Gap and we therefore expect from the outset to see only a few stars here. The stars in our sample range in mass from $0.7 $~M$_{\odot}\leq~$M$_{\star}~\leq~2.14$~M$_{\odot}$, effective temperatures from $4702$~K$~\leq~\Teff~\leq~6603~$K, and surface gravities from $2.70 \leq \logg \leq 4.69$.

Central to this work is a set of reliably measured stellar parameters. We choose those from B17, who used 1 dimensional LTE model spectra to fit to a star's observed spectrum to determine effective temperatures, metallicities, surface gravities, and elemental abundances. In conjunction with \emph{Hipparcos} parallaxes and V-band magnitudes, these spectral measurements were used to derive masses, radii, and luminosities\footnote{Although \emph{Gaia} DR2 is now available with updated parallaxes, the effect this has on the stellar properties is minimal as these are generally bright targets with well-measured \emph{Hipparcos} data. Among our sample, \emph{Gaia} and \emph{Hipparcos} distances agree to a median of 3\%. The largest offset comes from giant stars ($\logg < 3.5$, 156 stars in our sample) that are distant, faint, and cool where the median difference is 8\%. The spectroscopic parameters ($\Teff$, $\logg$, $v\sin{i}$, and abundances) calculated in B17 are all independent of the distance. The luminosity is dependent on distance, which affects the mass (and radius, but our analysis does not use radii). In general we find that the difference tends toward larger distances in the \emph{Gaia} catalog and therefore stars are more luminous than in \emph{Hipparcos}. The effect is that for these stars the median 8\% distance error leads to an observed 16\% luminosity error and subsequently an expected 16\% mass error (given the fixed $\logg$).}. Their iterative fitting technique and improved line list correct for systematic discrepancies in $\logg$ between spectroscopy and asteroseismology \citep{Huber2013,Bastien2014b} and their spectroscopic methods are now consistent with the values of $\logg$ obtained from asteroseismology for stars within the asteroseismic calibration range \citep{Brewer2015}. Outside of this calibration range, \citet{Brewer2015} have taken great care to show that they see no systematic trends with effective temperature or metallicity that had been seen in previous work \citep{Valenti2005}. Specifically for low mass stars, where asteroseismic data is unavailable, comparisons were made with a sample of stars with gravity constraints from transiting exoplanets.

A key portion of our analysis is an investigation of RV jitter trends by mass. There has been ongoing debate surrounding the reported masses of the intermediate mass subgiants targeted by RV surveys such as the ``reitred" A-star survey \citep{Johnson2006}. For a detailed summary of this debate, see \citet{Stello2017} and references therein. Indeed, \citet{Stello2017} measure asteroseismic masses for 8 stars and find that planet discoveries appear to have systematically overestimated the masses of stars above 1.6~M$_{\odot}$. Despite this worrisome conclusion, given the good agreement of the surface gravities from B17 with asteroseismology, we expect that the derived masses should also agree with asteroseismically-derived masses. Finally, more recent work by \citet{Ghezzi2018} show that despite a small overestimate in evolved star masses, it is not nearly as large as the 50\% overestimate as suggested by \citet{Lloyd2011,Lloyd2013}. 

\subsection{Spectra and Radial Velocity Measurements}\label{sec:rv_measurements}
Observations were taken at Keck Observatory using the High Resolution Echelle Spectrometer (HIRES) with resolution $R \approx 55,000$. The CPS employs a standard observing procedure for bright stars that ensures uniform SNR and instrumental/algorithmic velocity precision on all bright FGK targets. Typical values for a $\mathrm{V} = 8$ magnitude star is a signal to noise ratio of 190 at 5800 \r{A} for an exposure of 90 s. Radial velocities are calculated using the iodine-cell calibration technique and the forward-modeling procedure described in \citet{Butler1996} and later \citet{Howard2011}. Our velocity dataset is largely the same as the sample in \citet{Butler2017}, although the radial velocities were calculated using a different pipeline. 

For several stars that were known planet hosts, we included in the planet-fitting procedure non-Keck RV measurements as published with the initial discovery or most recent orbital analysis. \autoref{tbl:non_keck} lists the stars for which we included additional velocities as well as the telescopes and spectrographs where the measurements were taken.  We mainly use previously published radial velocities for only the stars for which we are unable to reproduce the published planetary models with the Keck data alone during our vetting process and so \autoref{tbl:non_keck} is therefore not an exhaustive list of all previously published velocities for all 617 stars in this sample.

\begin{deluxetable*}{c c c c c}
\tablecaption{Summary of Additional Non-Keck HIRES velocities\label{tbl:non_keck}}
\tabletypesize{\scriptsize}
\tablehead{
	 \colhead{Star} 	&  \colhead{Telescope}    & \colhead{Instrument}  	 & \colhead{N$_{\mathrm{obs}}$} & \colhead{Ref} }
	  \startdata
	 HD 1502 & Harlan J. Smith Telescope & TCS & 25 & \citet{Johnson2011} \\
	 HD 1502 & Hobby-Eberly Telescope & HRS & 20 & \citet{Johnson2011} \\
	 HD 159868 & Anglo-Australian Telescope & UCLES & 47 & \citet{Wittenmyer2012} \\
	 HD 192699 & Lick Observatory & Hamilton spectrometer & 34 & \citet{Johnson2007b} \\
	 HD 114613 & Anglo-Australian Telescope & UCLES & 223 & \citet{Wittenmyer2014} \\
	 HD 38801 & Subaru & High Dispersion Spectrograph & 11 & \citet{Harakawa2010} \\
	 HD 181342 & CTIAO 1.5~m & CHIRON & 11 & \citet{Jones2016} \\
	 HD 181342 & CTIAO 2.2~m & FEROS & 20 & \citet{Jones2016} \\
	 HD 181342 & Anglo-Australian Telescope & UCLES & 5 & \citet{Wittenmyer2011} \\
	 HD 5608 & OAO 1.88~m & HIDES & 43 & \citet{Sato2012} \\
	 HD 10697 & Harlan J. Smith Telescop & TCS & 32 & \citet{Wittenmyer2009} \\
	 HD 10697 & Hobby-Eberly Telescope & HRS & 40 & \citet{Wittenmyer2009} \\
	 HD 210702 & Lick Observatory & Hamilton spectrometer & 29 & \citet{Johnson2007} \\
	 HD 210702 & OAO 1.88~m & HIDES & 36 & \citet{Sato2012} \\
	 HD 214823 & Observatoire de Haute-Provence 1.93~m & SOPHIE & 13 & \citet{Diaz2016} \\
	 HD 214823 & Observatoire de Haute-Provence 1.93~m & SOPHIE+ & 11 & \citet{Diaz2016} \\
	 HD 12484 & Observatoire de Haute-Provence 1.93~m & SOPHIE & 65 & \citet{Hebrard2016} \\
	 HD 150706 & Observatoire de Haute-Provence 1.93~m & ELODIE & 48 & \citet{Boisse2012} \\
	 HD 150706 & Observatoire de Haute-Provence 1.93~m & SOPHIE & 53 & \citet{Boisse2012} \\
	 HD 16702 & Observatoire de Haute-Provence 1.93~m & ELODIE & 22 & \citet{Diaz2012} \\
	 HD 26965 & La Silla Observatory 3.6~m ESO telescope & HARPS & 229 & \citet{Diaz2018} \\
	 HD 28185 & Hobby-Eberly Telescope & HRS & 34 & \citet{Wittenmyer2009} \\
	 HD 28185 & 6.5~m Magellan II telescope & MIKE & 15 & \citet{Minniti2009} \\
	 HD 28185 & 1.2~m Leonhard Euler Telescope & CORALIE & 40 & \citet{Santos2001} \\
	 HD 45652 & Observatoire de Haute-Provence 1.93~m  & ELODIE & 14 & \citet{Santos2008} \\
	 HD 45652 & 1.2~m Leonhard Euler Telescope  & CORALIE & 19 & \citet{Santos2008} \\
	 HD 45652 & Observatoire de Haute-Provence 1.93~m  & SOPHIE & 12 & \citet{Santos2008} \\
	 HD 47186 & La Silla Observatory 3.6~m ESO telescope & HARPS & 66 & \citet{Bouchy2009} \\
	 HD 9446 & Observatoire de Haute-Provence 1.93~m & SOPHIE & 79 & \citet{Hebrard2010} \\
	 HD 95128 & Lick Observatory & Hamilton spectrometer & 208 & \citet{Gregory2010} \\
	 HD 125612 & La Silla Observatory 3.6~m ESO telescope & HARPS & 58 & \citet{LoCurto2010} \\
	 HD 142091 & Lick Observatory & Hamilton spectrometer & 46 & \citet{Johnson2008} \\
	 HD 1605 & Subaru & High Dispersion Spectrograph & 14 & \citet{Harakawa2015} \\
	 HD 1605 & OAO 1.88~m & HIDES & 61 & \citet{Harakawa2015} \\
	 HD 1666 & Subaru & High Dispersion Spectrograph & 11 & \citet{Harakawa2015} \\
	 HD 1666 & OAO 1.88~m & HIDES & 67 & \citet{Harakawa2015} \\
	  \enddata
\end{deluxetable*}

\subsection{The activity metric}\label{sec:activity_metric}
The primary activity metric used in this work is the Mount Wilson S-index, $S_{\mathrm{HK}}$, which measures the emission in the cores of the \ion{Ca}{2} H \& K lines relative to the nearby continuum \citep{Duncan1991}. When a star experiences increased magnetic activity, the flux in the cores of these lines measurably increases. Since the cores of these lines are formed in the chromosphere, $S_{\mathrm{HK}}$ measures the amount of chromospheric emission and is a well-studied index of chromospheric activity of a star. 

The wavelength coverage of Keck-HIRES contains the \ion{Ca}{2} H \& K lines. As a result, we benefit from having measurements of \SHK{} made simultaneously with each radial velocity observation\footnote{After the 2004 Keck/HIRES upgrade. There is an offset between the pre- and post-upgrade \SHK{} values that is different for each star. Since the majority of the observations in this sample come post-upgrade, we opt to use solely those values. See \autoref{sec:upgrade} for more details on the upgrade.} using the method outlined in \citet{Isaacson2010}. Our reported values of \SHK{} is simply the median of the \SHK{} time series. Further, by examining the \SHK{} time series, we can more closely investigate correlations between radial velocity and activity. Particularly, given the typical cadence of observations, we can identify stars with activity cycles which can be especially pernicious for planet hunters as they can induce planet-like RV variations. For this work, we wish to include the RV signals due to activity cycles as they are intrinsic stellar variability and it is therefore crucial to identify stars with apparent activity cycles to avoid subtracting out the stellar signal we seek by mistaking it for center-of-mass motion. We follow a similar philosophy for stars with evidence of activity signals similar to possible rotation periods. These are generally tougher to identify given the sampling of observations for the typical star in our sample.

We also briefly examined the $\logRHK$ values for these stars, also from B17. This activity metric accounts for the different continuum levels near the \ion{Ca}{2} lines for different spectral types as well as the base photospheric contribution \citep{Noyes1984}. However, it has not been calibrated for subgiant and giant stars and is therefore only a useful metric for the main sequence stars. Since a large portion of this work investigates the relation between RV jitter and the evolutionary stage of the star, we mostly ignore $\logRHK$ as an activity metric and instead use $S_{\mathrm{HK}}$. Values of $\logRHK$ were mainly used in conjunction with \SHK{} when vetting jitter measurements (see \autoref{sec:blind_fits}) because of known relations between $\logRHK$ activity RV jitter \citep[e.g.,][]{Wright2005}.

\section{Calculating RV jitter} \label{sec:jitter_calc}
Our primary goal in this work is to study the radial velocity variations induced by intrinsic stellar variability. As such, it is crucial that we remove any Keplerian signal caused by a companion that may be present in the data (``planetary noise"). To ensure that the reported RV jitter is in fact due to intrinsic variability and not other effects (companions, instrument errors, etc.), we found it necessary to undergo a vetting process on a star-by-star basis to remove any non-intrinsic variations, as described below. Therefore, each star with a jitter measurement here has cleared our vetting process and represents what we interpret to be astrophysical stellar jitter. Only a few stars in our initial sample contained RVs that led us to remove the stars from our sample (see \autoref{sec:removed} for a description of these stars). Because we expect that our corrections to the RV time series (subtracting a planet or long term linear trend, etc.) will not have captured all center-of mass motions that we wish to subtract off (i.e. there are as-yet-unidentified planets or simply poorly constrained planets contributing noise to our measurements), we certainly have not isolated \emph{true} stellar jitter for all of our stars and rather the measured jitter presented here represents our best \emph{estimate} of the true astrophysical stellar jitter. 

All Keplerian fits to RV data in this work were done using the IDL RVLIN package \citep{Wright2009a}, which is capable of performing multi-planet fits as well as incorporating RV data from multiple telescopes and solving for telescope offsets. Much of the vetting process involves subjective decisions to choose whether we believe a Keplerian fit or not. We do not impose an objective criterion (such as $\chi^2$ goodness of fit cutoff or false alarm probability of periodogram peaks), as we find that no one criterion can adequately establish whether the fit is indeed due to an orbital companion. We instead use several metrics (both statistical and astrophysical) in conjunction with each other to holistically judge if the fit from RVLIN could indeed be due to an orbital companion. 

We are deliberately conservative when deciding if RV variations are due to companions. In all cases where we subtract the best-fit RV signal of a companion, we list that companion and the best-fit orbital parameters in \autoref{tbl:companions}. Our conservative approach to subtracting companions means many stars will have an RV ``jitter" value here that is inflated by orbital companions that we did not deem sufficiently securely detected to remove.  As a result, it is the jitter \emph{floor} that we have robustly identified, and many stars lying above this floor may in reality be low-jitter stars with as-yet unannounced planetary systems that are inflating the measured jitter.

Our procedure for judging a fit is roughly as follows. First, we perform the by-eye test. We examine both the time series fit and the phase-folded fit to each star. The reduced $\chi^2$ value gives us a numerical value on which to anchor our judgment. However, the reduced $\chi^2$ can be severely affected by the number of data points as well as systematic errors and can be convincingly low in cases where the best-fit orbit actually traces a stellar activity cycle. For this reason, we also look at both the times series of the activity metric \SHK{}, described in \autoref{sec:activity_metric} and the median of the time series. Since more active stars are expected to have larger RV variations, we are generally more suspicious of active stars with planet fits. We also look at the $\logg$ of the star. From \citet{Bastien2014} and \citet{Kjeldsen1995} and their follow-up \citet{Kjeldsen2011} (also \citet{Wright2005,Dumusque2011a}), we expect RV jitter to increase as stars evolve during the subgiant phase. Finally, we look at the resulting RV jitter from the fit. Using our experience and intuition, we can piece all of these elements together to decide whether or not we believe a planetary fit. Again, we generally only subtract a companion if all evidence suggests the RV variations are due to center of mass Keplerian motion.

\subsection{Removing Known Planets}
We first take our sample and search for known planet hosts.For each planet-hosting star, we start with the published best-fit parameters for each planet. These serve as the initial parameters that we input to RVLIN, which calculates new orbital parameters. We choose not to simply subtract a Keplerian with the published best-fit orbital parameters because previously undetected planets can change the best-fit planet model or because we have additional RV observations taken after initial publication and so we expect our new best fit results to be slightly improved. In the majority of cases, we do not find a large change in orbital parameters, although there are a few that now have better-constrained periods, especially for long period companions. 

Once RVLIN has produced a best-fit for a system with a known planet or planets, we investigate the phase curves of each planet and examine the residuals (and their periodogram) after subtracting all planets from the system. If we feel based on the periodogram that there is a chance that an additional unpublished planet remains in the data, we revise the fit, adding a period guess from the periodogram for that planet. By comparing the $\chi^2$ goodness of fit, phase curves, and resulting RV jitter, we approve or disfavor the extra planets as needed. Note that in 15 cases our analysis revealed previously unpublished planets around subgiants \citep{Luhn2019}, several of which were additional companions to known planet hosts.

For planets that were not discovered with Keck, we sometimes do not have enough observations from the Keck data alone to detect the planet's signal. In these cases we combine the RV's in the initial discovery paper and the Keck RV's to determine the best-fit planet model (see stars listed in \autoref{tbl:non_keck}). In our final calculation of the RV jitter, we ensure consistency by only calculating jitter using the Keck velocities. 

\subsubsection{Transiting Planets}
For stars with known transiting planets, we use the transit time and period as fixed inputs to RVLIN and allow it to find the remaining best-fit orbital parameters. Since the planet is known to transit, the star \emph{must} have an embedded RV signal from that planet. In many cases where the planets -- and therefore the semi-amplitudes -- are small, the fit from RVLIN would not be convincing by our procedure defined above. However, in these cases, the RV jitter is of order the RV semi-amplitude, and so subtracting out the signal only affects the resulting RV jitter by a few percent. In these cases, it is likely that we are not subtracting out the \emph{true} RV signal from the transiting planet, however, since there \emph{must} be a signal in the data, the only signal we can subtract is the best-fit, regardless of how well it appears to fit. The fact that the RV jitter isn't largely affected in these cases means that our decision to subtract the signal most likely does not matter, but we subtract anyway for completeness.

For transiting planets, we also inspect velocities near the time of transit for any possible Rossiter-McLaughlin effects \citep{Gaudi2007} during transit. In only one case, HD 189733, is a clear RM signal present. In this case we remove the velocities taken during transit so that they do not artificially inflate the measured RV jitter.

\begin{deluxetable*}{c c c c c c c c c c c c c}

\tablecaption{Orbital Parameters of Signals Subtracted from RV Time Series \label{tbl:companions}}
\tabletypesize{\scriptsize}
\tablehead{
	 \colhead{Name} 	&  \colhead{Com}    & \colhead{msini}  	 & \colhead{P} & \colhead{a} & \colhead{T$_{\mathrm{p}}$} & \colhead{$e$} & \colhead{$\omega$} & \colhead{K} & \colhead{$\gamma$} & \colhead{dvdt}  & \colhead{Orbit Reference}\\[2pt]
	  \colhead{}	&	    \colhead{}		& \colhead{(M$_{\mathrm{Jup}}$)}  & \colhead{(days)} & \colhead{(AU)} & \colhead{ (JD)} & \colhead{} & \colhead{(deg)} & \colhead{(m/s)} & \colhead{(m/s)} & \colhead{(m/s)} & \colhead{}}
	  \startdata
            HD 1388 &   * &         28.179 &      9941.964 &   9.31 &  2448143.72 &    0.565 &    115.4 &      300.01 &   -153.33 &          0 &    This work   \\ [2pt]
            HD 1461 &   b &          0.028 &         5.772 &   0.06 &  2450366.22 &    0.229 &     26.4 &        3.11 &     -2.03 &          0 &              \citet{Rivera2010}   \\ [2pt]
                --- &   c &          0.033 &        13.508 &   0.11 &  2439940.70 &    0.477 &    204.4 &        3.10 &      0.00 &          0 &                \citet{Diaz2016}   \\ [2pt]
            HD 4208 &   b &          0.823 &       828.000 &   1.67 &  2451040.00 &    0.052 &    339.8 &       19.12 &     -4.65 &          0 &              \citet{Butler2006}   \\ [2pt]
            HD 4203 &   b &          1.774 &       437.128 &   1.16 &  2451913.88 &    0.519 &    331.1 &       52.10 &     13.44 &    0.00656 &                \citet{Kane2014}   \\ [2pt]
                --- &   c &          3.831 &      8865.852 &   8.65 &  2455823.98 &    0.075 &    175.3 &       35.33 &      0.00 &          0 &                \citet{Kane2014}   \\ [2pt]
            HD 4628 &   b &          0.016 &        14.728 &   0.11 &  2455764.15 &    0.403 &    313.3 &        1.72 &     -1.55 &          0 &    This work   \\ [2pt]
            HD 4747 &   * &         49.359 &     12077.336 &   9.64 &  2438393.51 &    0.730 &    266.8 &      704.07 &   -117.86 &          0 &               \citet{Crepp2016}   \\ [2pt]
            HD 6558 &   * &         17.393 &      7938.025 &   8.37 &  2451362.08 &    0.210 &     43.3 &      155.65 &     35.44 &          0 &    This work   \\ [2pt]
            HD 8574 &   b &          1.688 &       226.696 &   0.77 &  2453974.90 &    0.351 &     17.3 &       54.05 &    -10.75 &          0 &          \citet{Wittenmyer2009}   \\ [2pt]
         \enddata
         \tablenotetext{\dagger}{Stars listed with periods ``$> 36500$" have hit the maximum period limit in RVLIN. In these cases we believe the fit contains curvature but is a companion with period more than 100 years. We include the fit because we believe it to be subtracting center of mass motions, despite a poorly-fit and poorly-constrained period.}
         \tablecomments{\autoref{tbl:companions} is published in its entirety online in the machine-readable format. A portion is shown here for guidance regarding its form and content.}
\end{deluxetable*}

\subsection{Blind Fits}\label{sec:blind_fits}
For all other stars, we perform a blind single-planet fit using RVLIN. We do this to account for any planetary signals that may be present in the data, but have so far been missed by planet hunters and have not yet been published. We then follow up each fit and non-fit with our vetting process to ensure we are left with the most accurate stellar jitter possible. Most blind fits by this method result in rejection of a planet signal via our vetting process (338 out of 391). In many cases we were able to quickly discard the fits as spurious because they end up in portions of parameter space where false positives to blind Keplerian fits are common (i.e. fits with $e>0.9$ where the fitter ``chases" a single outlying point). However, some require more careful analysis. In the end, our approach was to only accept those fits which have coherent periodic signals that seem to demand subtraction. 

The other common result of the blind fit is to find spectroscopic binaries. Since many of these do not have catalogued orbital parameters (often because the period of the system is so much longer than the span of the observations), this often means we are rediscovering these binaries. Luckily, we are not in danger of missing these types of systems since the individual measurement uncertainty is orders of magnitude smaller than the observed RV variations. These systems are usually obvious by eye as having $\sim$ km/s variations, unlike planets, which can be difficult to disentangle from RV jitter in many cases, especially if the observations are spaced out over several years.

In several cases, our blind fit failed to converge on a best fit solution, but the RV time series and the periodogram showed evidence for potential long period trends or even sinusoidal variations. In these cases (and even some shorter period cases), we tried another blind fit but with an initial period guess to help RVLIN converge on a fit. The vetting process was then repeated as needed until we obtained a satisfactory fit, or were convinced by a lack of fit.

In all, we reemphasize that each star has been through our by-eye vetting process and has been manually confirmed, with many stars being visually inspected two to three times before we were able to conclusively rule out or accept a planet fit and definitively calculate RV jitter. Our sample therefore represents the most comprehensive set of RV jitter measurements from CPS data to date. 

Because of our stringent requirements for believing blind fits, the values of RV jitter presented in this work represent upper limits. It is likely (and expected) that many of the stars in our sample still have yet-unsubtracted orbital companions present in the RV time series. 

\subsection{Summary of Companion Subtraction Procedure}
In total, we have subtracted 335 companions from 267 stars. Of these 267 stars, 145 were known previously to host planets, leaving 121 ``new" systems, many of which are stellar companions. All subtracted companions are listed in \autoref{tbl:companions} with the final best-fit parameters that were used in the subtraction, as well as the reference for the orbital parameters used as initial guesses in the fitting procedure, if applicable. We again emphasize that we do not claim that every new planet subtracted is a confirmed planet, rather we have simply subtracted \emph{every strong Keplerian signal} that appears to be due to a companion and many are long-period stellar binaries. Because we are focused on the astrophysical interpretation of RV jitter, a rigorous investigation into the veracity of any planet-mass companions is beyond the scope of this work. Many of the new planets and stellar companions around the \emph{subgiant} stars have been analyzed in more detail in \citet{Luhn2019}.

\subsection{Jitter Calculation and Jitter Error}
Our calculation of RV jitter is a simple RMS calculation. Once we have obtained our best-fit model to subtract from the velocities, we are left with the residuals,
\begin{equation}
\epsilon_{i} = RV_{i} - RV_{\mathrm{fit}}(t_{i})
\label{eqn:residuals}
\end{equation}
In the case where no best-fit model was found or the best-fit model was rejected, the residuals used in \autoref{eqn:residuals} were simply the unaltered velocities, $\epsilon_{i} = RV_{i}$. The RV jitter is then simply
\begin{equation}
j \equiv \sigma_{\mathrm{RV}} = \sqrt{\frac{1}{(N-1)}\sum(\epsilon_{i} - \overline{\epsilon})^2},
\label{eqn:jitter}
\end{equation}
where $N$ is the total number of velocities for the given star. We note that this is not a true RMS in the strict sense and is instead a standard deviation uncertainty calculation. Because the 0 point for each RV time series is arbitrary, it is necessary to subtract off the mean rather than simply taking the square root of the sum of the squared residuals. In the cases where we have accepted a Keplerian fit and have a large number of observations, \autoref{eqn:jitter} is essentially the same as an RMS since the mean of a $\chi^2$ fit is defined to be 0. In fact in our sample the RMS and standard deviation agree with median absolute difference of 0.15 m/s and mean 0.44 m/s. We continue to refer to $\sigma_{RV}$ as an ``RMS." In past works, notably \citet{Wright2005}, the ``jitter" is found by subtracting the mean reported instrumental uncertainty, $\sigma_{instr}$, from the RMS term ($\sigma_{RV}$) in quadrature. We do not follow that approach in this work because we do not assume to know the instrumental systematics of Keck-HIRES. Subtracting the mean internal error for each star may correctly remove instrumental noise but it may also introduce or retain systematics that we do not fully understand. Instead we use the derived RV RMS as the reported RV jitter and compare it to the typical Keck-HIRES instrumental uncertainty of 1-2 m/s \citep{Butler2017}.

Additionally, since this work involves investigating trends with activity, we have many active stars in our sample, which are typically rotating more quickly, leading to broadened absorption features. In principle this is a concern for measuring precise velocities due to the lack of Doppler content in rapidly rotating stars and should add additional variability to the RV measurements. However, this is incorporated in the reported internal errors and despite seeing a gradual increase in the median reported single-measurement errors for stars as a function of $v\sin{i}$, this increase is well below the increase in RV jitter seen with $v\sin{i}$, indicating that we have not reached the rotation broadening floor for the stars in our sample.

For our analysis, we also wish to represent the uncertainty in our measurement of the RMS. In \autoref{sec:results}, we analyze RV jitter as a function of several stellar properties for our sample of stars. Since each star has a different number of observations, the uncertainty of the measured RMS will differ for each star. To that end, we also calculate the error in our measurement of RV jitter and include the calculation in \autoref{sec:jitter_error}. We note simply that the error bars do not account for individual RV measurement uncertainty, the goodness of fit of any subtracted companions, or the potential for the velocities to contain any additional companions.

\subsection{Keck Data Before 2004}\label{sec:upgrade}
In August 2004, the Keck-HIRES instrument went through upgrades and recommissioning, resulting in an improvement in precision from 4-5 m/s to 1-2 m/s in RV measurements post-2004 \citep{Butler2017}. In several cases, the errors on the RV's before and after this upgrade lead to visually different RV observations (either larger scatter or in some cases RV offsets between pre- and post-2004) and require being treated as observations from two separate telescopes. As mentioned before, RVLIN is capable of separately fitting RV's from multiple telescopes and solving for the offset between them. Because of the sometimes large difference in quality of data between pre- and post-2004 observations, we have taken several approaches to account for this, depending on the individual system's observations. 

\paragraph{Removing pre-2004 data from jitter calculation}
In most cases where the pre-and post-2004 data appear different (by comparing the reported errors), we only make use of the pre-2004 data to constrain a fit to the RV data and discard those observations in the final calculation of RV jitter. Our reason for this is the same as discarding other non-Keck observations in our calculation of RV jitter. Since the instrumental errors pre-2004 are significantly higher, they will inherently have larger scatter and will inflate our measurements of RV jitter. 

Note that this approach is only taken when we observe noticeable differences between the pre- and post-2004 data. In our sample, only 8 stars showed such a necessity (shown in \autoref{tbl:jitter}). For a large number of stars in our sample, we don't observe any obvious differences and so the pre-and post-2004 data are treated the same and are included in both the fitting and final jitter calculation. 

\paragraph{Removing pre-2004 data altogether}
In only two cases did we find the need to completely ignore the data before the 2004 upgrades (2 out of 617, HD 1205 and HD 101472). These systems has a large quantity of data after the upgrades such that completely removing the pre-upgrade data does not severely limit the number of observations used in the fit. We are not suspicious of any long period trends or companions for these stars and so including the pre-2004 data to maintain the long baseline is not necessary.

\paragraph{Pre- and post-upgrade offsets}
For 19 stars, we noticed that there appeared to be a reduction error when calculating the velocities before and after the upgrade, leading to a slight but noticeable offset in the radial velocities. By treating the pre- and post-upgrade velocities as coming from two separate telescopes, RVLIN is able to solve for the offset, which is typically no more than 15 m/s.

In only one case (HD 50639), errors in RV extraction have produced large (km/s) offsets between pre- and post-2004 RV's as a result of only containing a single observation in the immediate 2 years following the upgrade. This occurrence is obvious by-eye as a large discontinuity in the otherwise smooth RV curve. In this case, the pre- and post-2004 Keck data are so largely offset that RVLIN cannot solve for the offset and instead finds a long period, highly eccentric fit that manages to explain the discontinuity in 2004 as the periastron passage. To resolve this, we manually apply a first order offset of $\sim$1~km/s before inputting the data into RVLIN as two separate telescopes to find the exact offset that minimizes the $\chi^2$.

\subsection{Activity Cycles and Correlated Activity}
By examining the \SHK{} periodograms for every star, we notice activity cycles among many stars, which are listed in \autoref{sec:notes}\footnote{The classification of stars as having ``activity cycles" in this work is not rigorous. We refer to stars with periodic activity as those with activity cycles, with the periodicity determined by visually examining the strength of peaks in periodograms of activity.}. 

To examine the correlation between radial velocities and \SHK{}, we use a Pearson correlation coefficient
\begin{equation}
r=\frac{\sum \left(S_{\mathrm{HK}}(t_{i}) - \overline{S_{\mathrm{HK}}}\right)\left(RV(t_{i}) - \overline{RV}\right)}{\sqrt{\sum  \left(S_{\mathrm{HK}}(t_{i}) - \overline{S_{\mathrm{HK}}}\right)^2} \sqrt{\sum  \left(RV(t_{i}) - \overline{RV}\right)^2}},
\label{eqn:pearson}
\end{equation}
where $RV$ and $S$ represent the set of velocities and s-indices. For strongly correlated variables, the Pearson coefficient is near 1, and for strong anticorrelation $r$ is near -1. The ability to simultaneously extract \SHK{} and the radial velocity from the same stellar spectrum is what allows this correlation to be measured. For stars that show $|r| > 0.5$, we are particularly suspicious of activity-induced jitter and make a special note of them.

In general, we find that among the stars that show a correlation between the RVs and \SHK, the majority show a positive correlation (125 stars) as opposed to a negative correlation (19 stars), also seen in \citet{Lovis2011}. If we interpret activity index \SHK{} as a proxy for surface starspots and faculae, it follows that as activity increases, the number of spots and faculae also increases, which leads to suppressed convection in those regions. Since the stellar surface has a net convective blueshift from convective granulation, the suppressed convection results in a redward shift, toward positive radial velocities, leading to the positive correlation between activity and radial velocity. 

However, not all stars with activity cycles show evidence of correlated radial velocities. Similarly, many stars show radial velocities that are highly correlated with non-cyclical, stochastic activity. Despite most correlated RVs showing a positive correlation with activity, we observe a wide range of features among stars with activity cycles and activity-correlated radial velocities. That is to say, correlated RVs are not always indicative of a cycle, and cycles are not always indicative of a correlated RVs\footnote{In the case of a star showing an activity cycle but no correlated RV's, this could be explained by having RV's that are significantly rotationally modulated, such that the rotationally modulated RV's no longer correlate with the overall activity cycle.}. The relation between activity and how it manifests in the radial velocities remains an open question. A detailed discussion of these features is beyond the scope of this paper and for now we describe individual stars in \autoref{sec:notes}. 

To summarize, our efforts to subtract companions in order to retain the stellar jitter means that we must ensure that any periodic signals present in the RVs are not in fact due to an activity cycle that is correlated with the RVs. Since stellar astrophysical jitter includes cycles, we have made the effort to examine the correlations between activity and RVs on a star-by-star basis.

\subsection{Outliers}
In many stars we see radial velocity observations that appear to be obvious outliers in the data, usually in one of two ways. In the majority of cases we notice that the reported errors for a given observation are several times larger than the typical errors for that star. Usually this also occurs with velocities themselves that appear to be significantly displaced from the mean. As a general rule, we remove these points if the errors are larger than 2.5 times the typical errors. In other cases where the velocity rather than the error is what identifies it as an outlier, we also investigate the reported $\chi^2$ of the fit to the stellar spectrum as reported by the RV measurement pipeline. This indicates observations where the extraction of a radial velocity was more difficult and is not always represented in the velocity error. We describe instances of outlier removal on individual systems in \autoref{sec:notes}.

\begin{deluxetable*}{c c c c c c c c c c c c c c c c}

\tablecaption{RV Jitter and Analysis \label{tbl:jitter}}

\tabletypesize{\scriptsize}
\tablehead{
	 \colhead{Name} 	&  \colhead{Jitter}    & \colhead{$\sigma_{j}$}  	 & \colhead{N$_{obs}$} & \colhead{$\logg$} & \colhead{\SHK} & \colhead{BFF} & \colhead{PF} & \colhead{N$_{p,p}$} & \colhead{N$_{p,u}$} & \colhead{LTF}  & \colhead{Outlier} & \colhead{RMS$_{pu}$} & \colhead{Removal} & \colhead {Offset} & \colhead{Nothing} \\[2pt]
	  \colhead{}	&	    \colhead{(m/s)}		& \colhead{(m/s)}  & \colhead{} & \colhead{} & \colhead{} & \colhead{} & \colhead{} & \colhead{} & \colhead{} & \colhead{} & \colhead{} & \colhead{} & \colhead{}}
	  \colnumbers
	  \startdata
	              HD 105 &            50.546 &            11.260 &                14 &              4.53 &              0.38 &        \checkmark &                 - &                 - &                 - &                 - &                 - &                 - &                 - &                 - &          \checkmark            \\ [2pt]
            HD 166 &            17.681 &             2.118 &                40 &              4.51 &              0.42 &        \checkmark &                 - &                 - &                 - &                 - &                 - &                 - &                 - &                 - &          \checkmark            \\ [2pt]
            HD 377 &            53.023 &             5.715 &                64 &              4.46 &              0.38 &        \checkmark &                 - &                 - &                 - &                 - &                 - &                 - &                 - &                 - &          \checkmark            \\ [2pt]
            HD 691 &            22.473 &             6.326 &                17 &              4.48 &              0.56 &        \checkmark &                 - &                 - &                 - &                 - &                 - &                 - &                 - &                 - &          \checkmark            \\ [2pt]
           HD 1388 &             5.615 &             0.715 &                51 &              4.32 &              0.16 &                 - &                 1 &                 0 &                 1 &                 - &                 - &                 - &                 - &        \checkmark &                   -            \\ [2pt]
           HD 1461 &             4.023 &             0.193 &               593 &              4.34 &              0.16 &                 - &                 1 &                 2 &                 0 &                 - &        \checkmark &                 - &                 - &                 - &                   -            \\ [2pt]
           HD 4208 &             5.048 &             0.786 &                55 &              4.50 &              0.19 &                 - &                 1 &                 1 &                 0 &                 - &                 - &                 - &                 - &                 - &                   -            \\ [2pt]
           HD 4203 &             3.320 &             0.259 &                49 &              4.08 &              0.15 &                 - &                 1 &                 2 &                 0 &        \checkmark &                 - &                 - &                 - &                 - &                   -            \\ [2pt]
           HD 4307 &             4.047 &             0.472 &                83 &              4.05 &              0.15 &        \checkmark &                 - &                 - &                 - &                 - &                 - &                 - &                 - &                 - &          \checkmark            \\ [2pt]
           HD 4628 &             2.589 &             0.192 &               188 &              4.54 &              0.19 &                 - &                 1 &                 0 &                 1 &                 - &                 - &                 - &                 - &                 - &                   -            \\ [2pt]
	 \enddata
	 \tablecomments{To save space, we have used the following column header abbreviations. Col (1) lists the star name as given in B17. Col (2) is the calculated RV jitter for the star and Col (3) is the uncertainty in that calculation as given in \autoref{eqn:uncertainty}. Col (4) is the number of observations used in the jitter calculation. Note that while our criterion is that stars have more than 10 observations, we do not apply this criterion to the actual jitter calculation, where we occasionally remove the observations before the Keck upgrades in 2004 from the jitter calculation. This applies to a total of 6 stars. Cols (5) and (6) are the surface gravities and activity measure used in Figures \ref{fig:all_rms_logg}-\ref{fig:rms_s2}. Col (7) is a flag indicating if a blind fit was applied to the system (Blind Fit Flag). Col (8) is a Planet Flag to indicate if we have subtracted a companion from the system. Cols (9) and (10) indicate the number of published planets (N$_{p,p}$) and unpublished planets (N$_{p,u}$) for each system. Col (11) is the Linear Trend Flag is a linear trend was subtracted. Col (12) indicates if any outliers were removed. Col (13) contains a flag for when only the post-upgrade observations from Keck were used in the RMS calculation. Col (14) indicates the systems where the pre-upgrade observations were discarded altogether. Col (15) indicates the systems where the pre- and post-upgrade Keck velocities were treated as separate telescopes with an offset between them. Finally, Col (16) is a flag that indicates stars for which no alterations were made to the RVs. Note that because we first try a blind fit to every star without a published planet, this flag is equivalent to having BFF = 1 with no other flags checked. We include this column to explicitly indicate stars for which the raw RVs were used to calculate the RV RMS (no subtractions or removals). }
	 \tablecomments{\autoref{tbl:jitter} is published in its entirety online in the machine-readable format. A portion is shown here for guidance regarding its form and content.}
\end{deluxetable*}

\subsection{Summary of total sample statistics}
In all, we applied some sort of correction to the RVs (outlier rejection, companion subtraction, etc) for more than half of our sample, with only 303 of the 661 stars having an RV jitter simply calculated as the RMS of the unaltered velocities, highlighting the need for our careful approach. 158 stars had velocities that produced a successful Keplerian fit by RVLIN but did not pass our vetting procedure and resulted in rejected fits (that is, we did not alter the velocities). \autoref{tbl:jitter} gives the calculated jitter for each star and lists what changes, if any, have been made to the raw RVs of each system. Note that our criterion that stars have more than 10 observations is used for judging possible companions. We do not apply this criterion to the actual jitter calculation, where we occasionally remove the observations before the Keck upgrades in 2004 from the jitter calculation. In these cases the pre-upgrade observations are enough to confirm or reject a planet, but will inflate the measured RV jitter if included in the jitter calculation. This applies to a total of 6 stars (HD 1388, HD 8765, HD 30708, HD 35974, HD 191876, and HD 216275). Detailed notes on individual systems can be found in \autoref{sec:notes}.

\subsection{Theoretical Calculations of Convective Components of RV Jitter}
The previous sections all dealt with the empirical measurement of jitter for our sample of stars. The following two sections deal with calculating a theoretical RV jitter for the two convective components we account for in this work: stellar oscillations and granulation. The theoretical calculations will later be used to compare with the empirical results.

\subsubsection{Theoretical Oscillation Component of RV Jitter}\label{sec:oscillations}
\citet{Kjeldsen2011} provide a theoretical scaling relation for the velocity amplitude of p-mode oscillations at $\nu_{\mathrm{max}}$, the frequency at which the oscillation power peaks,
\begin{equation}
A_{\mathrm{vel}} \propto \frac{L \tau^{0.5}}{M^{1.5} \Teff^{2.25}},
\label{eqn:oscillation_scaling}
\end{equation}
where $\tau$ is the mode lifetime, for which scaling relations have not been solidly established (See discussions in \citet{Kjeldsen2011, Kallinger2014}). For this work, we choose the mode lifetime scaling relation found in \citet{Corsaro2012}, 
\begin{equation}
\tau \propto \exp{\left(\frac{5777~K - \Teff^{}}{T_0}\right)},
\label{eqn:tau_scaling}
\end{equation}
where $T_{0}=601$~K. Putting this all together and scaling it to measured solar observations \citep{Kjeldsen1995}, we get
\begin{equation}
A_{\mathrm{vel}} = 0.234~\mathrm{m/s} \, \left(\frac{\Teff}{T_{\mathrm{eff,\odot}}}\right)^{1.75} \left(\frac{R}{R_{\odot}}\right)^{-1} \left(\frac{g}{g_{\odot}}\right)^{-1.5} \tau^{0.5}.
\label{eqn:oscillation_amp}
\end{equation}
However, \autoref{eqn:oscillation_amp} gives the \emph{amplitude} of p-mode oscillations. To derive a scaling relation for the \emph{RMS} of this velocity, we assume oscillation manifests as a single sinusoid with $\nu = \nu_{\mathrm{max}}$, whereby the RMS is $0.7087 A_{\mathrm{vel}}$, which gives
\begin{equation}
\sigma_{\mathrm{osc}}= 0.166~\mathrm{m/s} \, \left(\frac{\Teff}{T_{\mathrm{eff,\odot}}}\right)^{1.75} \left(\frac{R}{R_{\odot}}\right)^{-1} \left(\frac{g}{g_{\odot}}\right)^{-1.5} \tau^{0.5}.
\label{eqn:oscillation_RMS}
\end{equation}
While choosing a single sinusoid with amplitude $A_{\mathrm{vel}}$ is a simplistic view of stellar oscillations, it should only affect the scaling constant and should still capture the evolutionary trends we seek to observe across the sample.

\subsubsection{Theoretical Granulation Component of RV Jitter}\label{sec:granulation}
The RV jitter due to granulation is a scaling relation that follows from the granulation size and the number of convective cells on the surface of the star. The proportionality comes again from \citet{Kjeldsen2011},
\begin{equation}
\sigma_{\mathrm{gran}} \propto \frac{H_{\mathrm{p}} c_{\mathrm{s}}}{R} \propto \frac{L^{0.5}}{M \Teff^{0.5}},
\label{eqn:granulation_RMS_proportion}
\end{equation}
where $H_{\mathrm{p}}$ is the pressure scale height, which is the characteristic size of a granular region, and $c_{\mathrm{s}}$ is the sound speed on the surface of the star. These relations assume a constant mean molecular weight, $\mu$. To derive a proper scaling relation, we require a value for the sun's RV RMS due to granulation, a difficult quantity to measure. However, \citet{Meunier2015} derives an expected value of 0.8 m/s based on simulations of granulation and supergranulation. More recent work by \citet{Milbourne2019} has used the HARPS-N spectrograph to continuously observe the sun as it would appear as a star (described in \citet{Dumusque2015}) and find an RV RMS of 1.2 m/s after accounting for the suppression of convective blueshift by bright magnetic regions. Given the general agreement between these two values, we choose to split the difference between the two and adopt a simple value of 1 m/s. Our scaling relation is then
\begin{equation}
\sigma_{\mathrm{gran}} = 1~\mathrm{m/s} \left(\frac{L}{L_{\odot}}\right)^{0.5} \left(\frac{M}{M_{\odot}}\right)^{-1} \left(\frac{\Teff}{T_{\mathrm{eff},\odot}}\right)^{-0.5}
\label{eqn:granulation_RMS}
\end{equation}
We note that the granulation term in this work does not distinguish between the signal from the three scales of granulation: granulation, mesogranulation, and supergranulation. Each of these has different physical scales (1 Mm, 5 Mm, 30 Mm), different flow velocities (1 km/s vertical flow, 0.06 km/s vertical flow, 0.4 km/s horizontal flow), and have different lifetimes (0.2 hr, 3 hr, 20 hr) which makes their individual contributions to RV jitter difficult to study \citep{Rast2003}. However, based on the arguments in \citet{Kjeldsen2011} and \citet{Meunier2019}, we expect that these three granulation effects all follow the same scaling relation. Therefore by roughly scaling it to the solar values we are merely scaling the magnitude of the combined granulation, mesogranulation, and supergranulation effects, despite the fact that in practice stars will have variability due to each of these effects, which operate on different timescales. This is justified given the observing cadence of planet-search stars, which is too infrequent to resolve these individual components. The total RV RMS due to convection (both granulation and oscillation) is found by summing the two terms (Equations \ref{eqn:oscillation_RMS} \& \ref{eqn:granulation_RMS}) in quadrature.

\section{Empirical Analysis of RV jitter} \label{sec:results}
After applying our vetting process to our large sample of stars, we can perform our analysis, examining how RV jitter correlates with stellar parameters. We begin by examining bulk trends in the entire sample. However, first we wish to briefly summarize some key findings of previous works that also investigated the California Planet Search Stars monitored with Keck-HIRES.

\citet{Wright2005} examined a subsample of the CPS stars without known planets and used an activity metric {\ensuremath{\Delta F_{\mbox{\scriptsize\ion{Ca}{2}}}}}, which accounts for the minimum activity of stars as a function of $B-V$ \citep{Rutten1984}, finding that RV jitter increases with this activity metric. Importantly, \citet{Wright2005} notes that K and G type stars show slightly lower levels of RV jitter. \citet{Isaacson2010} performed a similar analysis but used $\Delta$\SHK{} to observe trends with activity. One of the key results of this work was that the RV jitter of K dwarfs showed little dependence on magnetic activity and showed the overall lowest levels of RV jitter, representing a ``sweet spot" for exoplanet searches.

Our analysis builds on these previous works in several ways. First is our thorough approach to calculating RV jitter. \citet{Wright2005} removed known planet hosts from the sample, and \citet{Isaacson2010} did not give special treatment to planet hosts or possible companions. The resulting jitter values certainly contained dynamical velocities, which is why they simply investigated the jitter floor. By accounting for planets and other companions in a consistent, conservative manner, we have brought many of the artificially high points down to the jitter floor, thereby strengthening its significance. Further, we have the benefit of several years' worth of additional observations that give us a better handle on the long-timescale RV variability as well as better constraints on long period planets and other long-term trends. Tied to this is the years of observations on stars that were not previously in the California Planet Search sample, or had very few observations at the time of publication of either \citet{Wright2005} or \citet{Isaacson2010}. As mentioned previously, this includes the sample of ``Retired" A stars \citep{Johnson2007} as well as a sample of young, active stars which had been observed as part of the \emph{Spitzer} Legacy Program, Formation and Evolution of Planetary Systems \citep{Meyer2006}. These two samples are crucial for this work as they give us much stronger leverage on both activity and evolution. Finally, we restrict ourselves to stars with updated stellar properties from B17, which allows us to investigate more clearly how stellar RV jitter manifests for different stellar types in a more precise manner than previously possible.

\autoref{fig:all_rms_logg} shows the RV RMS of our sample in two different ways. First, as a function of $\logg$ and colored by their activity, \SHK. In this panel, we have cut off the y-axis to exclude stars that have large levels of jitter that are likely due to additional companions that cannot fully be constrained and to better show the trends among the low jitter stars. The second panel shows the same data but plotted as a function of activity, \SHK, and color-coded by their evolutionary state, $\logg$.

The first panel in \autoref{fig:all_rms_logg} has two immediately noticeable features: a vertical pileup of stars at high surface gravities ($\logg \sim 4.5$), and a horizontal pileup among the low surface gravity stars. By noting the color-coding, it is clear that the vertical pileup contains the active stars and the inactive stars are contained in the horizontal pileup. We can therefore easily see the two regimes of RV jitter --- activity-dominated and convection-dominated. The second panel in \autoref{fig:all_rms_logg} does not make as clear of a distinction between these two regimes. Instead, we see a general trend where jitter decreases with decreasing activity while stars are still in the main sequence (yellow-shaded points). Upon leaving the main sequence (orange/red points), we see that decreasing activity results in an increase in RV jitter. We note that because we are looking at the full sample spanning several spectral types, a clear relation between \SHK{} is not expected. Upon closer examination, \autoref{fig:all_rms_logg} (in particular the top panel) paints a general astrophysical picture of RV jitter evolution, as follows:

\begin{figure*}
\centering
\includegraphics[width=0.75\textwidth]{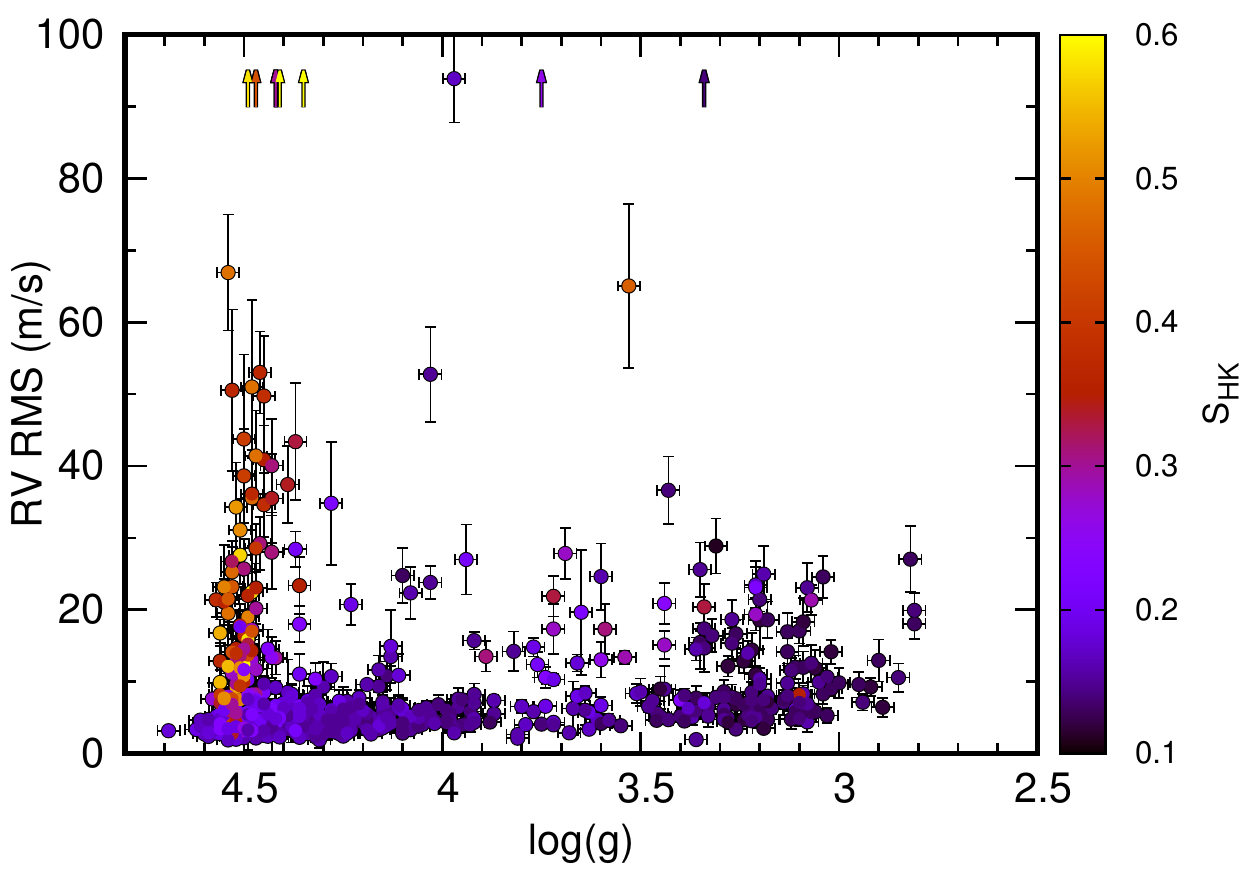}
\includegraphics[width=0.75\textwidth]{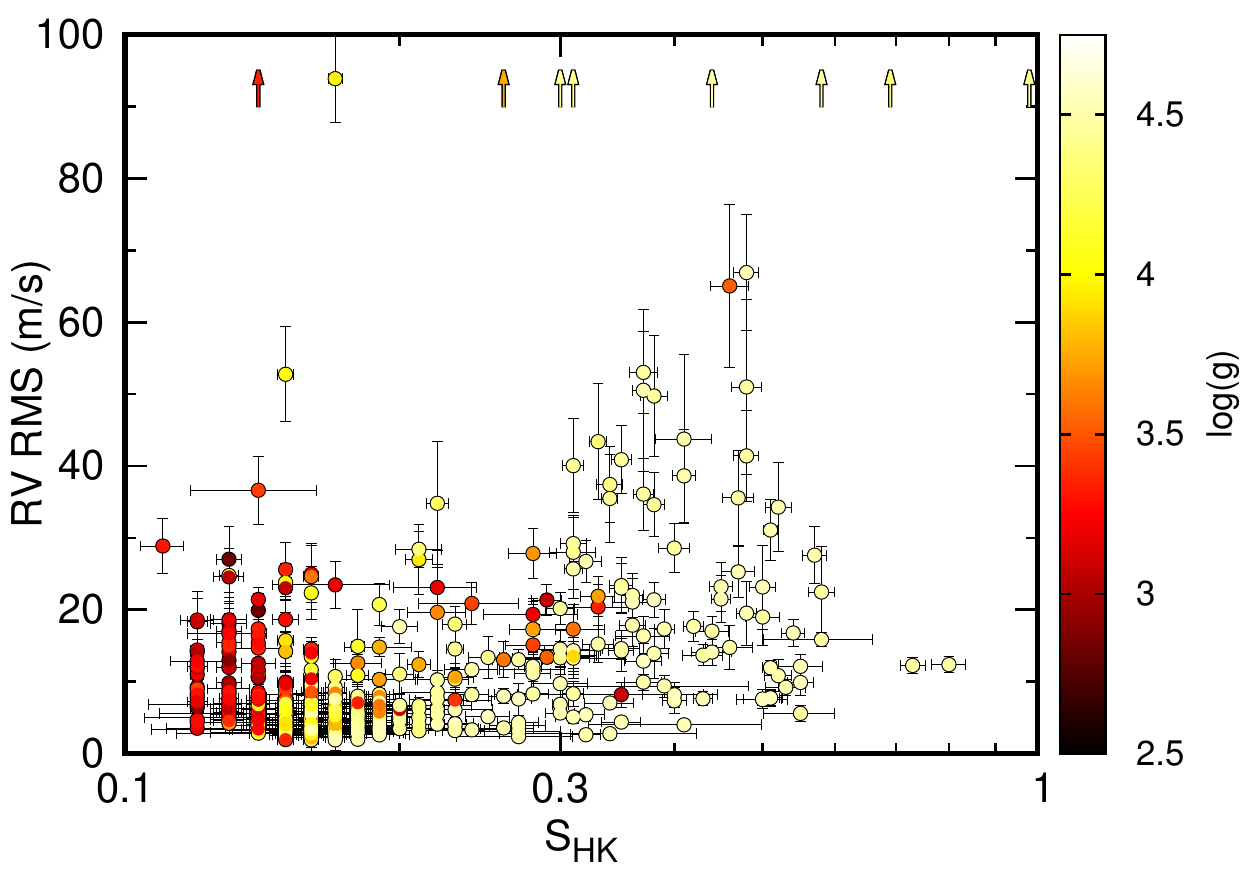}
\caption{\emph{Top}: RV jitter as a function of $\logg$ for our entire CPS sample. The y-axis has been capped at 100 m/s to show only those stars whose jitter are likely dominated by astrophysical sources rather than additional unsubtracted companions (upward-pointing arrows indicate the 8 stars with RV RMS above 100 m/s, two of which overlap in this plot with identical $\logg$ values of 4.42). The vertical error bars are calculated from \autoref{eqn:uncertainty} as described in \autoref{sec:jitter_error}. The horizontal error bars are uniformly 0.028 dex in $\logg{}$ as described in B17. The color bar shows activity metric \SHK{} as defined in \autoref{sec:activity_metric}. As a typical star evolves, it will begin at the top left of this plot with large surface gravity, high activity, and high jitter; it then moves vertically downward due to main sequence spin-down; as it evolves off the main sequence it transitions from activity-dominated jitter to convection-dominated jitter where the jitter shows a gradual increase with evolution. \emph{Bottom}: RV RMS as a function of \SHK{} for the stars in our sample. Horizontal error bars come not from individual measurement uncertainty for \SHK{} but instead are the standard deviation of the \SHK{} time series, effectively indicating how variable a star is in activity (typically from cycles). Points are colored by their surface gravity. Again, the 8 stars with RV RMS greater than 100 m/s are shown as upward arrows. As a star evolves it will begin on the right side of this plot with high jitter; it then moves diagonally down to the left as it spins down, eventually increasing again as an inactive giant star, when convective phenomena dominate the RV jitter.} 
\label{fig:all_rms_logg}
\end{figure*}

\begin{figure*}
\centering
\includegraphics[width=0.75\textwidth]{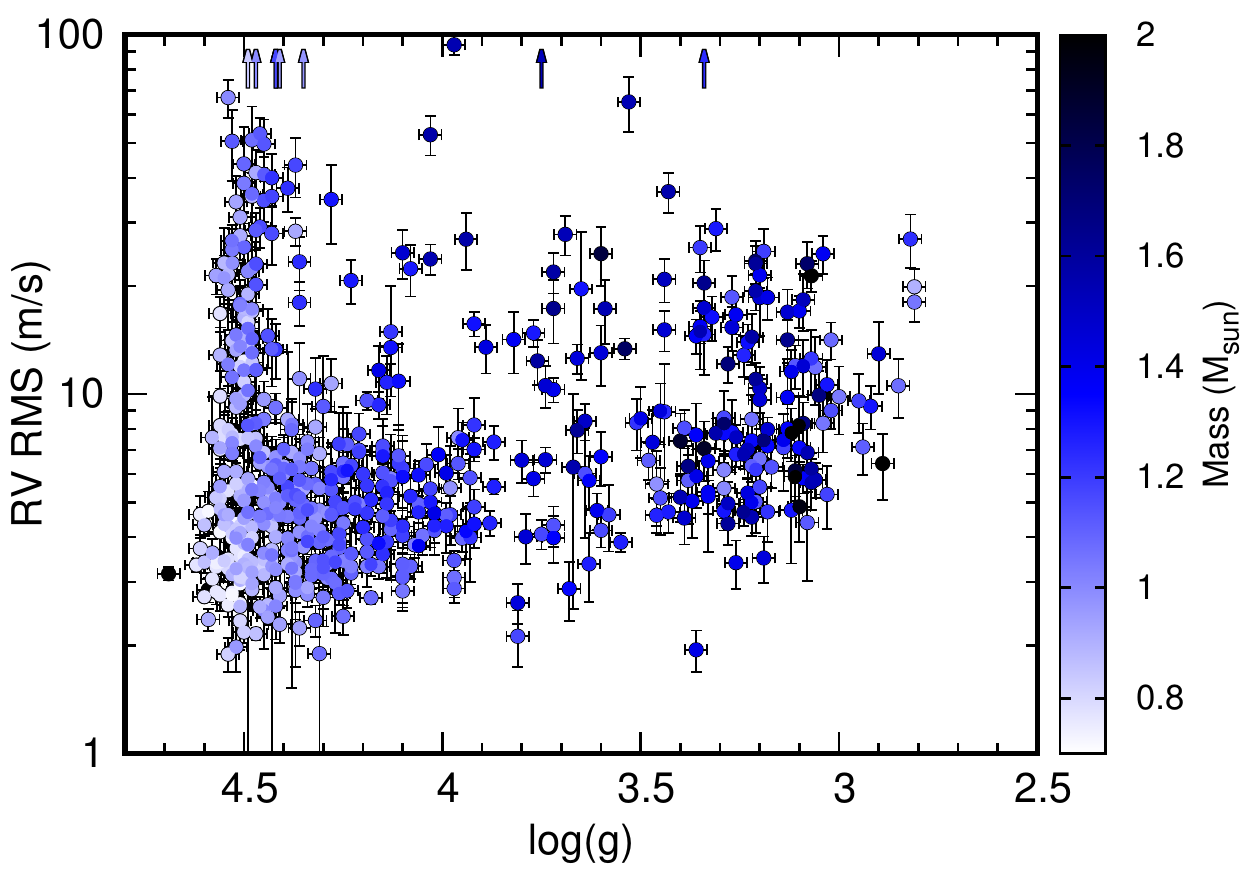}
\includegraphics[width=0.75\textwidth]{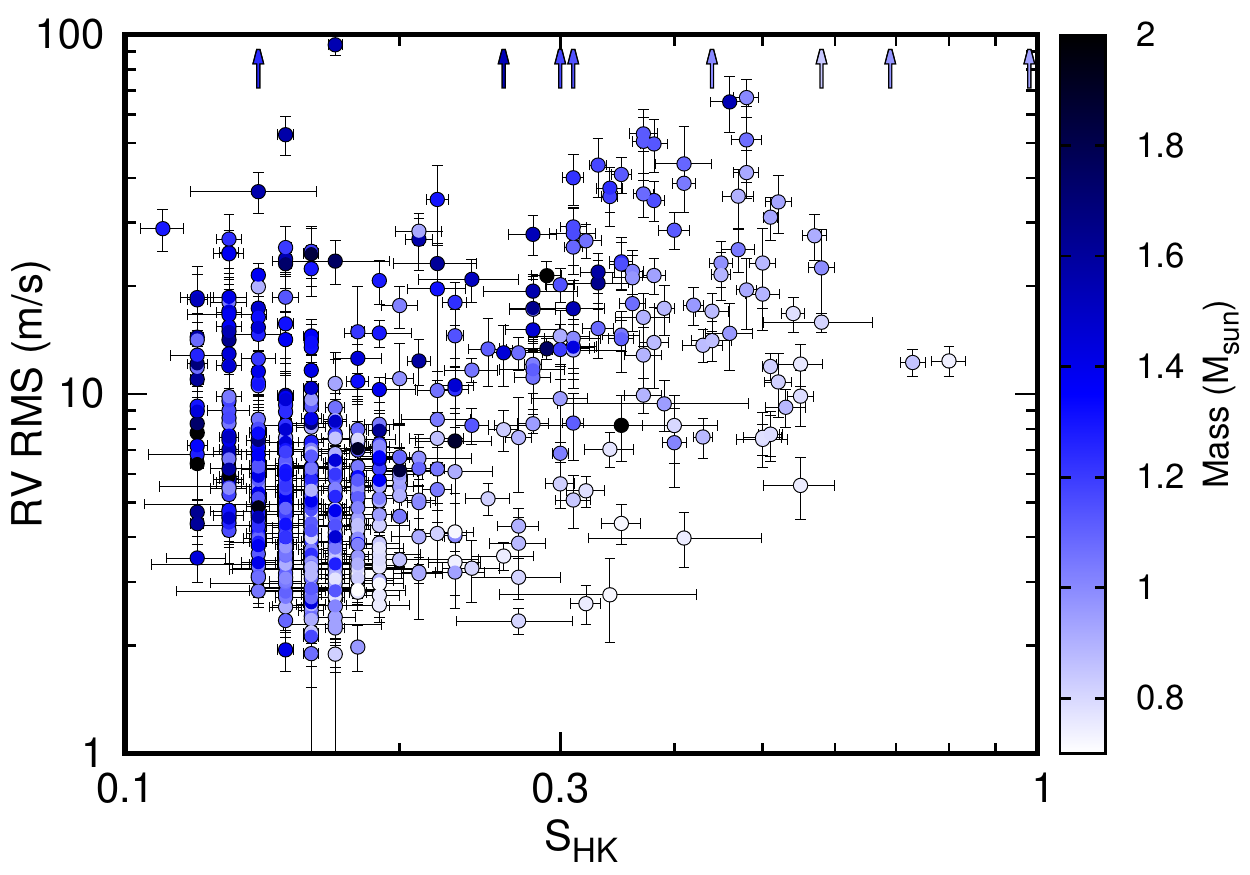}
\caption{Same as \autoref{fig:all_rms_logg}, but with the following changes: the y axes have been transformed to logscale and the points in both plots are now color coded by mass. The mass color-coding highlights two effects: 1) the shift in toward lower $\logg$ and higher RV RMS with higher mass star in the top panel, and 2) among active stars (above $\sim$0.25), the higher RV jitter associated with higher mass seen in the bottom panel. These plots motivate the analysis by mass bin to see trends more clearly as a function of mass.} 
\label{fig:all_rms_logg_2}
\end{figure*}

\paragraph{The Main Sequence}
The main sequence is easily identified by the vertical pileup in the left of the upper panel of \autoref{fig:all_rms_logg}, at $\logg \sim4.5$. These main sequence stars are further distinguished by the large fraction of active stars (\SHK $\,\gtrsim0.25$) in this portion of the plot. Given the known correlation between RV jitter and activity \citep{Campbell1988,Saar1998,Santos2000,Wright2005,Isaacson2010}, this is unsurprising based on general main sequence evolution \citep[e.g.,][]{Mamajek2008}. We expect that stars start out on the top left of this diagram (at the top of the main sequence here), as rapidly rotating, active stars with high surface gravity characteristic of the zero age main sequence (ZAMS). As they live out their lives on the main sequence, they lose angular momentum to magnetic winds, spin down, and become less active. As a result, they become quieter in RV observations and are seen to have less jitter. Therefore, a given star's path on the main sequence is to drop vertically as it becomes less active and less jittery. Eventually, as stellar evolution progresses, its surface gravity drops and it enters the subgiant and giant regime where it tends to be inactive \citep{Wright2004b}. A typical surface gravity for a terminal age main sequence (TAMS) Sun-like star is $\logg \approx 4$. We emphasize that a portion of the inactive horizontal floor seen in the upper panel of \autoref{fig:all_rms_logg} is during a star's final main sequence evolution. Therefore we find it useful to distinguish between the main sequence (the time in which the star is burning hydrogen in its core) and the phase in which the star is in the vertical portion of \autoref{fig:all_rms_logg}. We introduce the term ``active main sequence" when we wish to refer specifically to the vertical pileup of active stars. Stars on the ``active main sequence" are stars whose jitter is dominated by magnetic activity.

\paragraph{Subgiants and Giants}
The remainder of a star's life (at least until into the early giant phase, beyond which we cannot probe with this sample) is spent moving mostly horizontally to the right in the upper panel of \autoref{fig:all_rms_logg}. These stars are no longer active, as they have spun down to the point that they have very weak magnetic fields and therefore little chromospheric emission. 
However, we see a noticeable increase in RV jitter as stars evolve and their convective power increases. Among the giants and subgiants, stars with lower $\logg$ show higher levels of RV jitter than do stars with higher $\logg$, a result expected and seen by \citet{Wright2005}, \citet{Kjeldsen2011} and \citet{Bastien2014}. Since these are almost entirely inactive stars that have fully spun down, their jitter is dominated by convection, through a combination of granulation and oscillations. The increase in RV RMS with decreasing $\logg$ can be more easily seen when the y-axis is plotted in log-scale, as seen in \autoref{fig:all_rms_logg_2}, where we now color-code by mass.

\vspace{12pt}
We therefore see from \autoref{fig:all_rms_logg} that RV jitter tracks stellar evolution as stars transition from active to inactive stars and then exhibit increased convective power as they continue to evolve. From \autoref{fig:all_rms_logg_2}, we see color gradients that indicate strong mass dependencies, namely the lower $\logg{}$ and higher RV jitter with increased mass and the increase in RV jitter with mass among the most active stars. To examine these trends more closely, we divide our sample into mass bins. 

\begin{figure*}
\centering
\includegraphics[width=\columnwidth]{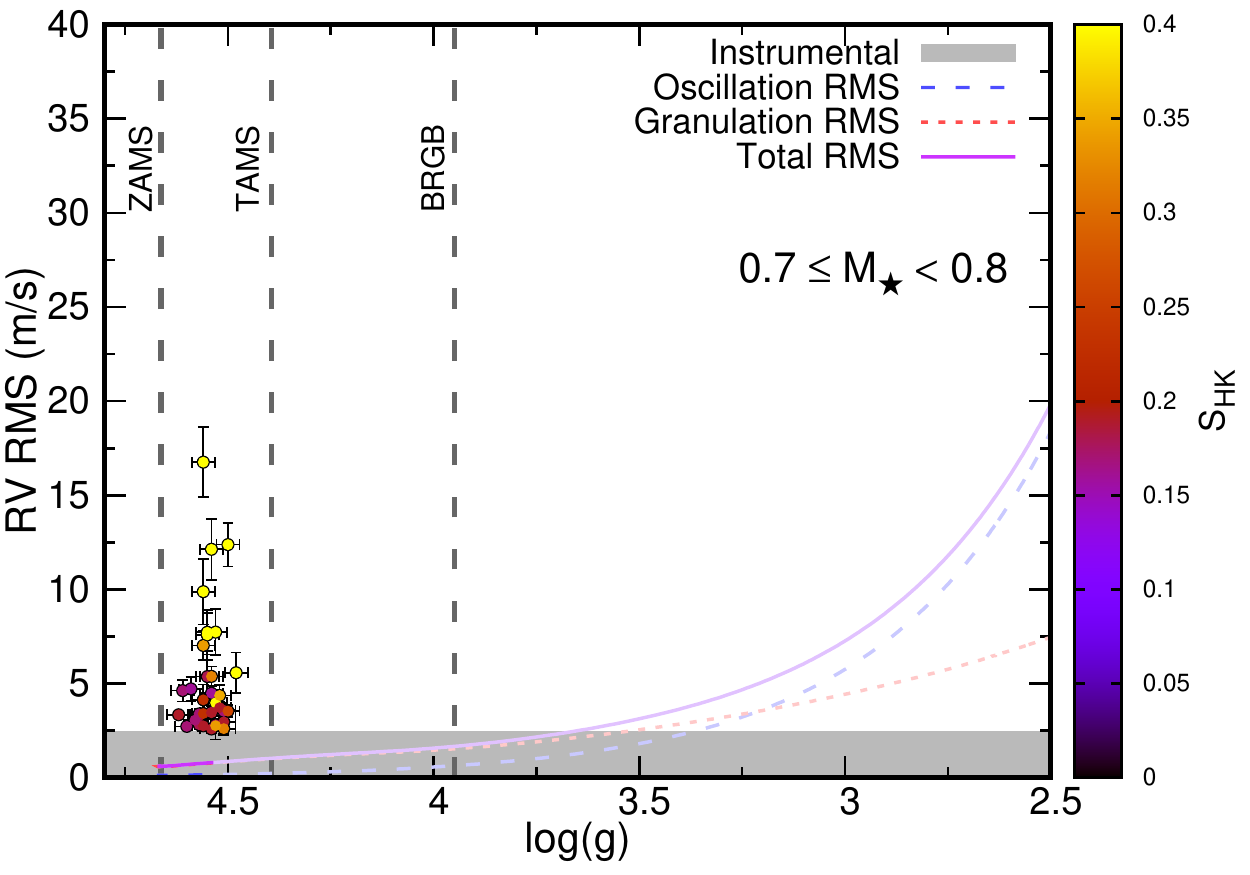}
\includegraphics[width=\columnwidth]{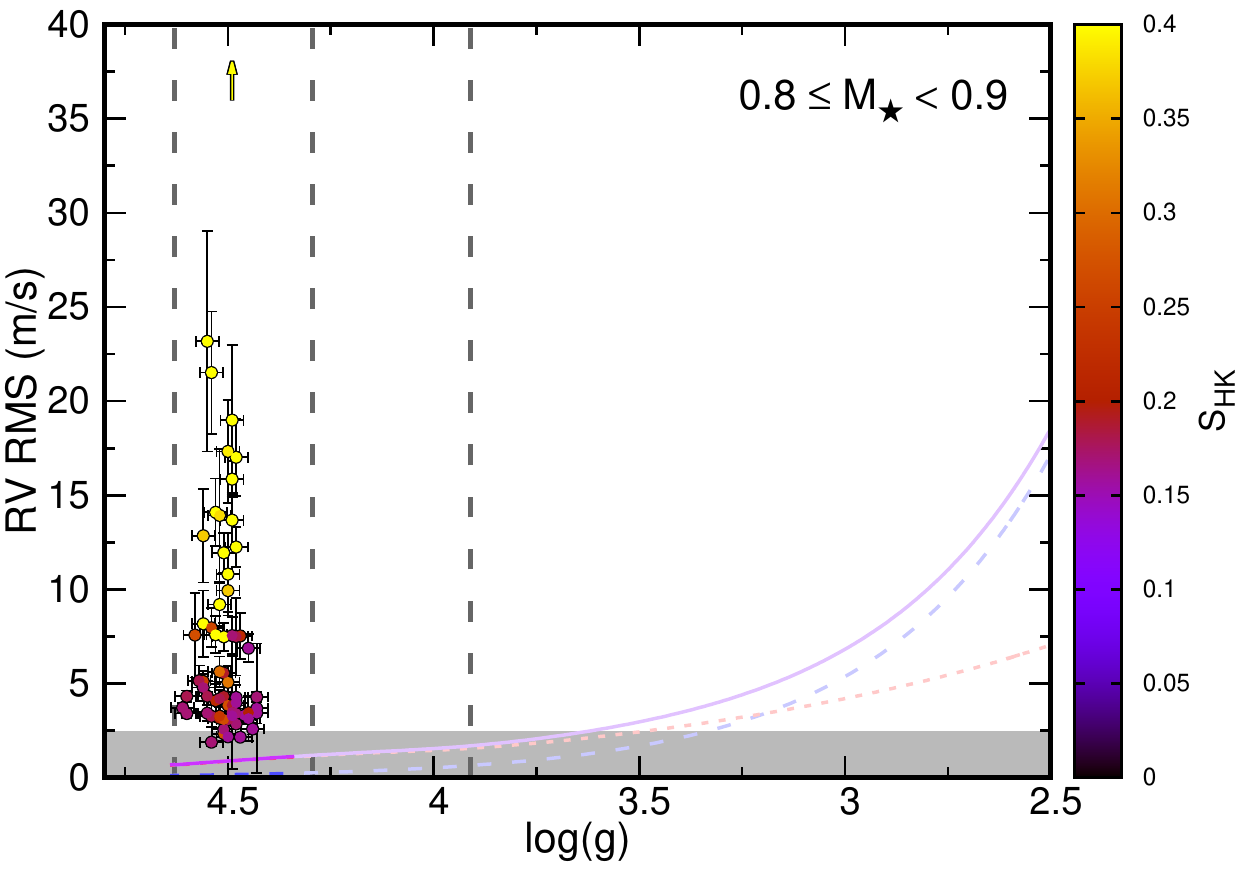}
\includegraphics[width=\columnwidth]{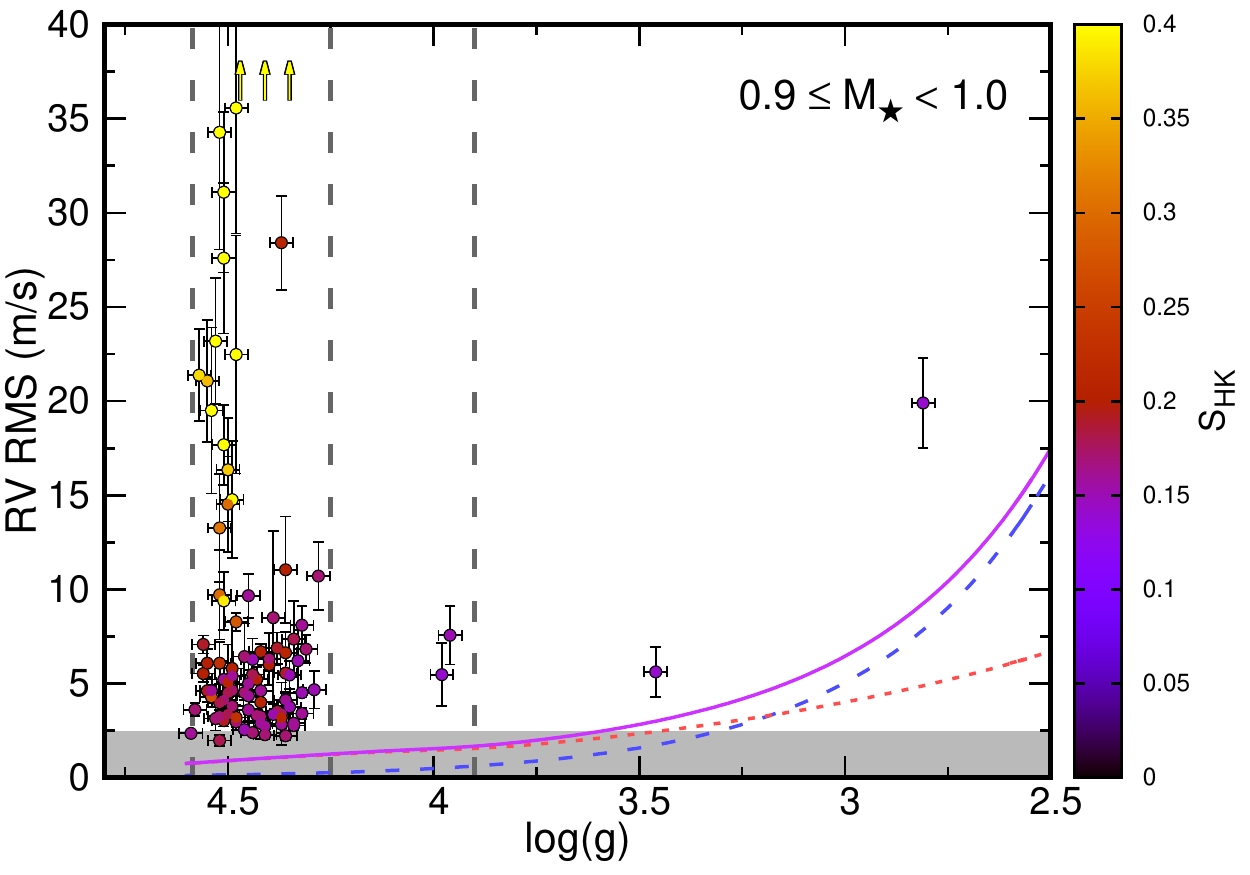}
\includegraphics[width=\columnwidth]{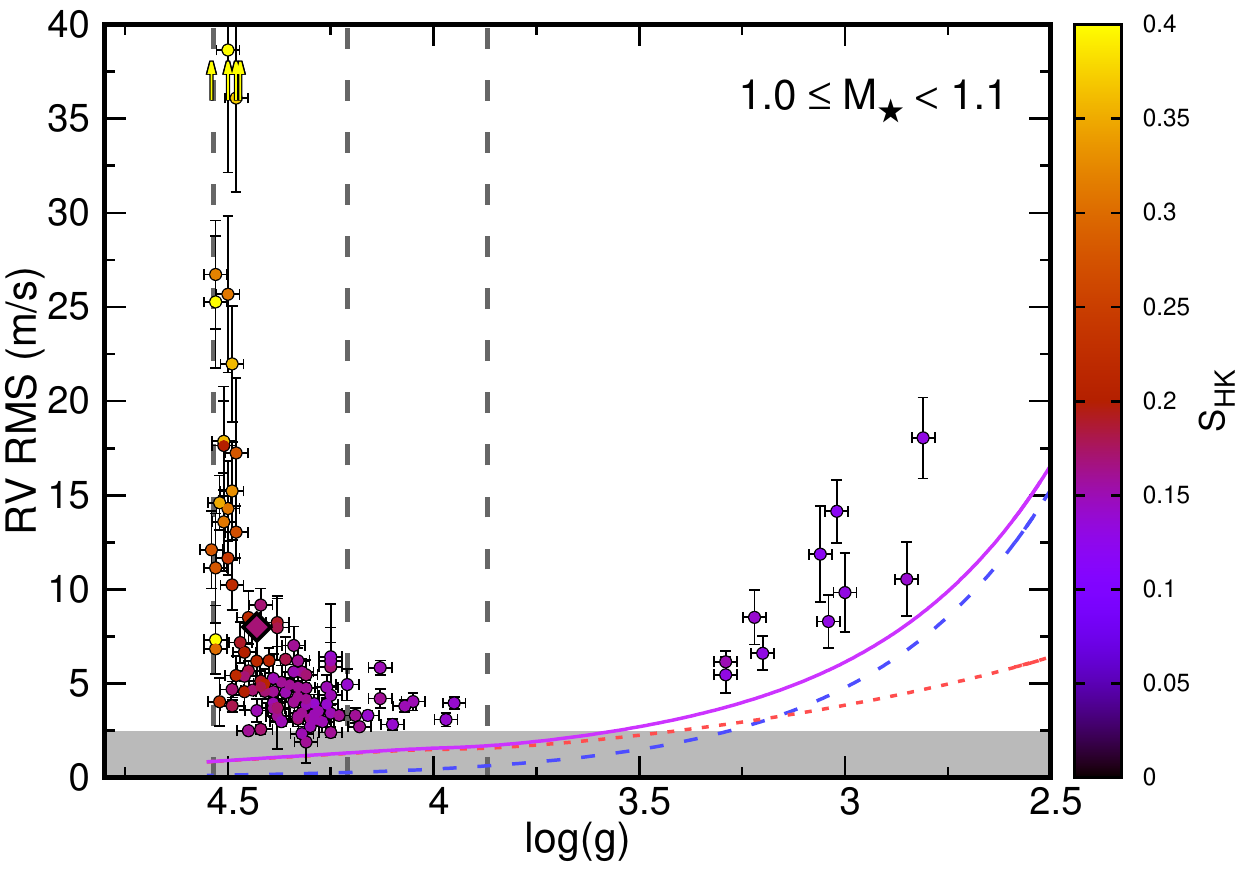}
\includegraphics[width=\columnwidth]{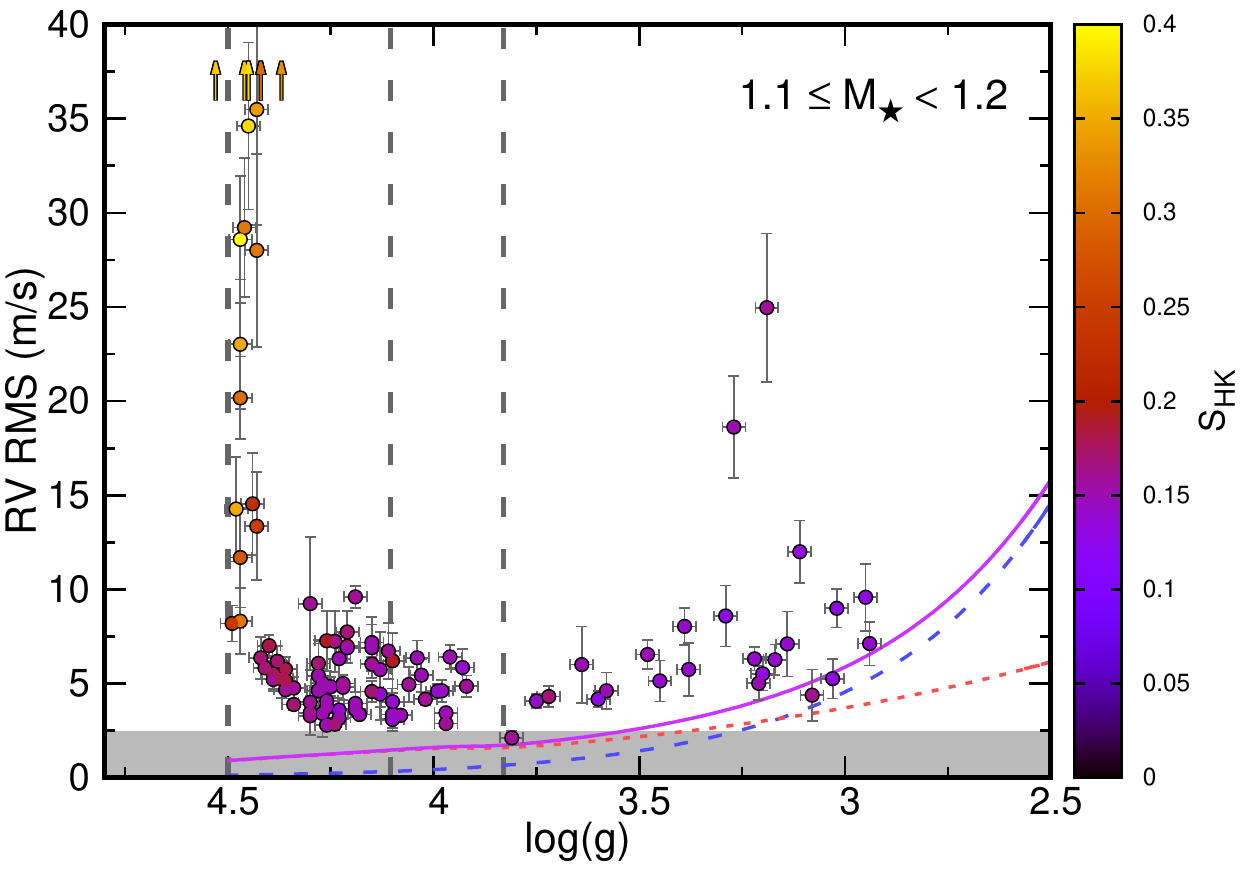}
\includegraphics[width=\columnwidth]{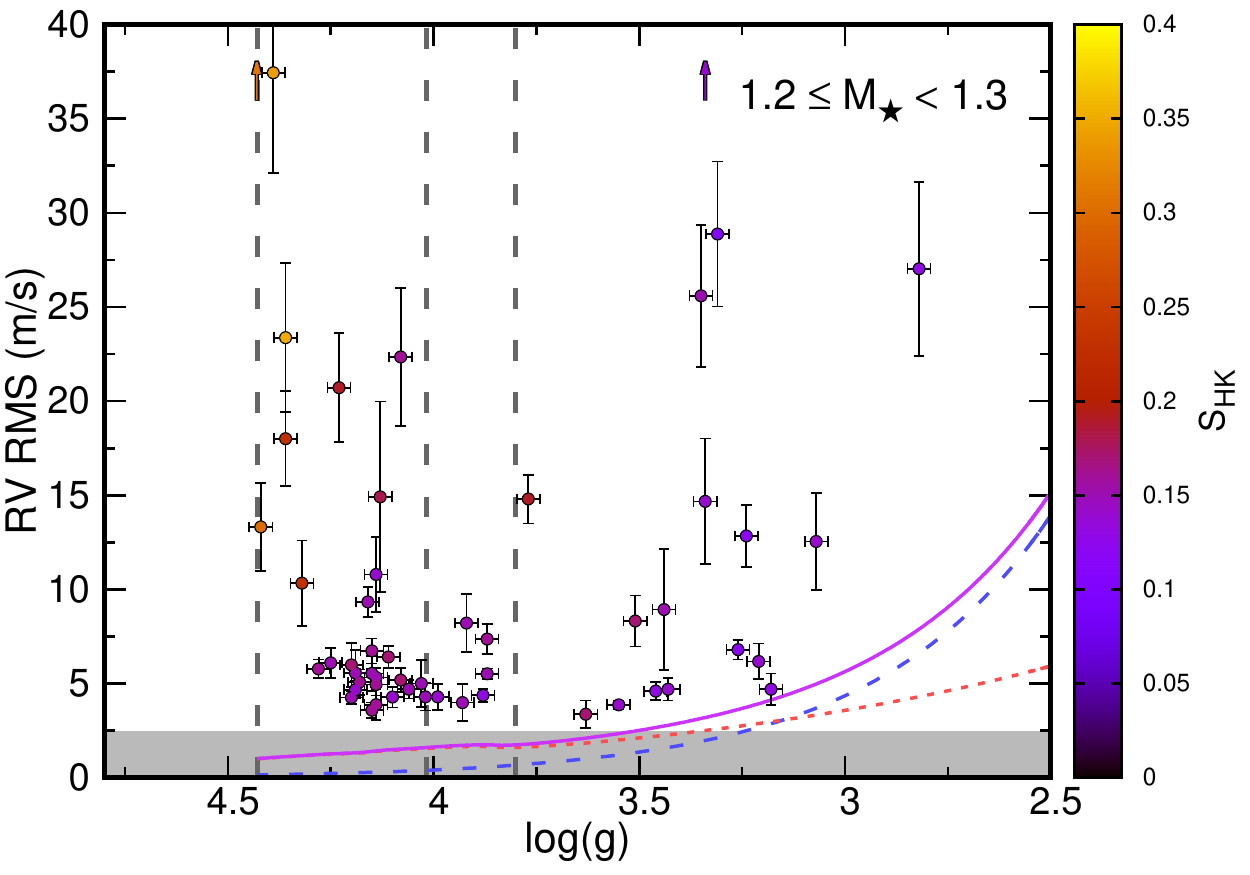}
\caption{Same as first panel of \autoref{fig:all_rms_logg} but separated by mass. The y-axis covers a smaller range to highlight the slope with $\logg$. Further, the color scale has been decreased to highlight the transition from the vertical activity-dominated regime to the horizontal convection-dominated regime. Solar values are plotted in the 1.0 to 1.1~M$_{\odot}$ plot as a large diamond ($\logg = 4.43$, RV RMS = 8 m/s). We use the solar RMS found in \citet{Meunier2010} and the activity-averaged \SHK{} from \citet{Egeland2017}. The vertical dashed lines plot the theoretical zero-age main sequence (ZAMS) for the lowest mass star and the theoretical terminal-age main sequence (TAMS) for the highest mass star in a given bin. The shaded region at the bottom of each plot shows the typical Keck-HIRES instrumental uncertainty, which has not been subtracted out from our calculation of RV jitter. The dashed blue and red lines show the RV jitter components for oscillation and granulation, respectively, as given by Equations \ref{eqn:oscillation_RMS} and \ref{eqn:granulation_RMS}. Their total contribution to RV jitter (added in quadrature) is shown by the purple line. The theoretical lines in the first two panels have been grayed out for surface gravities where the stellar model is older than the age of the universe. See \autoref{fig:rms_logg2} for the remaining high mass bins. }
\label{fig:rms_logg1}
\end{figure*}

\begin{figure*}
\centering
\includegraphics[width=\columnwidth]{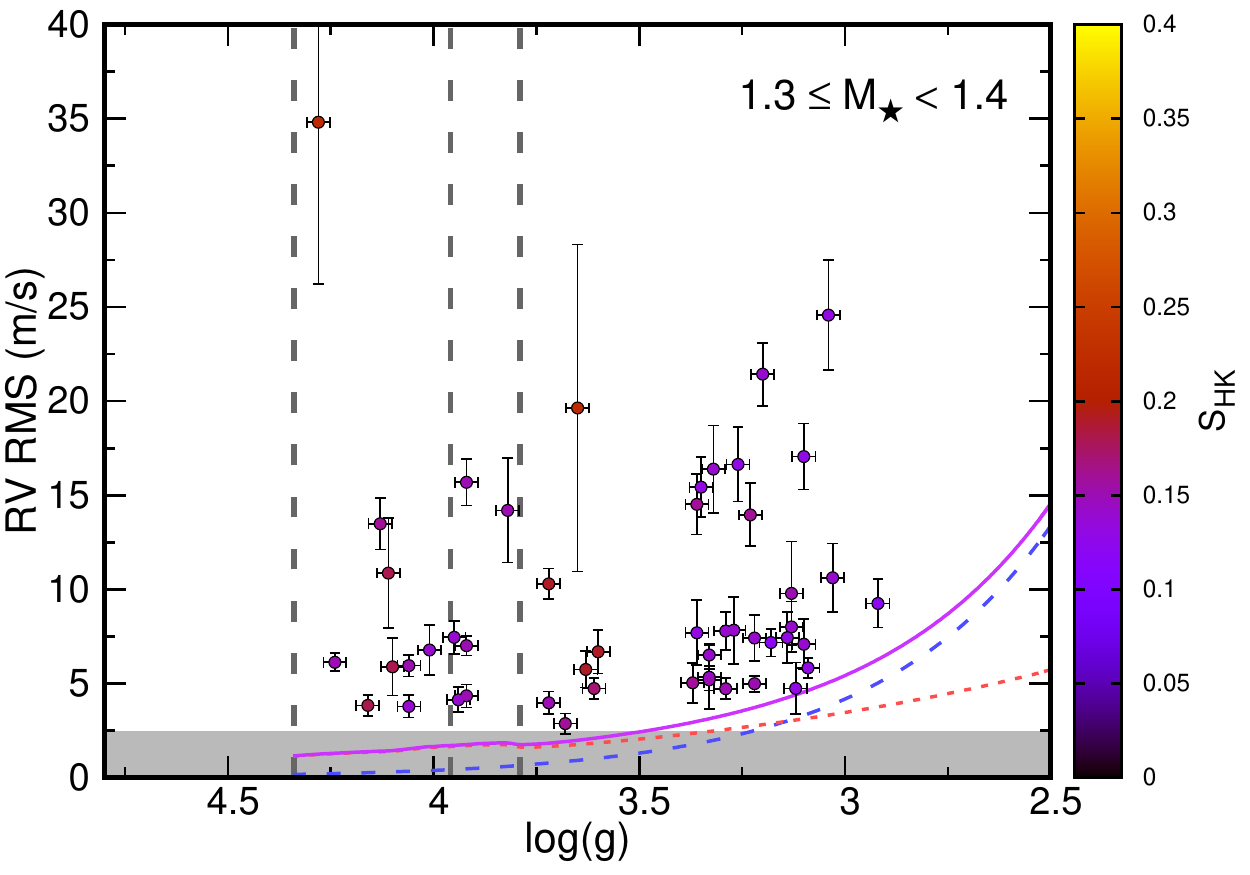}
\includegraphics[width=\columnwidth]{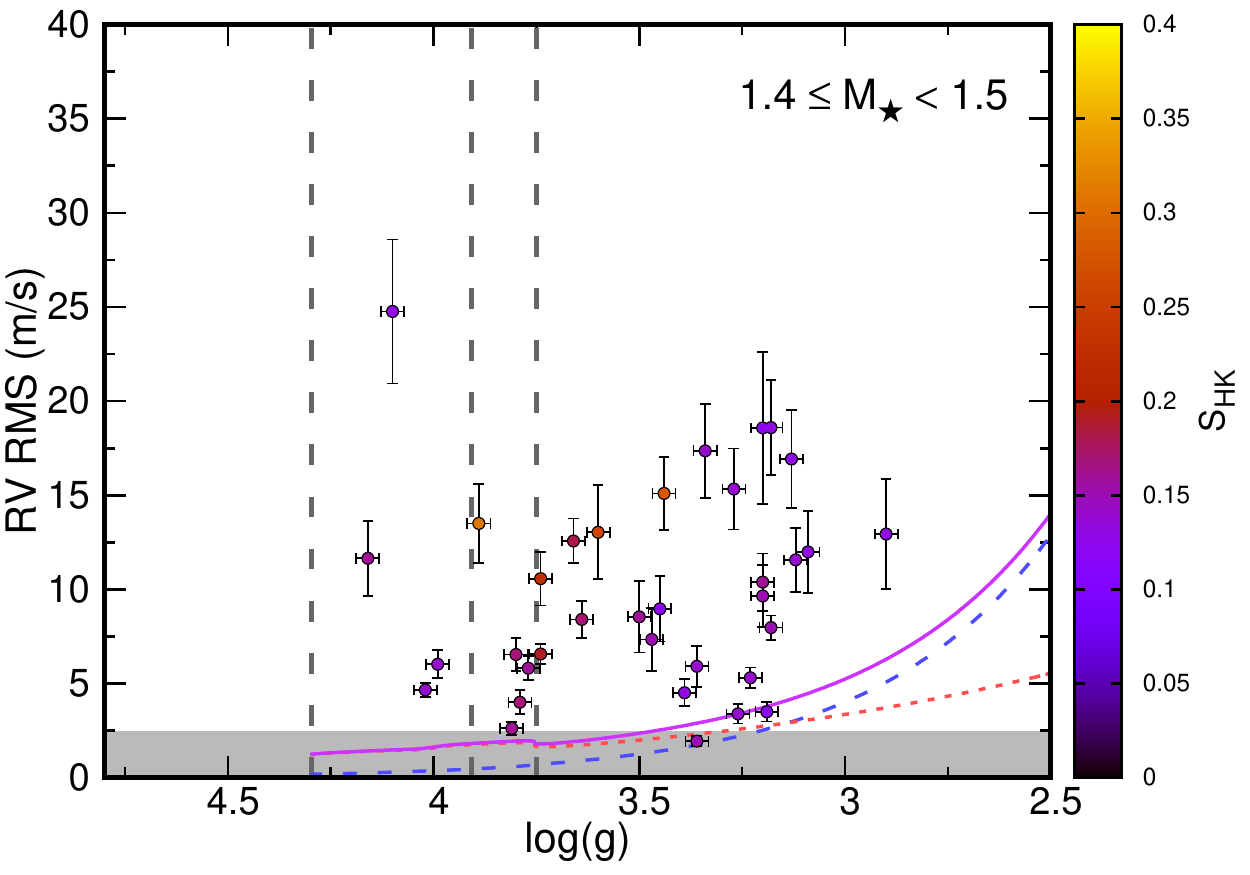}
\includegraphics[width=\columnwidth]{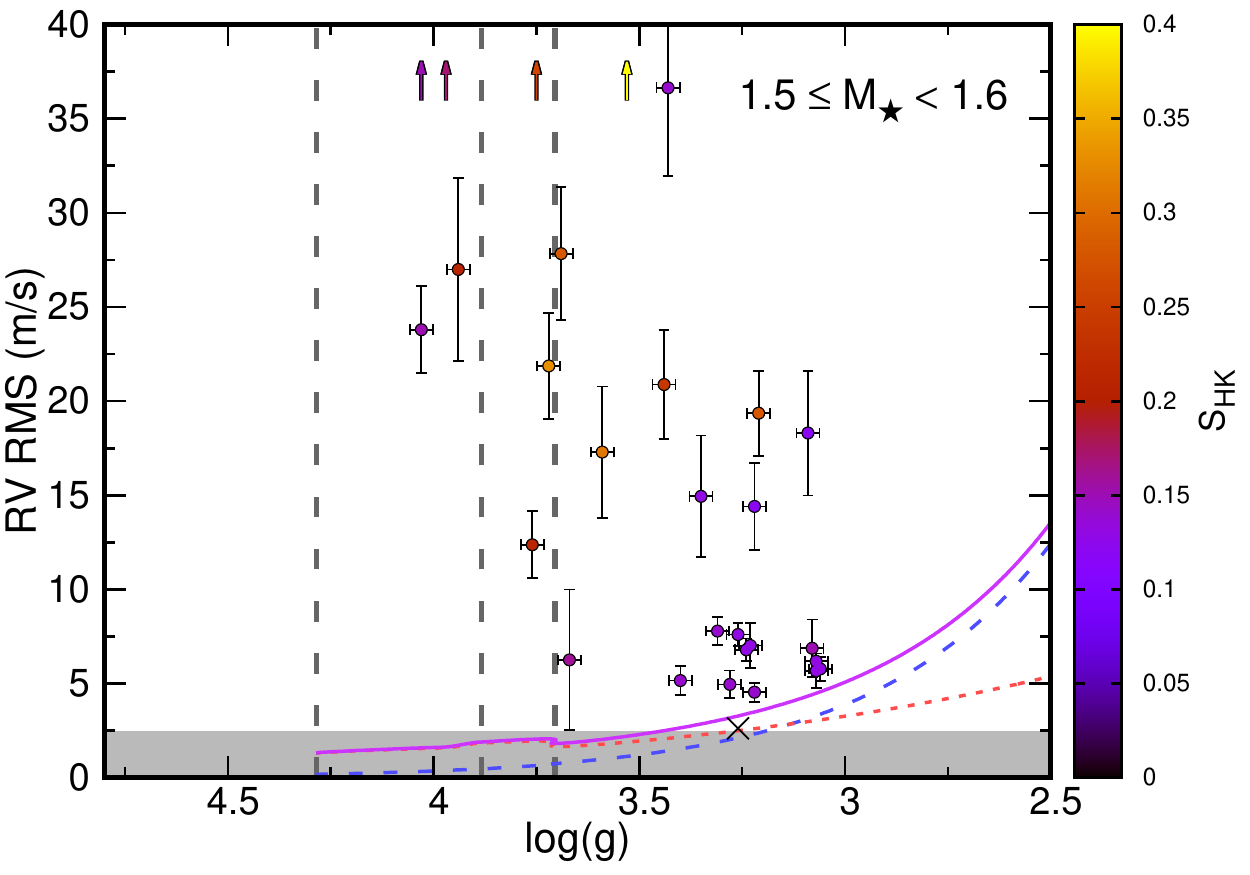}
\includegraphics[width=\columnwidth]{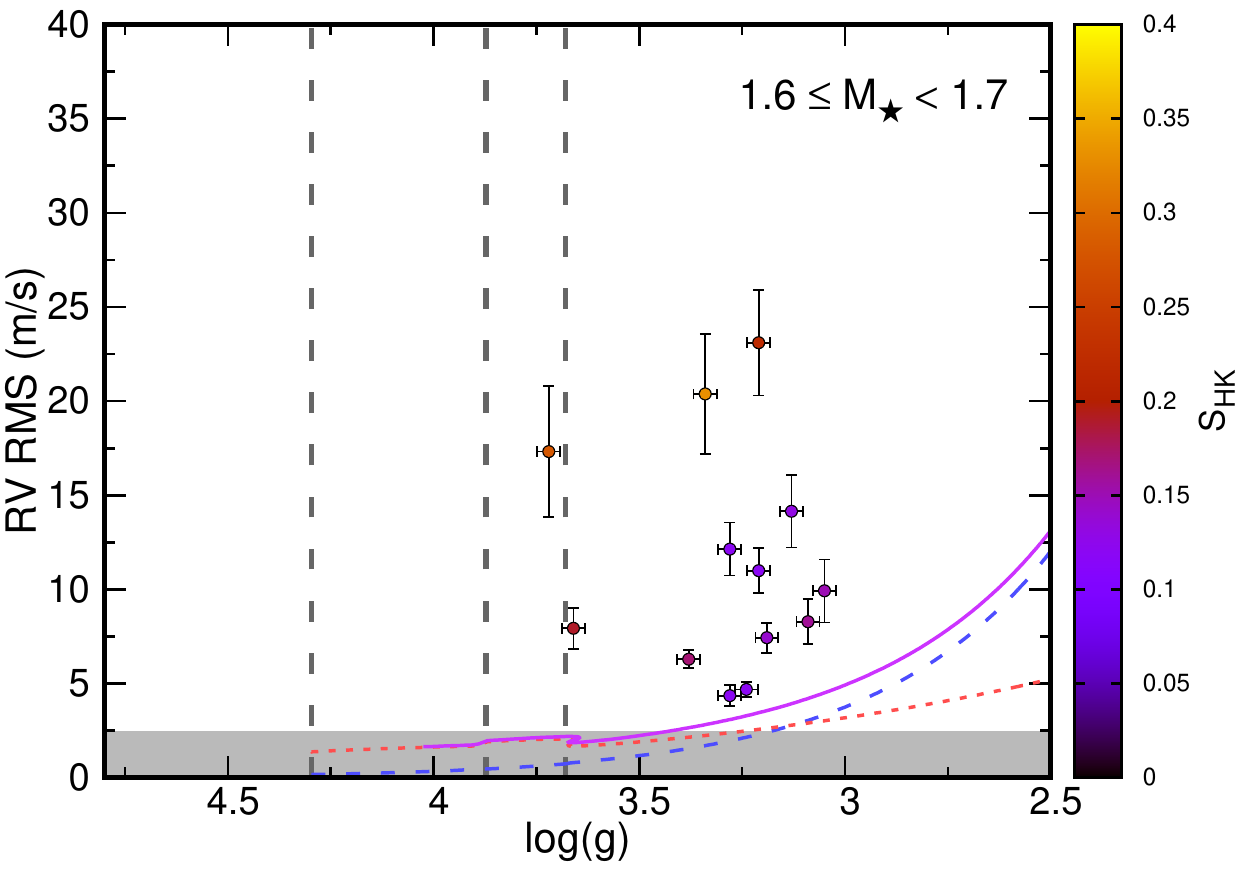}
\includegraphics[width=\columnwidth]{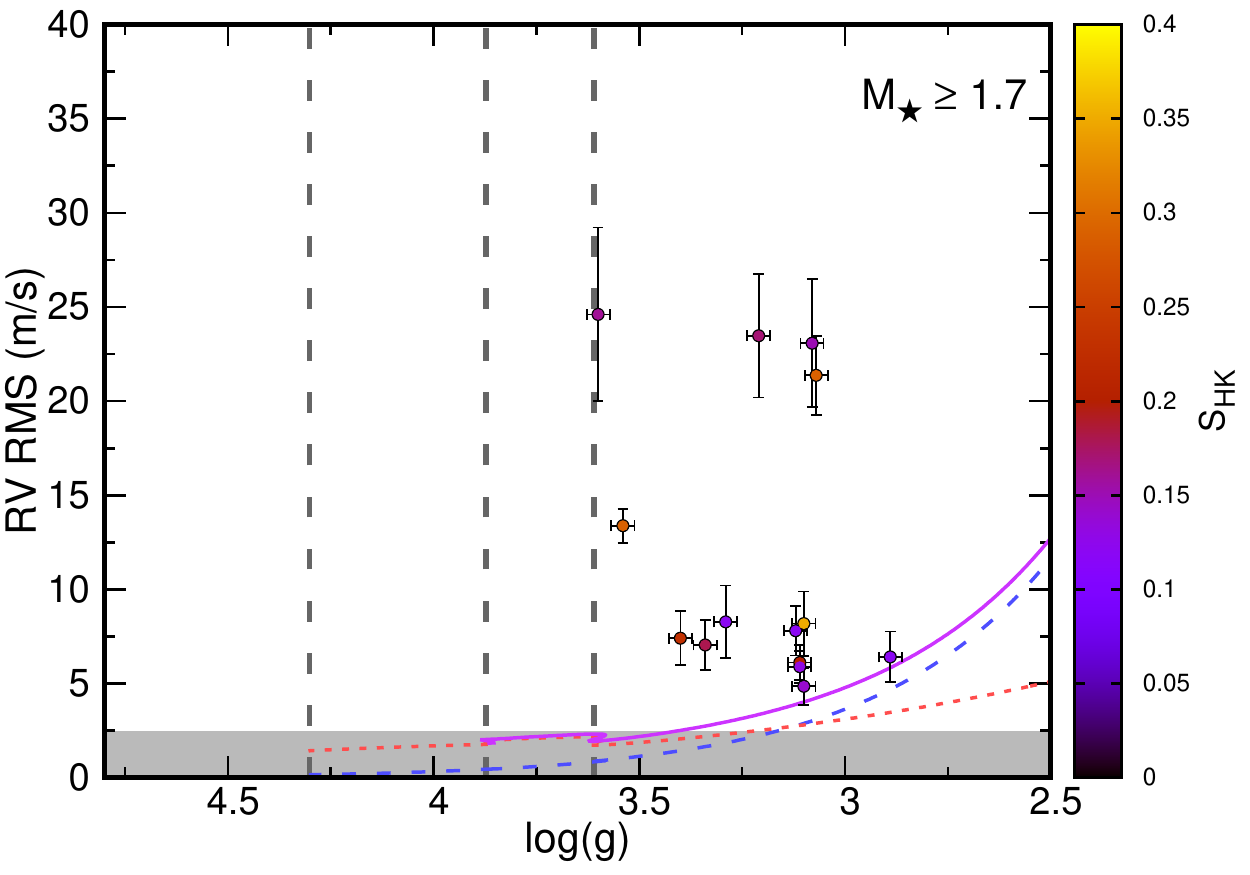}
\caption{Same as \autoref{fig:rms_logg1}, for larger masses. We note the `X' in the 1.5 to 1.6 M$_{\odot}$ bin, which shows the RV RMS of two nights of targeted observations of HD 142091 to observe stellar p-mode oscillations. This serves as a validation of the theoretical scaling relation for the oscillation component (blue dashed line) of the RV RMS and is described in more detail in \autoref{sec:oscillation_validation}.}
\label{fig:rms_logg2}
\end{figure*}

\begin{figure*}
\centering
\includegraphics[width=\columnwidth]{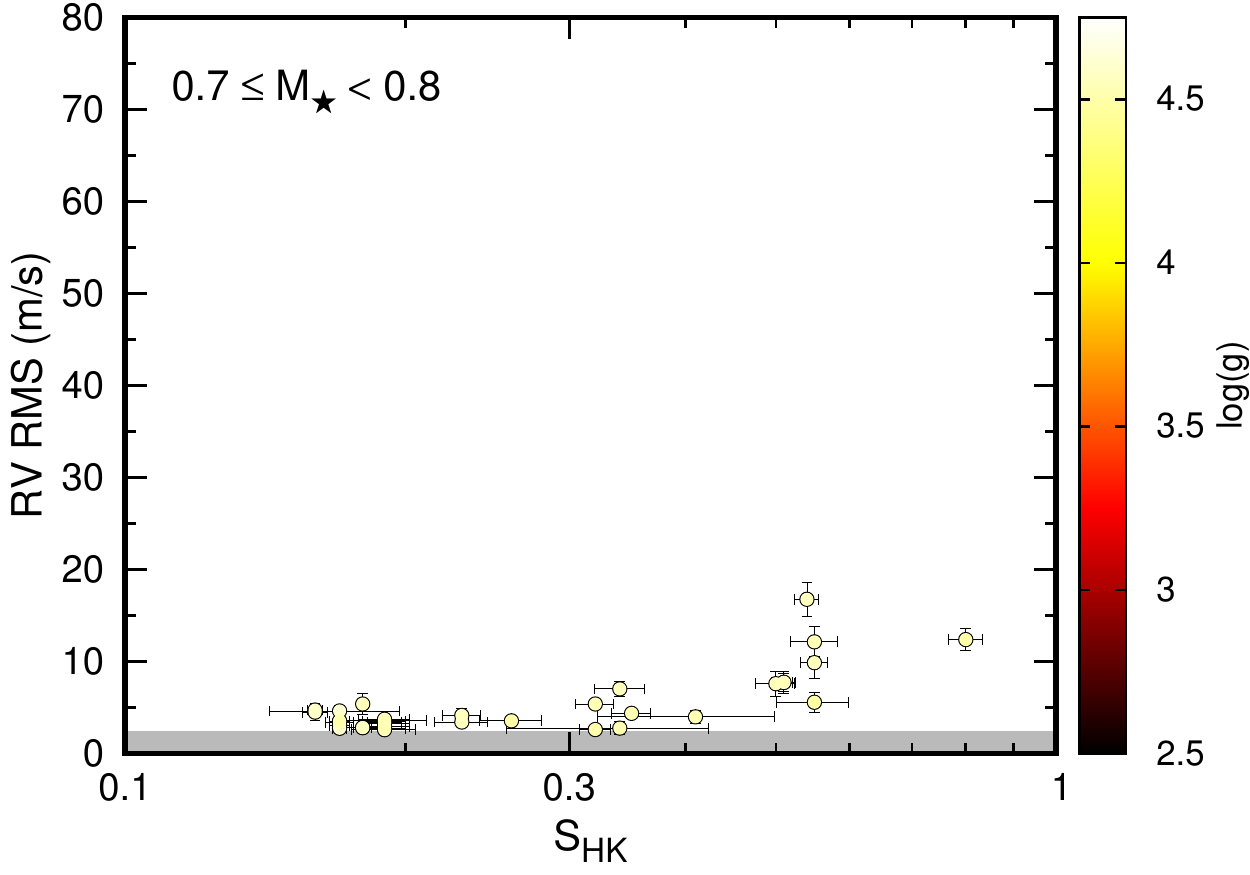}
\includegraphics[width=\columnwidth]{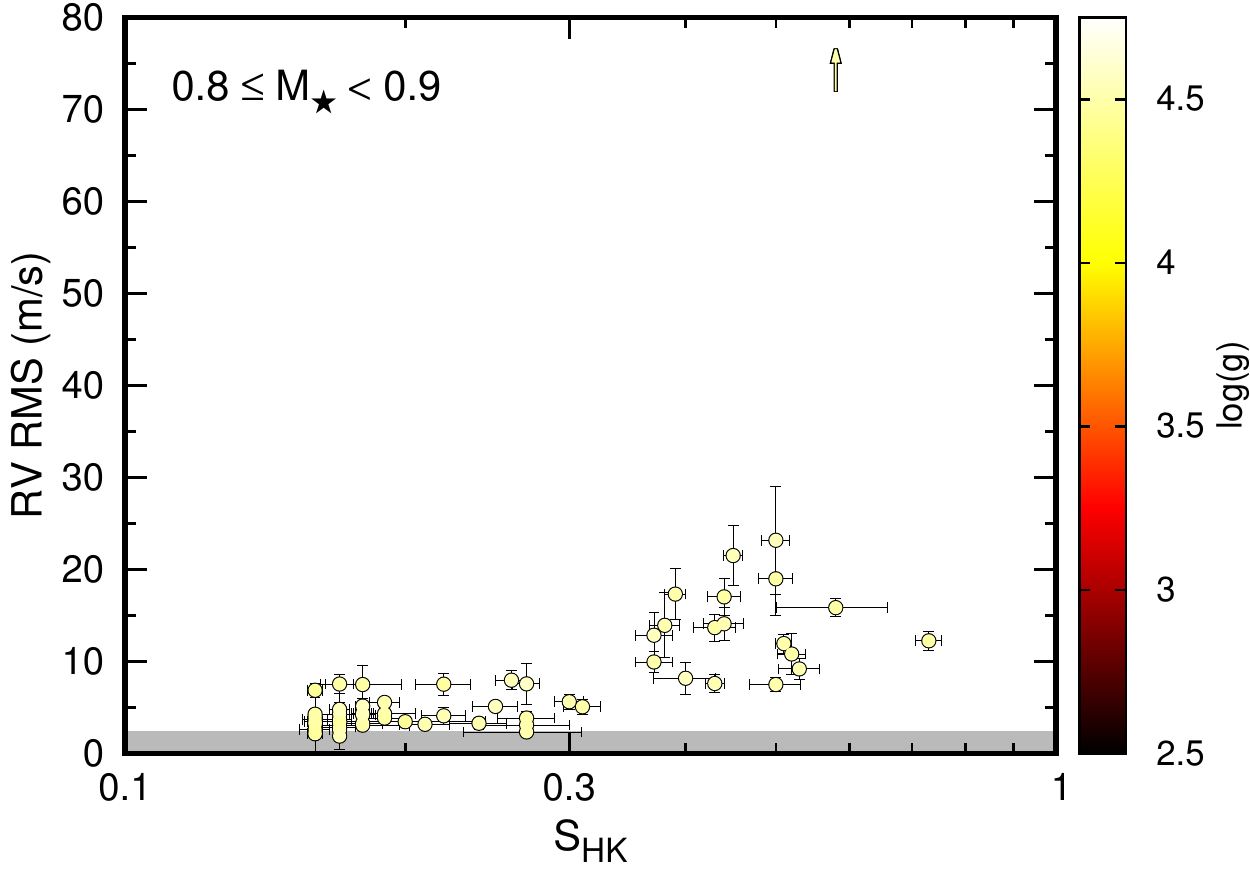}
\includegraphics[width=\columnwidth]{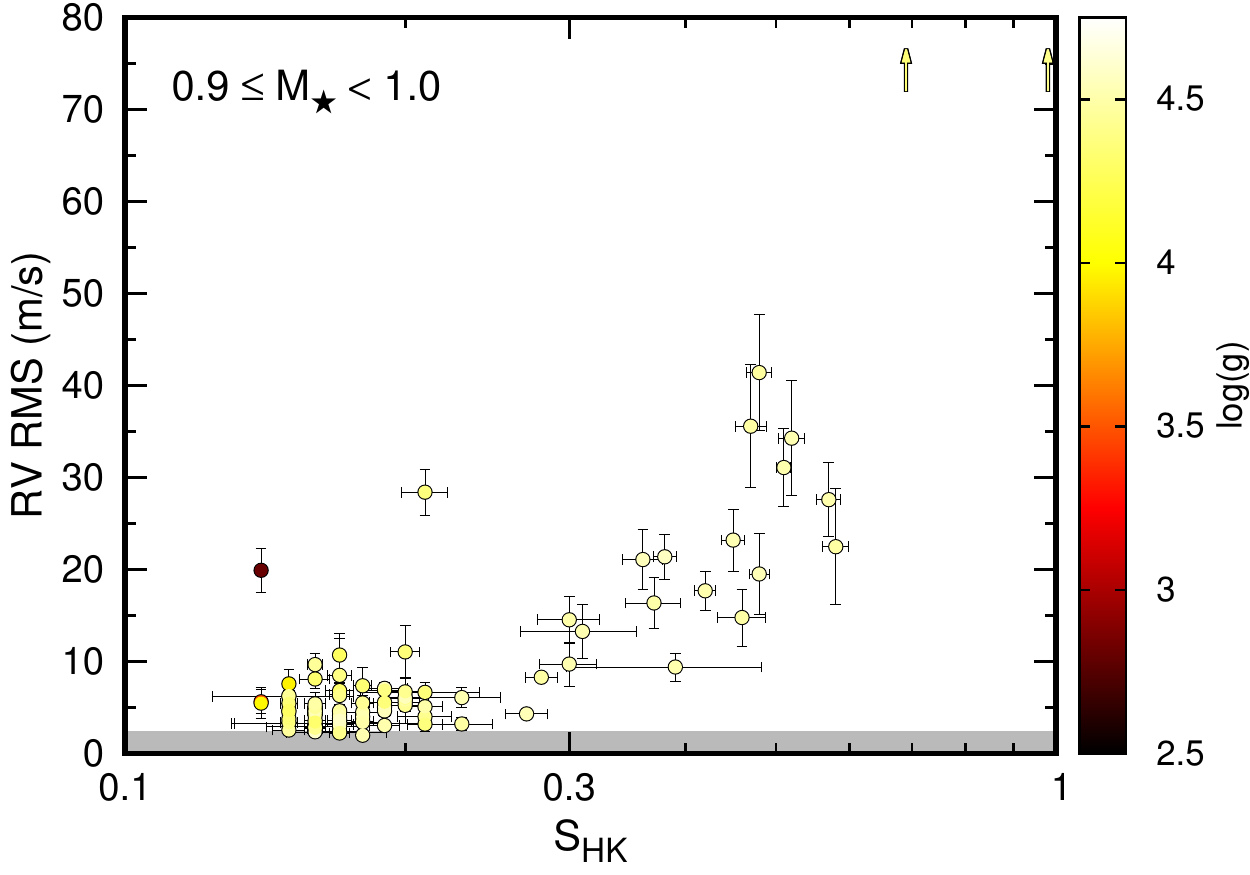}
\includegraphics[width=\columnwidth]{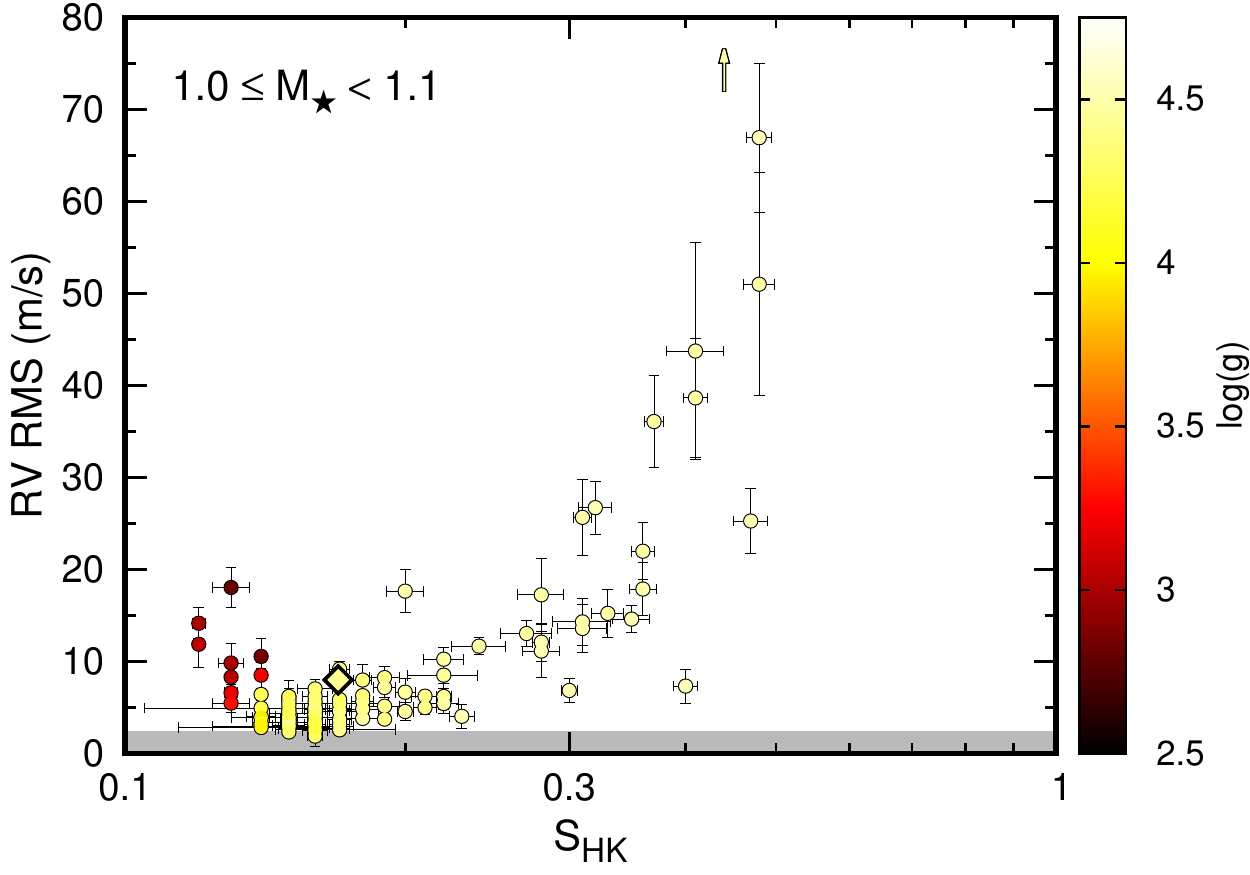}
\includegraphics[width=\columnwidth]{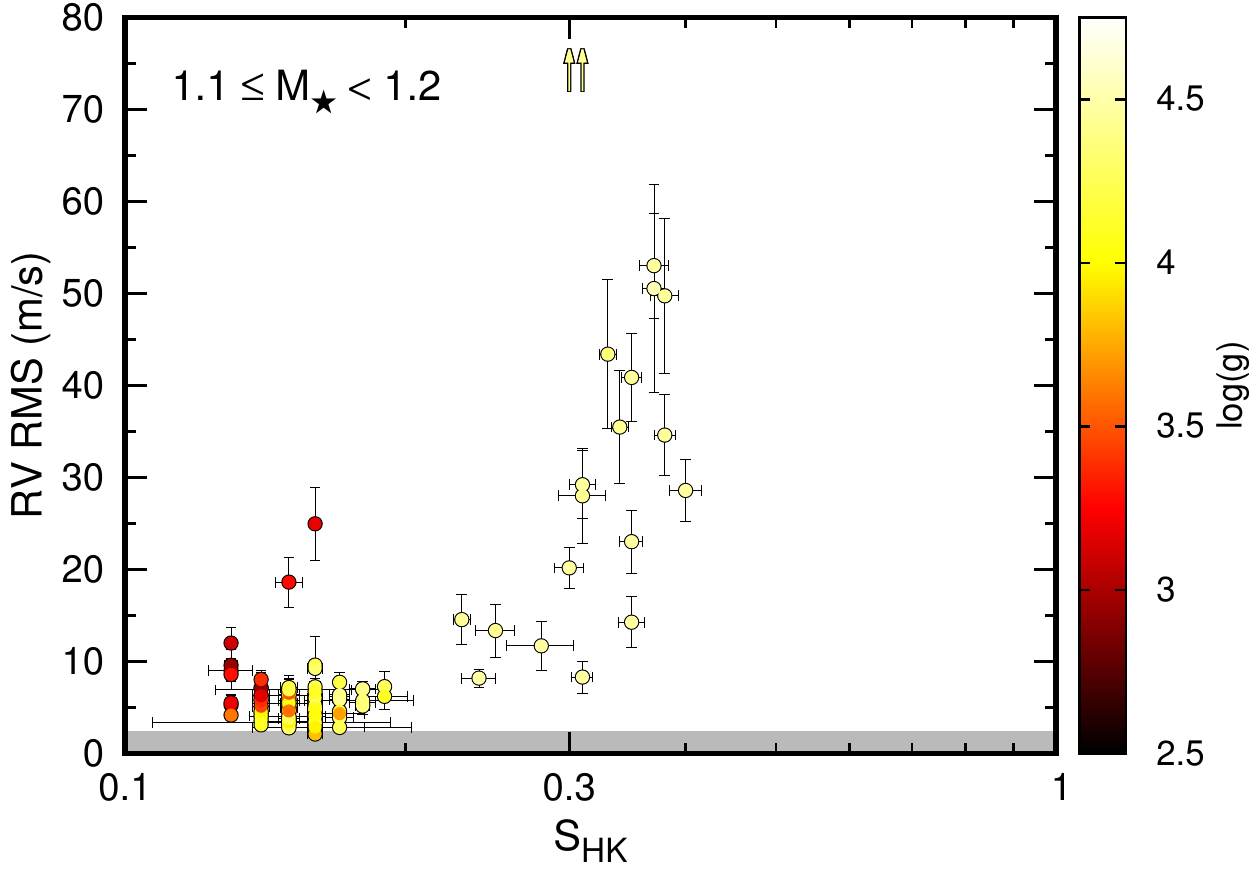}
\includegraphics[width=\columnwidth]{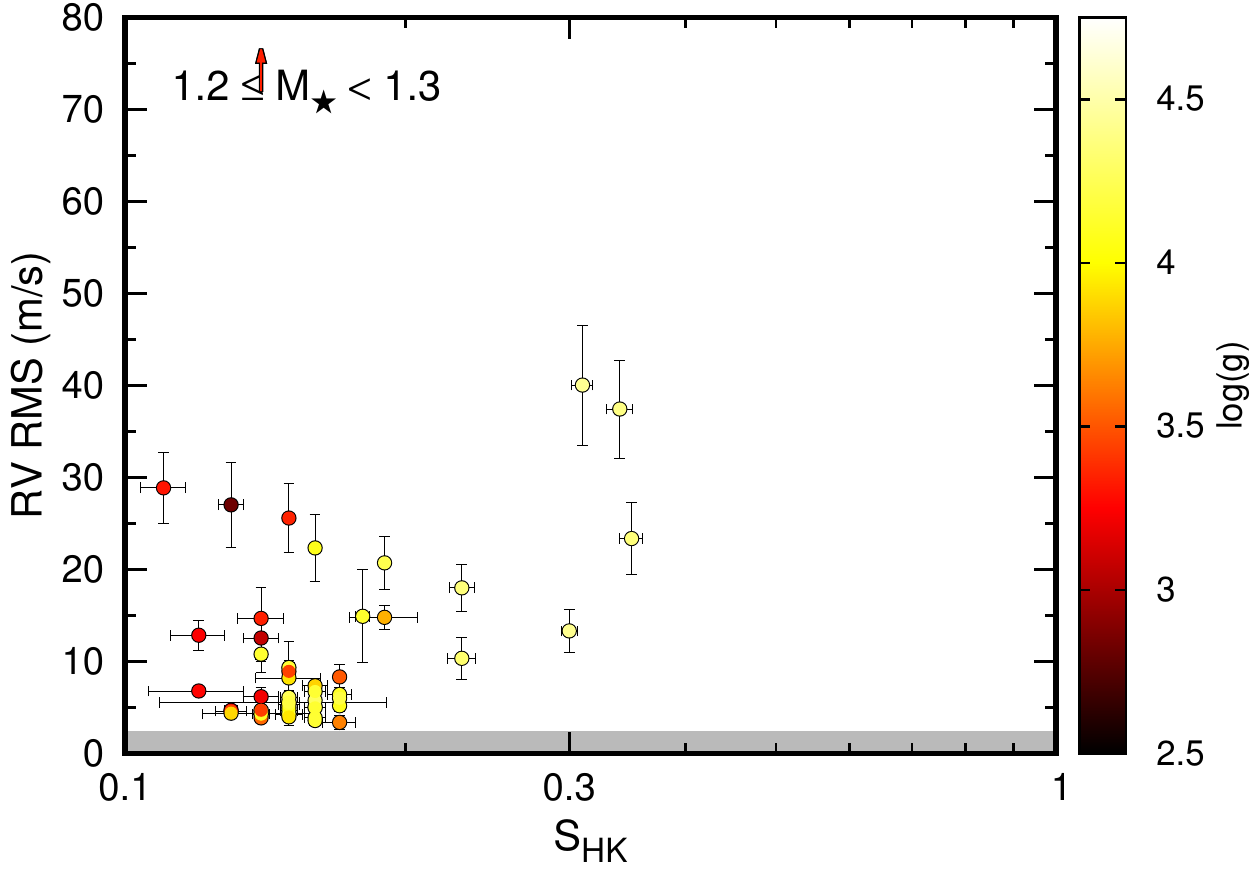}
\caption{Same as the second panel of \autoref{fig:all_rms_logg} but separated by mass bin. The highest mass bins in our sample are seen in \autoref{fig:rms_s2}. Solar values are plotted as a diamond in mass bin 1.0 $ \leq$ M$_{\star} < 1.1$~M$_{\odot}$, as in \autoref{fig:rms_logg1}. The shaded region at the bottom of the plots shows the typical Keck-HIRES instrumental uncertainty.}
\label{fig:rms_s1}
\end{figure*}

\begin{figure*}
\centering
\includegraphics[width=\columnwidth]{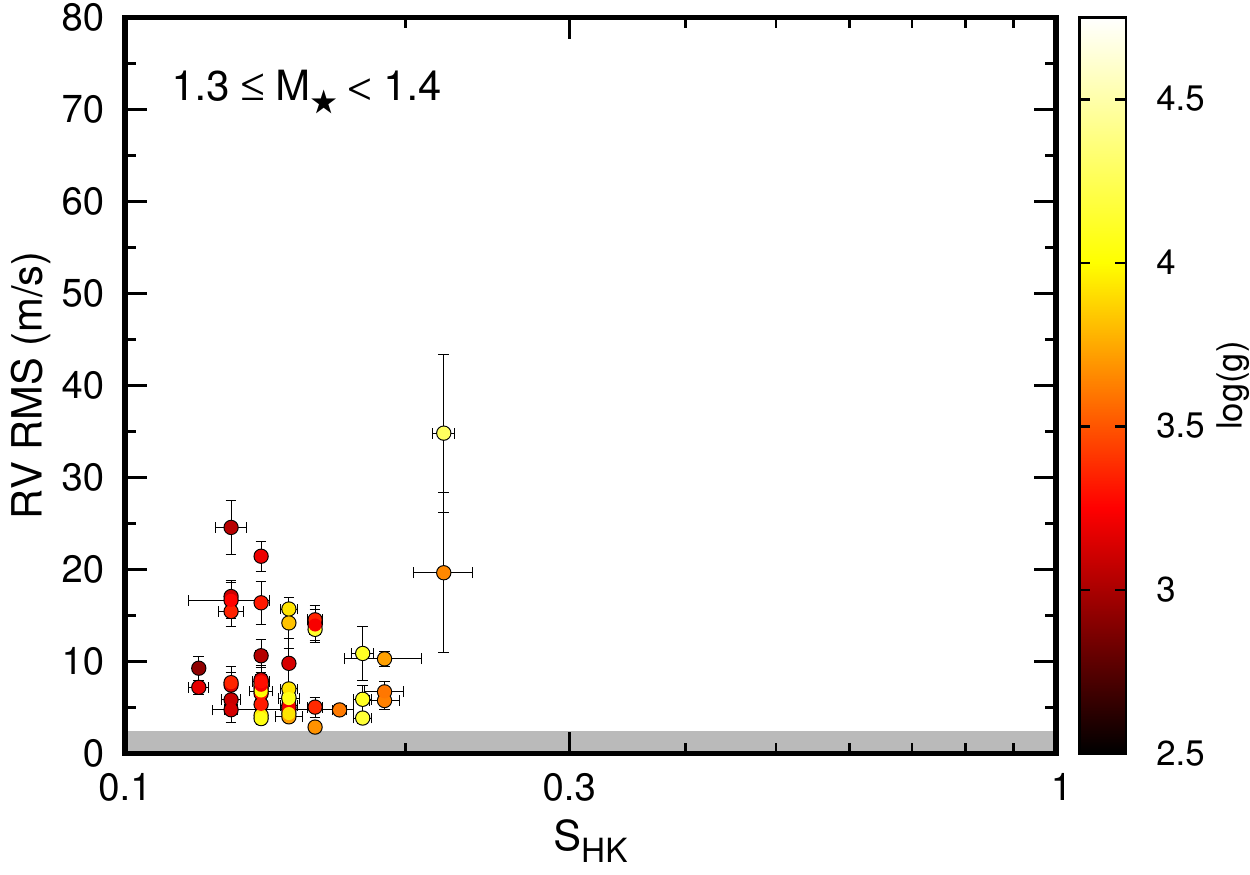}
\includegraphics[width=\columnwidth]{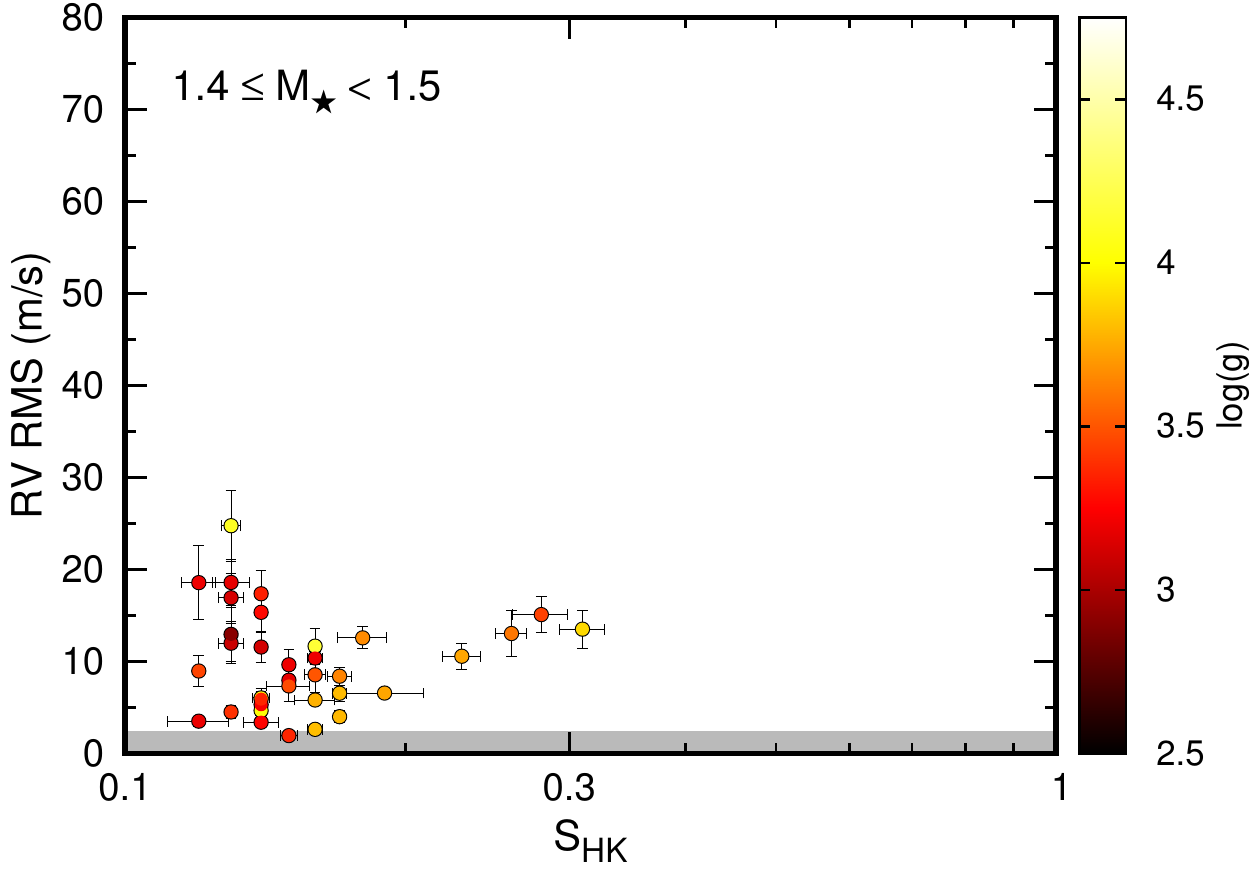}
\includegraphics[width=\columnwidth]{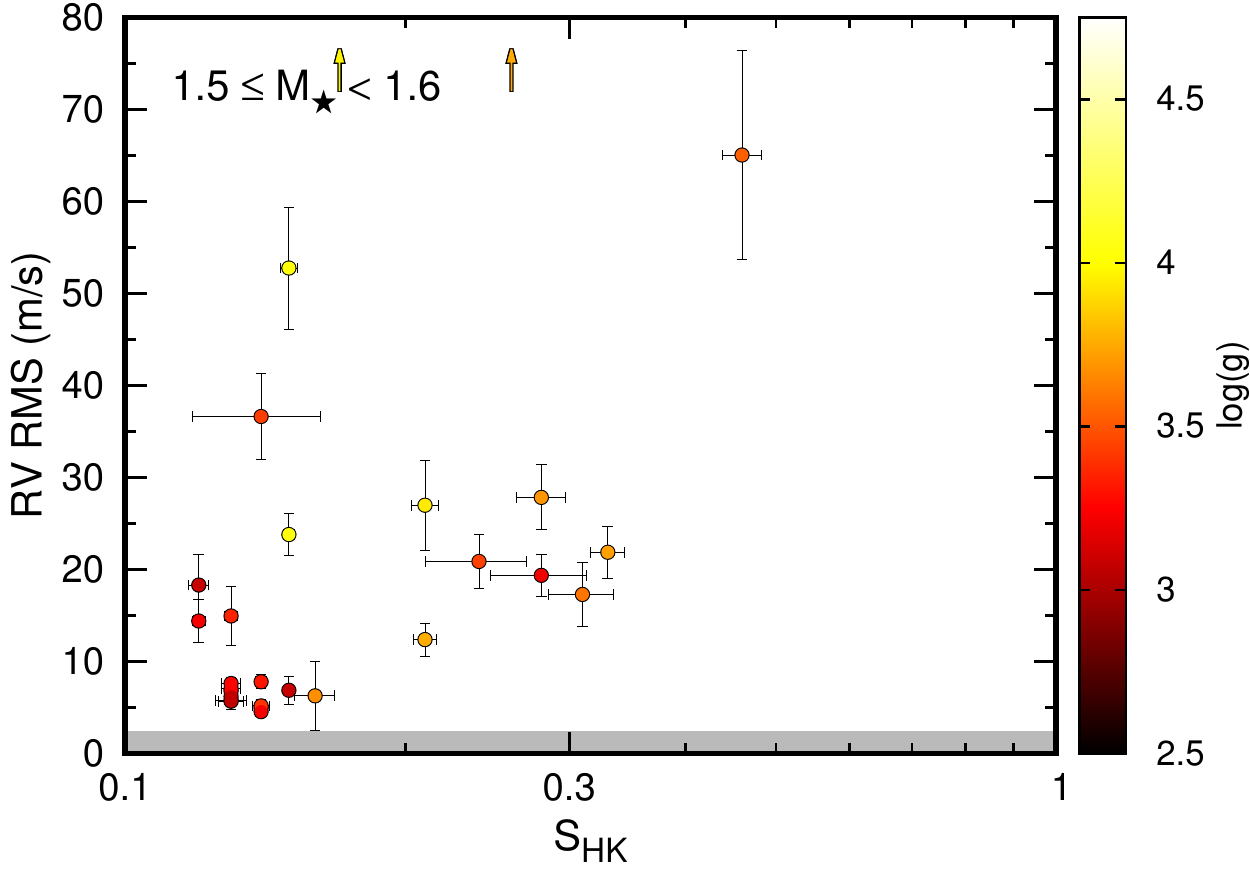}
\includegraphics[width=\columnwidth]{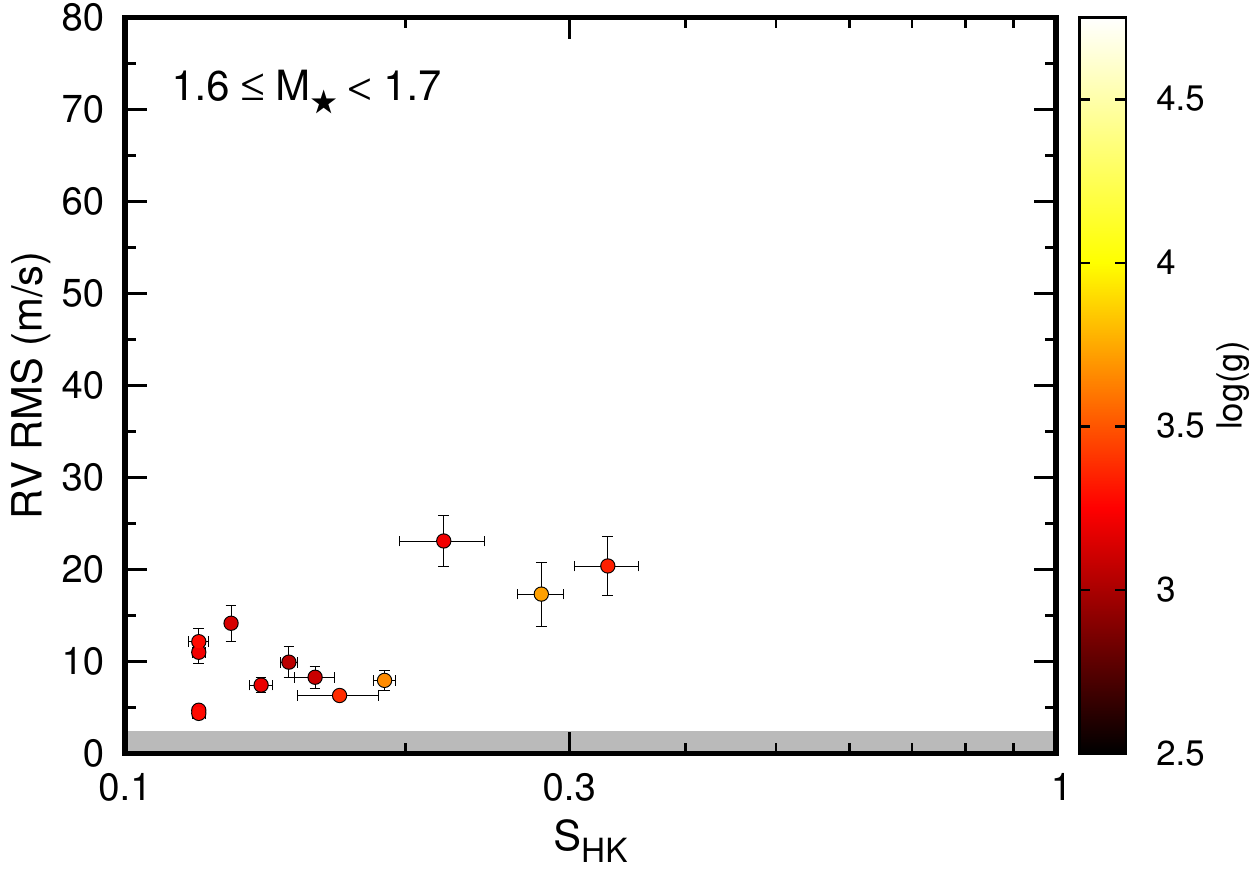}
\includegraphics[width=\columnwidth]{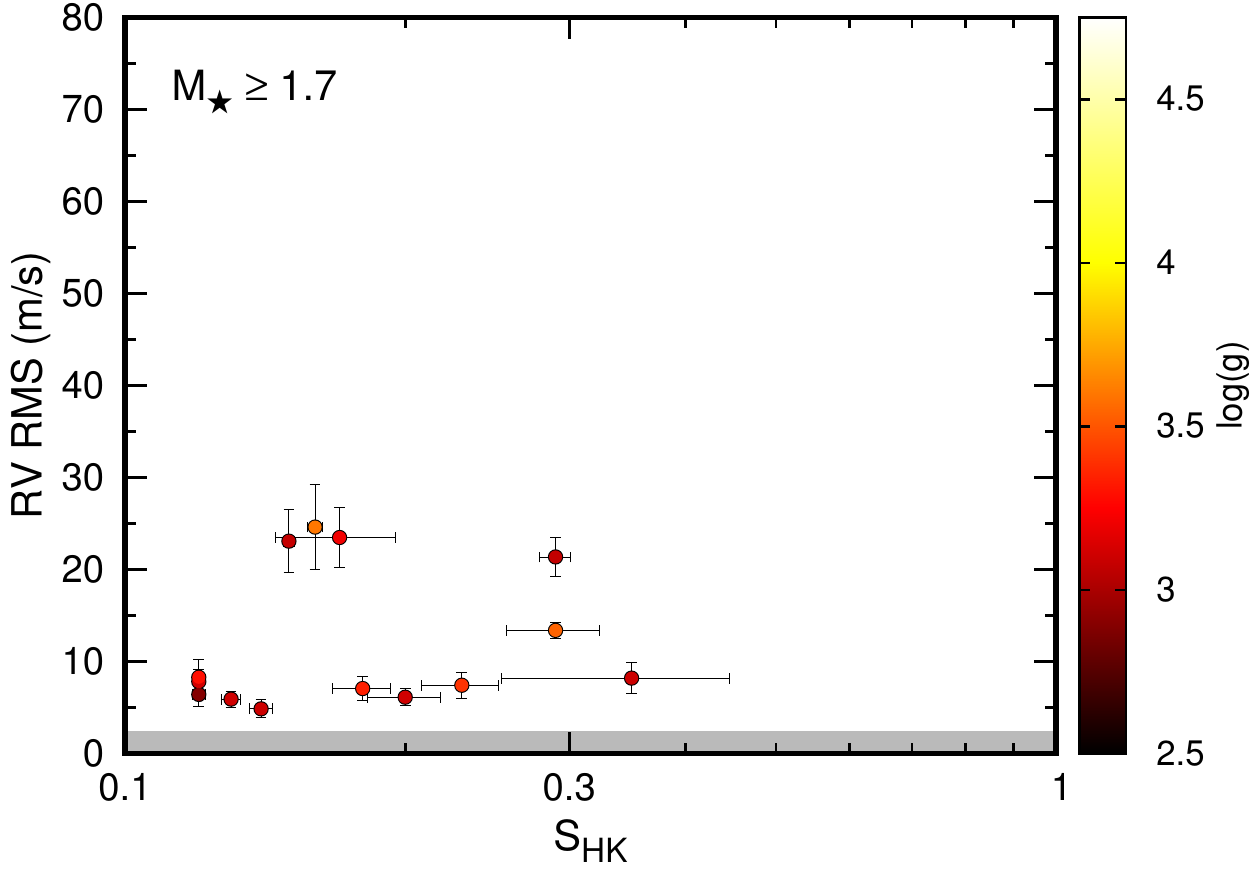}
\caption{Same as \autoref{fig:rms_s1}, for the larger mass bins}
\label{fig:rms_s2}
\end{figure*}

Figures \ref{fig:rms_logg1} \& \ref{fig:rms_logg2} are the same as the first panel in \autoref{fig:all_rms_logg} i.e., RV RMS vs. $\logg{}$) and Figures \ref{fig:rms_s1} \& \ref{fig:rms_s2} are the same as the second panel (i.e., RV RMS vs. \SHK) but now they have been broken into mass bins, each 0.1 $M_{\odot}$ wide\footnote{We note that our mass bins are smaller than the typical uncertainty in the mass from B17. The median mass uncertainty in our sample is $0.16$~M$_{\odot}$ ($\sim15\%$). As such, it is likely that a number of our stars do not actually fall within the mass bin to which we have assigned them. We have settled on bin widths of 0.1~M$_{\odot}$ after comparing several different bin widths, and seeing that our conclusions hold for both larger and smaller bin widths. The bin width we used is both convenient numerically while being small enough to track the differences across mass bins but large enough to contain enough stars to clearly show trends within a bin.}. For reference, we have added the sun to the plot of stars 1.0 $ \leq$ M$_{\star} < 1.1$~M$_{\odot}$, plotted as a large diamond symbol\footnote{We use the measured disk-integrated solar RV RMS of 8 m/s (over the solar cycle) from \citet{Meunier2010}. The solar \SHK{} is 0.1694 averaged over the solar activity cycle \citep{Egeland2017}.}.

For these plots, it is useful to compare trends as they relate to astrophysical checkpoints in their evolution. For that reason, in each mass bin, we plot three vertical lines that correspond to the surface gravities at three points in the evolution of a star: the zero-age main sequence (ZAMS), the terminal age main sequence (TAMS) and the base of the red giant branch (BRGB). These values come from a simple MESA \citep{Paxton2013} stellar evolution model where we set the TAMS by determining the age at which the hydrogen core fraction falls below $X_{c}$ = 0.0002 and the BRGB by determining the local minimum on the HR diagram \citep{VanSaders2013}. These three lines break each plot into the three basic phases of evolution that they cover: main sequence, subgiant, and giant phases. The inclusion of the TAMS line also highlights the fact that stars are known to evolve while on the main sequence \citep{Mamajek2008} and illustrates the effects this has on RV jitter. A star's main sequence lifetime is spent between the ZAMS and TAMS lines and we can use these as a quick way to estimate spin-down timescales (as indicated by the decreasing activity and RV jitter)  as they compare to the main sequence lifetime for a given mass. 

\begin{figure}
\includegraphics[width=\columnwidth]{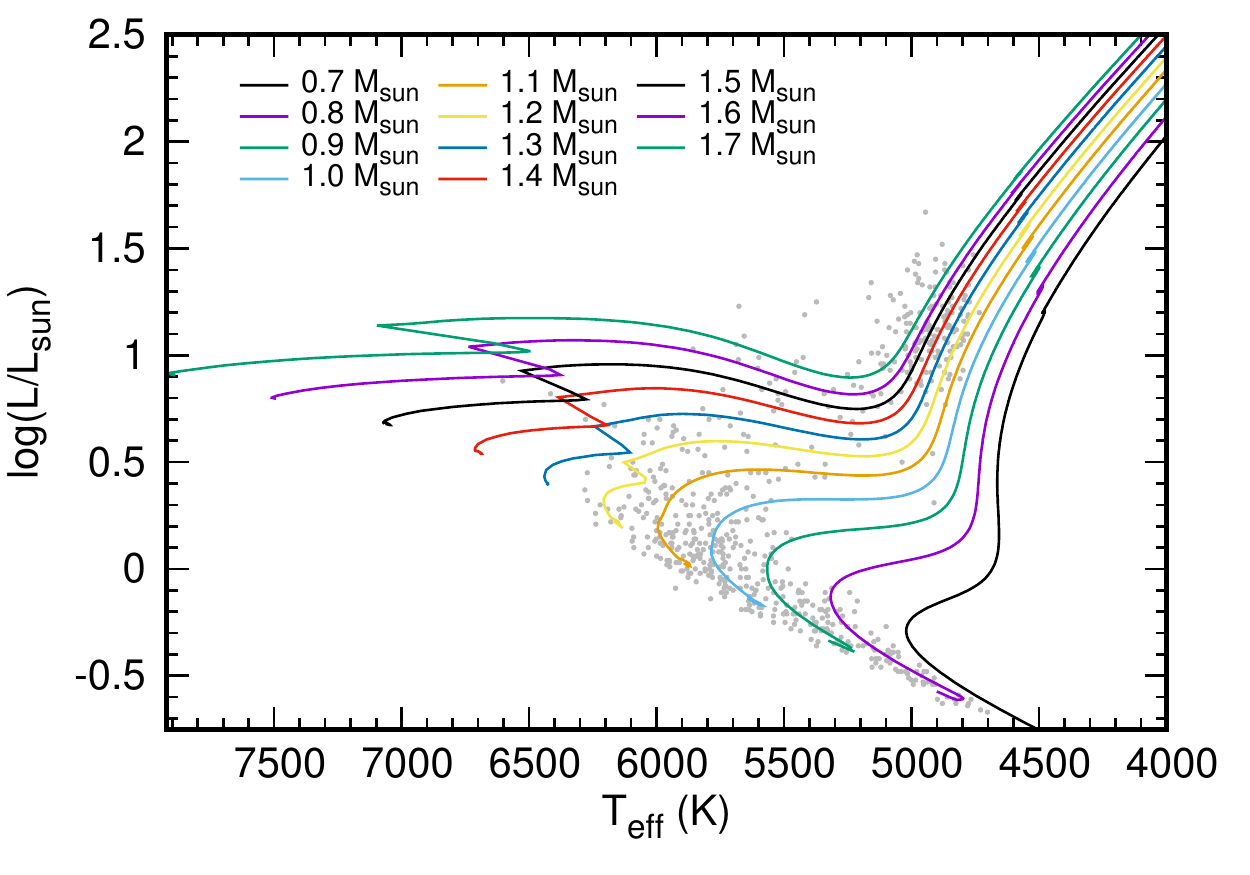}
\caption{MESA evolutionary tracks used for the theoretical granulation and oscillation components of RV jitter (Equations \ref{eqn:granulation_RMS} \& \ref{eqn:oscillation_RMS}) as a function of surface gravity for each mass bin in \autoref{fig:rms_logg2}. The CPS sample in this work is shown in light gray points for reference.}
\label{fig:evolution}
\end{figure}

Additionally, we have included theoretical curves from relations found in \citet{Kjeldsen2011}. These curves plot the expected scaling relation of RV jitter with $\logg$ due to the two components of convectively driven RV jitter: stellar oscillations and stellar granulation, as discussed previously in Sections \ref{sec:oscillations} and \ref{sec:granulation} and given in Equations \ref{eqn:oscillation_RMS} and \ref{eqn:granulation_RMS}, respectively. To plot them as a function of $\logg$, we use MESA  to generate an evolution track for each mass range. For each mass bin we evolve a solar-metallicity star with the mass of the lowest mass in the mass bin from ZAMS to the tip of the red giant branch, since we are only concerned with evolving a star until $\logg = 2.5$. We can then use the stellar parameters at each point in the evolution in the theoretical relations, which depend not only on the surface gravity, but also depend on temperature and radius at a given point in the star's evolution. The evolution tracks used in this work can be seen in \autoref{fig:evolution}. 

\subsection{Two Regimes of RV Jitter: The Transition from Activity-dominated to Convection-Dominated}
We have empirically identified two major regimes of RV jitter: magnetic activity-dominated and convection-dominated. As a reminder, we have treated the granulation and oscillation components as one combined regime where the RV jitter is driven by convection and increases with evolution (as opposed to the decrease with evolution seen in the activity-dominated regime). For this sample, we are not as concerned with determining whether a star is granulation-dominated or oscillation-dominated but rather the transition from activity-dominated to convection-dominated. In fact, based on the theoretical scaling relation, our sample barely contains stars that have evolved enough to probe the transition from granulation-dominated to oscillation-dominated. However, the fact that the floor of the observations seems to increase as theoretically expected, especially in stars with masses 1.2-1.3 M$_{\odot}$, suggests that the oscillation component dominates for stars below $\logg \sim 3$, and we use this as further validation of the scaling relation for the oscillation component. Further, despite good agreement with the data, the theoretical granulation component appears to follow the instrumental uncertainly threshold and so we are reluctant to make strong claims about the relative strengths of RV variations due to granulation and oscillations. For most stars we have $\sim$10-30 observations over a span of years with variable cadence depending on the star, which does not allow us to say much about the physical phenomena with timescales of minutes to hours (granulation, oscillation). It is clear however that these theoretical predictions convincingly describe the observations (whereas activity does not since the S-value is low despite elevated jitter in some cases). More work will be needed to distinguish different convection-driven processes.

The following paragraphs discuss the transition from activity-dominated to convection-dominated jitter as it depends on mass. First it is useful to define the ``jitter minimum", the point in a star's lifetime where its RV jitter is lowest, which, as our data suggest, occurs at the transition between the activity-dominated and convection-dominated regimes.

\subsubsection{Low-mass stars: M$_{\star} < 0.9~\mathrm{M}_{\odot}$}\label{sec:lowmassregime}
For low-mass stars in our sample we have very few evolved stars due to their long main sequence lifetimes. As we decrease in stellar mass, the main sequence lifetime increases and at some point exceeds the age of the universe. All Sun-like stars below a certain mass, ($\sim$0.9~M$_{\odot}$) are therefore activity dominated, and our data suggest that there is a fundamental mass limit where stars have not evolved enough to reach their astrophysical jitter minimum. 

The following trends with $\logg$ can be seen in the first two panels of \autoref{fig:rms_logg1}. Stars less massive than the sun are born as young, active stars and magnetic activity (manifested as spots/plages/etc.) dominates the RV jitter. As they continue to evolve on the main sequence, they spin down and become less active, which results in lower RV jitter amplitudes. Since they are still early in their MS lifetimes, they have not changed their structure much, and remain near their initial $\logg$. Their primary movement on a plot of RV jitter vs. $\logg$ is therefore vertically downward from their zero-age-main-sequence location. We note that for most of our mass bins in this regime we are unable to fully probe the minimum jitter value due to the instrumental uncertainty. We see some evidence in the stars between 0.9 and 1.0~$\mathrm{M}_{\odot}$ that the jitter floor for these stars occurs when these stars have begun to evolve toward the end of their main sequence lifetimes.

When looking at trends with \SHK{} in \autoref{fig:rms_s1}, we see the same story in a slightly different way. First, there is a lack of a convection-dominated regime, which would be indicated by an increase in jitter for the least active stars in these plots. These stars are therefore activity-dominated, and because they have reached our instrumental floor, it is unclear whether they have lived long enough to have spun down to their jitter minimum. We further note by comparing to other mass bins that the activity dependence is diminished for the lower masses. That is to say that even the most active stars in this group only have RV jitter of 10-20 m/s. Although activity and RV jitter are still strongly \emph{correlated}, the \emph{dependence} is much weaker, with a slope 6 times shallower than in the 1.0 to 1.1 M$_{\odot}$ bin (slopes of 15 and 98). This echoes the results of \citet{Isaacson2010}, who reported this ``sweet spot" for spectral type K stars, which exhibit relatively low levels of RV jitter across all measured activity levels.

We wish to note that for this group of stars there appears to be a discrepancy between the theoretical ZAMS and the ZAMS one would infer from looking at the vertical pileup of the ``active main sequence". That is, the observed surface gravities appear to be systematically lower than predicted by stellar models. We see this only in this sample of low-mass stars and we attribute this as due in part to the calibration of $\logg$ in B17, which used asteroseimic surface gravities to calibrate the measured spectroscopic surface gravities. As such, the asteroseismic sample was mostly evolved (and therefore massive enough to have evolved within the lifetime of the universe) stars. Therefore, there are very few calibration points for lower mass dwarfs, and it is expected that the measured surface gravities would be less certain. Despite the larger uncertainties and tendency to underestimate the surface gravities for the lowest mass dwarfs in our sample, our results hold. We note that this effect is reduced when using the isochrone-derived surface gravities given in B17, which are otherwise disfavored.

We are limited in our sample of low mass stars by the lower temperature limit of B17. We expect similar trends to hold (i.e., low jitter regardless of activity) for stars below 0.7 M$_{\odot}$, and therefore expect them to continue to be a ``sweet spot" for planet searches. However, the inability to have fully spun down on the timescale of the age of the universe means that these stars may only be able to spin down to a level of RV jitter that could be above what we see in our lowest mass bins. Further, we expect different manifestations of activity for stars that are fully convective and so we avoid speculating about the RV jitter of such stars. New infrared spectrographs (Carmenes, HPF, SPIROU, iSHELL) will provide better studies for the behavior of RV jitter for these fainter M dwarfs.

We remind the reader that we see no evolved stars in this set of stars due to the main sequence lifetimes being longer than the age of the universe (as is indicated by the washed out lines in \autoref{fig:rms_logg1} that show the theoretical convection component).

\subsubsection{Solar-mass stars: $0.9~\mathrm{M}_{\odot} \leq \mathrm{M}_{\star} < 1.5~\mathrm{M}_{\odot}$}\label{sec:solarmassregime}
Although the highest mass in this range would typically not be considered ``solar mass", we find that the RV jitter of stars in this range of masses behaves very similarly. Stars roughly solar mass stars up to 1.5 M$_{\odot}$ exhibit the following trends in $\logg$ (seen in \autoref{fig:rms_logg1} and \autoref{fig:rms_logg2}). These stars start as active stars that then move vertically down the ``active main sequence" as they spin down. However, it is clear from these plots that stars in this mass range do not reach their jitter minimum before beginning to evolve to lower $\logg$. The ``jitter minimum" instead occurs for stars that have measurably evolved, but before the stars get so big that RV variations caused by convection become dominant\footnote{This is less clear in the 0.9 to 1.1 M$_{\odot}$ mass bins since we are unable to fully probe the jitter minimum for these stars, given the instrumental uncertainty. However, it is still suggested in the plot that the jitter minimum occurs at lower $\logg$ values among the main sequence stars in this bin.}. This transition region from activity-dominated jitter to convection-dominated jitter comes mostly from the loss of magnetic activity. The color gradient seen in the stars in \autoref{fig:rms_logg1} in the mass range 1.0 to 1.1~$M_{\odot}$ near the jitter minimum at $\log \sim$ 4.3 clearly shows this transition as stars become magnetically quiet. From there, a star follows the general path shown in the purple line, with RV jitter dominated by convection: first, we predict, by granulation and later by oscillations. 

In terms of activity, \autoref{fig:rms_s1} and \autoref{fig:rms_s2} show the same trends in a different manner. In these plots, it is easy to see the relation between activity and RV jitter. We clearly see when comparing the 0.9 to 1.0 M$_{\odot}$ mass bin with the 1.0 to 1.1 M$_{\odot}$ mass bin that the most active stars in the higher mass bin have higher levels of RV jitter. We expect this to hold generally: the most active stars of higher masses have higher RV jitter, continuing with what we saw in the lower mass bins in \autoref{sec:lowmassregime}. This is confirmed by the increasing slopes with activity from 0.9 to 1.0 M$_{\odot}$ bin (slope of 62) to 1.0 to 1.1 M$_{\odot}$ bin (slope of 99), to 1.1 to 1.2 M$_{\odot}$ bin (slope of 132)\footnote{The increasing slope is also affected by the fact that \SHK{} is not normalized between spectral types (e.g., the most active stars in the 1.1 to 1.2 M$_{\odot}$ bin are all below 0.4 whereas the highest in the mass bin below it are below 0.5).}. In other words, despite the fact that the range of \SHK{} values decreases with increasing mass, the range of RV jitter values is observed to increase substantially as well, with 97th percentiles in RV jitter of (12.38 m/s, 18.99 m/s, 31.09 m/s, 36.09 m/s, and 43.402 m/s for the first 5 bins of \autoref{fig:rms_s1}).  However, our ability to probe this trend for intermediate mass stars ($\sim1.3$ to $1.5~$M$_{\odot}$) is hampered by selection effects, outlined below. The convection-dominated regime is not as clearly defined as it is in the plots of $\logg$. In general we see that below a certain activity level (different for each mass bin), stars exhibit high levels of RV jitter again.

From the plots for these stars, it is clear that both activity \emph{and} evolutionary state of the star are useful for selecting stars that are RV quiet as we see a well-defined transition between the activity-dominated regime and the convection-dominated regime.

We wish to quickly discuss the inclusion of the intermediate mass stars in this grouping. Stars of intermediate mass (1.3 to 1.5 M$_{\odot}$) on the main sequence are not suitable for RV observations. Their hot temperatures produce few absorption lines in their spectra and their rapid rotation broadens any absorption features they may have to the point where precise RV measurements are not possible. Therefore, to study planets around intermediate-mass stars, surveys like the ``Retired A-Star Survey" \citep{Johnson2006} have examined their evolved stages where they have both cooled and spun down, allowing for precise RV measurements. Therefore, we do not see many stars in the ``zero age main sequence" portion of these figures for stars above 1.3 M$_{\odot}$. We also must point out that this mass is not surprisingly near the Kraft break \citep{Kraft1967}. Above this mass range, dwarf stars rotate too rapidly for precise RV measurements. The Kraft break also pinpoints the region where stars begin to have very thin (or nonexistent) convective envelopes. If they lack a convective envelope, this further complicates our analysis because they would no longer be magnetically active, since convection is a required condition for our current understanding of magnetic dynamo \citep{Parker1955}. Instead, spin down for these stars occurs after the main sequence as they gain a deepening convective envelope \citep{Kippenhahn2012,VanSaders2013}. Any attempts at obtaining radial velocities of intermediate mass stars on the main sequence would probably contain large amounts of RV jitter, likely dominated by pure uncertainty in measuring a precise velocity from the rotationally-broadened absorption features\footnote{The A-F stars studied in \citet{Galland2005} were seen to exhibit RV uncertainty in the range of 50-300 m/s. Additional work has shown that RV uncertainty can be as much as $\sim$km/s in O-type stars \citep{Williams2013}.} rather than magnetic activity. 

Instead, intermediate mass stars show up in our sample once they have appreciably evolved and become amenable for precision radial velocities. \citet{VanSaders2013} argue that the stars above the Kraft break are able to spin down rather quickly post main sequence due to the rapid rotation during the onset of the magnetic winds coupled with the fact that these stars are substantially expanding and increasing their moments of inertia (moreso than in lower mass stars). As such, the massive stars are able to spin down from their main sequence rotations ($\vsini > 70$~km/s)\footnote{Based on \citet{Kraft1967}} to velocities more amenable for radial velocity measurements ($\vsini < 20$~km/s) in a relatively short amount of time. Despite the rapid spin down, their descent toward lower RV jitter is not as vertical as seen in the lower mass stars. Presumably this is because the subgiant lifetime is shorter than the timescale to spin down and so these stars then stay active down to very low $\logg$. They therefore travel along a diagonal path downward and to the right in the $\logg{}$ plots as they leave the main sequence. 

When looking at the $\logg$ plots for the intermediate mass stars (first two panels of \autoref{fig:rms_logg2}) there is evidence for an ``active main sequence" where stars are spinning down and traveling vertically downward as the decrease in RV jitter. However, there are only a few points that indicate the presence of an active main sequence. When we look at the \SHK{} plots, there is very little evidence of a trend with activity for the active stars, casting doubt on the any evidence of an active main sequence for these stars. Given the low significance, the appearance of an ``active main sequence" in the $\logg$ plots can be explained away by measurement uncertainties ($\logg$, $M$, RV RMS). However, these stars still clearly show the rising floor of the convection-dominated regime as they evolve through the subgiant and giant phases. Despite expecting stars in the range of 1.3 to 1.5 M$_{\odot}$ to have different trends than the solar-mass stars, it is clear from the $\logg$ plots that these stars follow the trends seen among the solar-mass stars instead of the trends seen in the higher mass stars.

We are limited in this sample by several selection effects. Probing the full effect of activity is challenging because of the nature of exoplanet searches, which have predominantly searched around inactive stars, rightly expected to have lower levels of RV jitter. Added to this is the effect of rotational broadening. At higher masses, stars have thinner and thinner convective envelopes and are unable to fully spin down on the main sequence. We only see the initial effects of this in the stars in this mass range. The effects are strongest in the next mass range.

\subsubsection{High-mass stars: $\mathrm{M}_{\star} > 1.5~\mathrm{M}_{\odot}$}\label{sec:highmassregime}
Astrophysically, the stars in this set follow the same ideas of transitioning from activity-dominated to convection-dominated, but with two distinct differences as seen in the plots of $\logg$. First, as mentioned above, these stars are not able to spin down until they gain a convective envelope, which occurs in the subgiant phase for these stars. Second, their evolutionary timescales are considerably shorter, such that the spin-down timescale is of order their evolutionary timescale. This means they are seen to be moving diagonally downward to the right as they spin down, decrease in jitter, and expand into a giant star. Evidence of this trend was seen in the previous set of stars, but it becomes very clear when looking at the behavior of this set. The transition from activity-dominated to convection-dominated is therefore spread out in $\logg$ and in fact the most evolved stars in our sample have only just hit the theoretical convection limit. This means that these stars are still spinning down and \emph{activity} becomes the dominant predictor of RV jitter rather than evolution.

We see slight evidence of this when looking at the plots of \SHK{} for this set of stars when compared with the first two bins of \autoref{fig:rms_s2}. Because the active and rapidly rotating main sequence stars in the first two bins (1.3 to 1.4 M$_{\odot}$ and 1.4 to 1.5 M$_{\odot}$) are avoided in target selection, there appears \emph{negative} correlation between \SHK{} and RV RMS since the stars are in the convective-dominated regime where RV jitter increases with time as they continue their final spin-down. For the stars in mass bins above 1.5 M$_{\odot}$ the sign flips and we see a stronger positive trend again, most likely because they evolve more quickly and were more rapidly rotating to start with, so they are still. Unfortunately our data are too sparse in these bins to fully support these claims. Instead, the data appear to be merely consistent with the framework we have adopted for the the evolution of RV jitter as seen in the lower mass bins and extrapolated to these bins. At least in the 1.6 to 1.7 M$_{\odot}$ bin, it seems that activity is a better predictor of RV jitter.

An important caveat to remember is that this sample does not contain many evolved giant stars. It is reasonable to expect that if we observed giant stars with $\logg < 3$ we would see the rise of the convection-dominated floor and therefore see the negative trend with \SHK{} below about 0.2. It is therefore only true that activity is the better predictor of RV jitter for this specific set of stars in our sample. Additionally, as mentioned above, main sequence stars of these masses were avoided in RV observations and so we continue to lack stars close to the ZAMS, and in this set of the highest mass stars we barely have any stars before the TAMS.

\vspace{12pt}
The two regimes of RV jitter as they relate to low mass stars (\autoref{sec:lowmassregime}), solar mass (\autoref{sec:solarmassregime}), and high mass (\autoref{sec:highmassregime}) can be briefly summarized as follows. Low mass stars evolve very little during the transition from activity-dominated to convection-dominated RV jitter. It is clear that for these stars, the spin-down timescale is much less than their main sequence lifetimes, as expected \citep{VanSaders2013}. Higher mass stars evolve quite dramatically before they complete the transition from activity-dominated to convection-dominated RV jitter. The two reasons for this are that 1) higher mass stars (above the Kraft break) are unable to spin-down while on the main sequence and so all spin down occurs in the subgiant and later evolutionary stages, and 2) higher mass stars evolve more quickly. The key result from this is that \emph{the jitter minimum occurs at later stages of evolution for higher mass stars}.

\subsubsection{Location of the jitter minimum}
Given this picture for the stellar evolution of RV jitter as a function of mass, we provide a rough estimate of the value and location (in terms of $\logg$) of the jitter minimum for each mass bin in \autoref{tbl:jitter_minima}. These estimates are performed by eye. For each mass bin we simply follow the RV jitter evolution in the $\logg{}$ plots, starting first at the active main sequence and following that sequence downward as the stars spin down. For the lowest mass stars, they hit the instrumental floor in a vertical strip that makes it quite easy to estimate the location of the jitter minimum. As they appear to be at the instrumental floor, we can only say that their jitter is likely below 2.5 m/s. For the mass bins in the solar range, we follow the spin down of the active main sequence in increasingly gradual transitions, as suggested by the floor of the data. When they reach the instrumental floor or appear to essentially flatten out, we again estimate the $\logg{}$ at which this first appears to happen (towards higher $\logg{}$.). For the highest mass stars, we follow the diagonal path downward until it roughly reaches the modeled convective floor.

\begin{deluxetable}{c c c c c}
\tablecaption{Estimated RV Jitter Minimum by Mass \label{tbl:jitter_minima}}
\tablehead{ \colhead{Mass bin} & \colhead{$j_{\mathrm{min}}$} & \colhead{Location of $j_{\mathrm{min}}$ } & \colhead{ZAMS} & \colhead{TAMS} \\
\colhead{(M$_{\odot}$)} & \colhead{(m/s)} & \colhead{$\logg$} & \colhead{$\logg$} & \colhead{$\logg$}}
	 \startdata
	 0.8 -- 0.9 & $\leq$2.5  & 4.5 & 4.629 & 4.294 \\
	 0.9 -- 1.0 & $\leq$2.5 & 4.4 & 4.587 & 4.251 \\
	 1.0 -- 1.1 & $\leq$2.5 & 4.35 & 4.535 & 4.210 \\
	 1.1 -- 1.2 & $\leq$2.5 & 4.25 & 4.500 & 4.104 \\
	 1.2 -- 1.3 & 3.5 & 4.2 & 4.429 & 4.017 \\
	 1.3 -- 1.4 & 4 & 4 & 4.339 & 3.959 \\
	 1.4 -- 1.5 & 4 & 3.8 & 4.297 & 3.907 \\   
	 1.5 -- 1.6 & 5 & 3.3 & 4.284 & 3.844 \\
	 1.6 -- 1.7 & 5 & 3.25 & 4.297 & 3.873 \\
	  $>$ 1.7 & 5.5 & 3.1 & 4.302 & 3.873 \\
	 \enddata
\end{deluxetable}

\subsection{Validation of Oscillation Scaling Relation}\label{sec:oscillation_validation}
\begin{figure*}
 \includegraphics[width=\columnwidth]{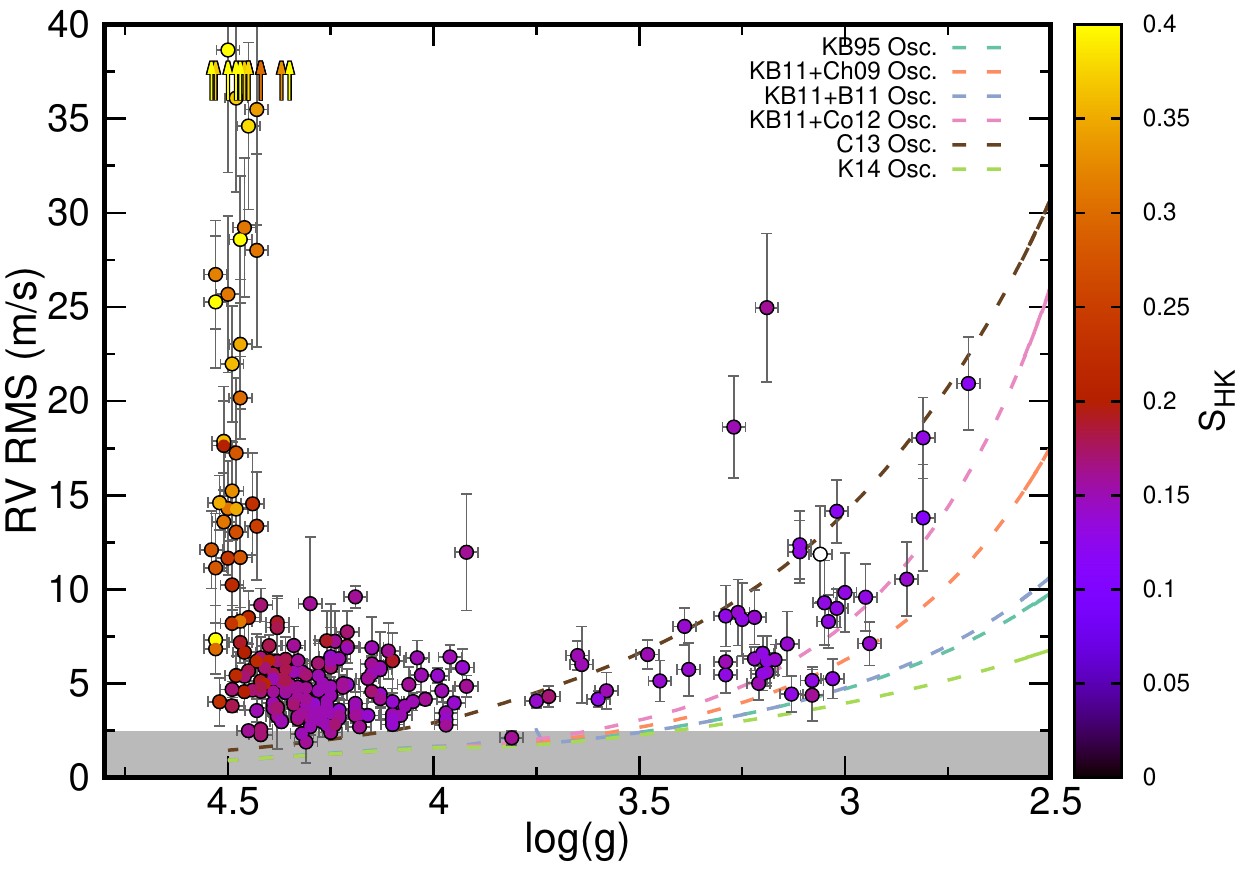}
  \includegraphics[width=\columnwidth]{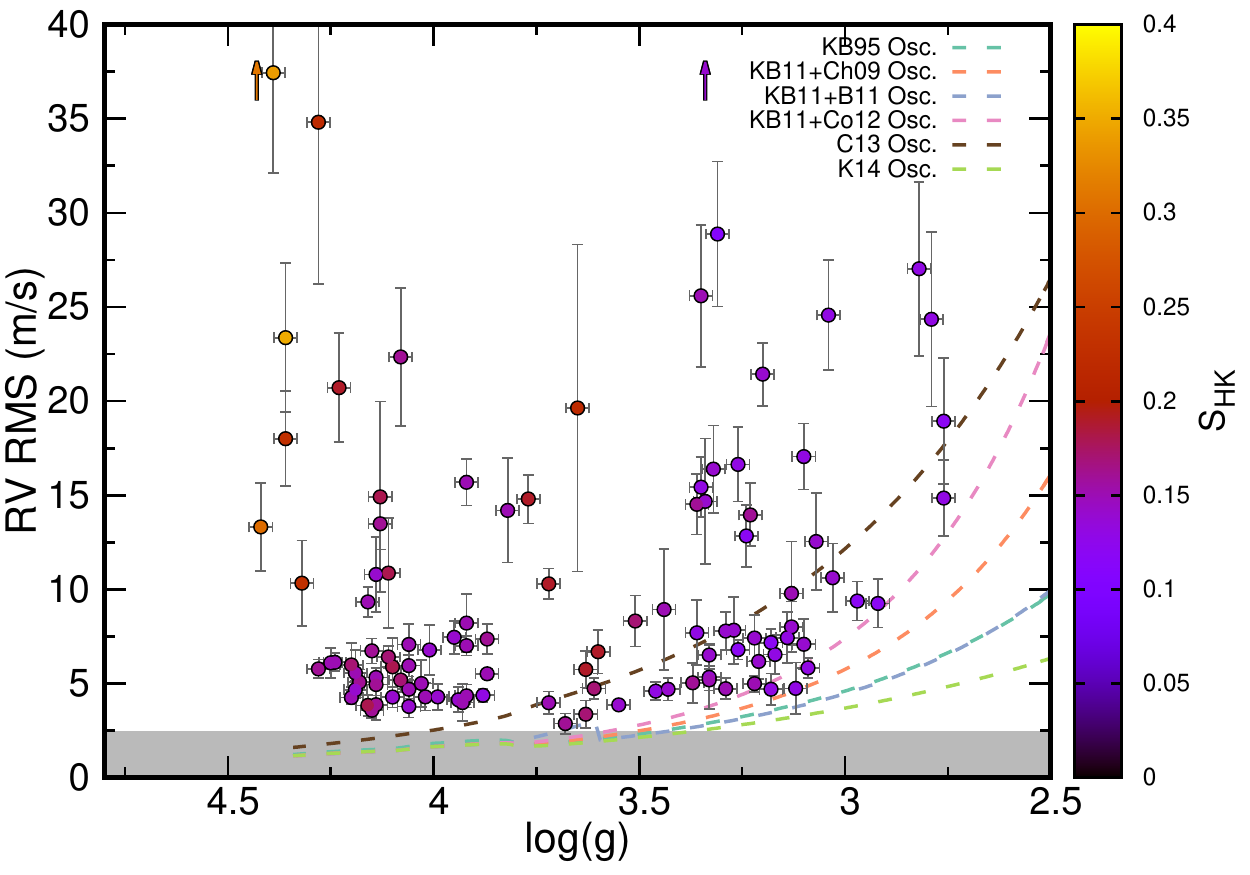}
 \caption{Comparison between various theoretical and empirical scaling relations for the oscillation component of RV jitter for stars between 1.0 and 1.2 M$_{\odot}$ (\emph{left}) and stars between 1.2 and 1.4 M$_{\odot}$ (\emph{right}). Points are color-coded as in \autoref{fig:rms_logg1}. Scaling relations are abbreviated as follows: KB95 for \citet{Kjeldsen1995}, KB11+Ch09 for \citet{Kjeldsen2011} with \citet{Chaplin2009} mode lifetime, KB11+B11 for \citet{Kjeldsen2011} with \citet{Baudin2011} mode lifetime, KB11+Co12 for \citet{Kjeldsen2011} with \citet{Corsaro2012} mode lifetime, C13 for \citet{Corsaro2013}, and K14 for \citet{Kallinger2014}.}
 \label{fig:theoretical_comparison}
 \end{figure*}

Now that we have discussed the major trends and observational limitations, we take an aside to comment on the validation of the scaling relation used for the oscillation component of RV jitter. We tested several different scaling relations for the oscillation component of RV jitter that can be found in the literature. A detailed account of each relation would detract from the purpose of this paper. We find that many of the relations agree with the general trends, and we cannot with confidence claim that one relation is preferred by the data. To compare all of the different scaling relations, we check using two wider mass bins, 1.0-1.2~M$_{\odot}$ and 1.2-1.4~M$_{\odot}$ by examining how well the resulting scaling relation followed the jitter floor. We find that the effects are seen most clearly toward the end of the subgiant regime and into the giant regime, which is why we use these bins where our sample is large and the floor is fairly well defined. All tested relations can be seen in \autoref{fig:theoretical_comparison}. We find that  KB11+Co12 (\citet{Kjeldsen2011} oscillation scaling relation with the \citet{Corsaro2012} mode lifetime scaling) shows the best agreement to the floor of our data by eye, and we use this as light empirical evidence in favor of these relations. As can be seen, the other relations either showed too sharp or too shallow of an increase with $\logg$. We note that KB11+Ch09 (\citet{Kjeldsen2011} relation with the \citet{Chaplin2009} mode lifetime) also does a decent job fitting the RV jitter floor. We choose the \citet{Corsaro2012} mode lifetime over the \citet{Chaplin2009} lifetime for its ability to better fit the floor at the lowest surface gravities.

\begin{figure}
\includegraphics[width=\columnwidth]{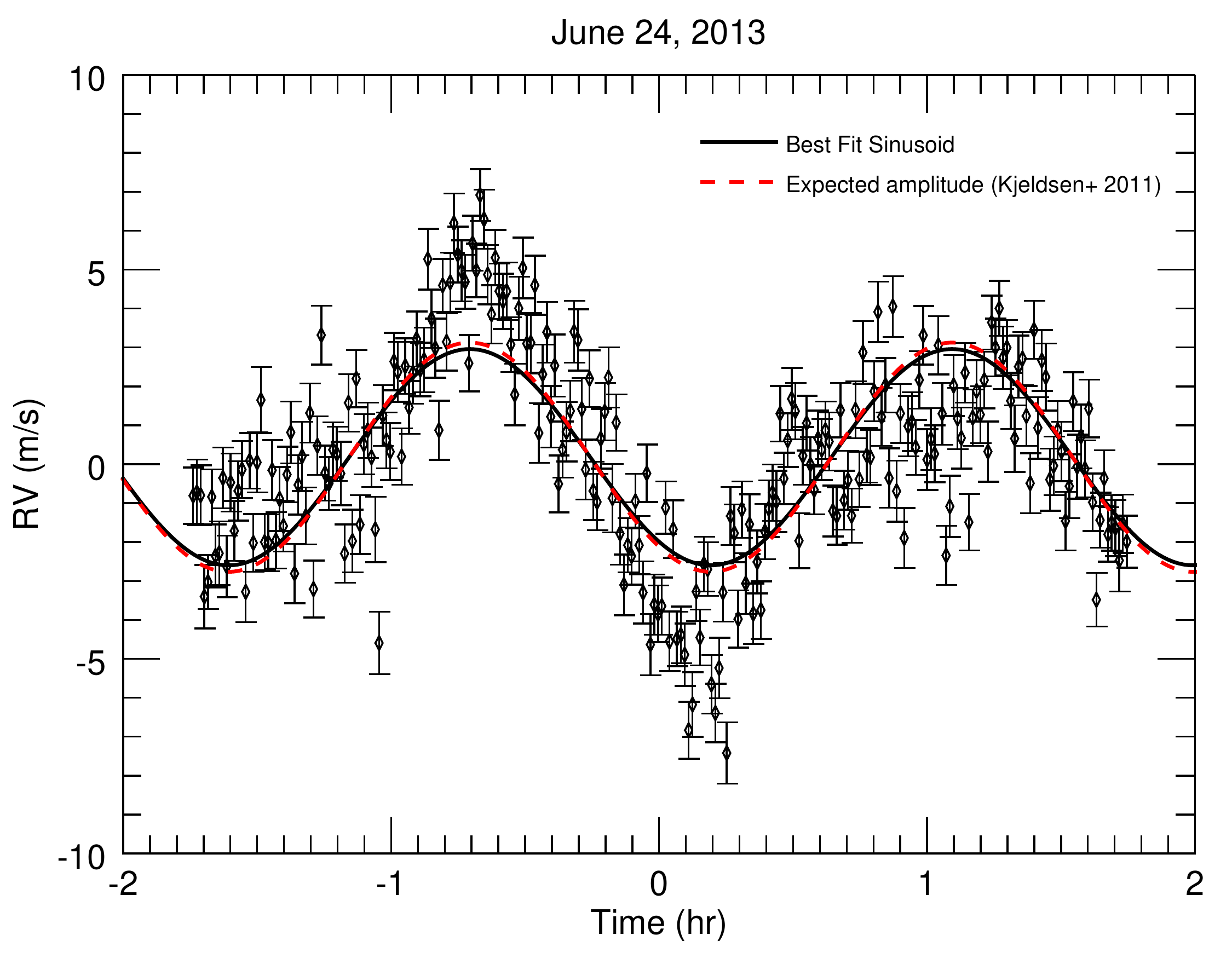}
\includegraphics[width=\columnwidth]{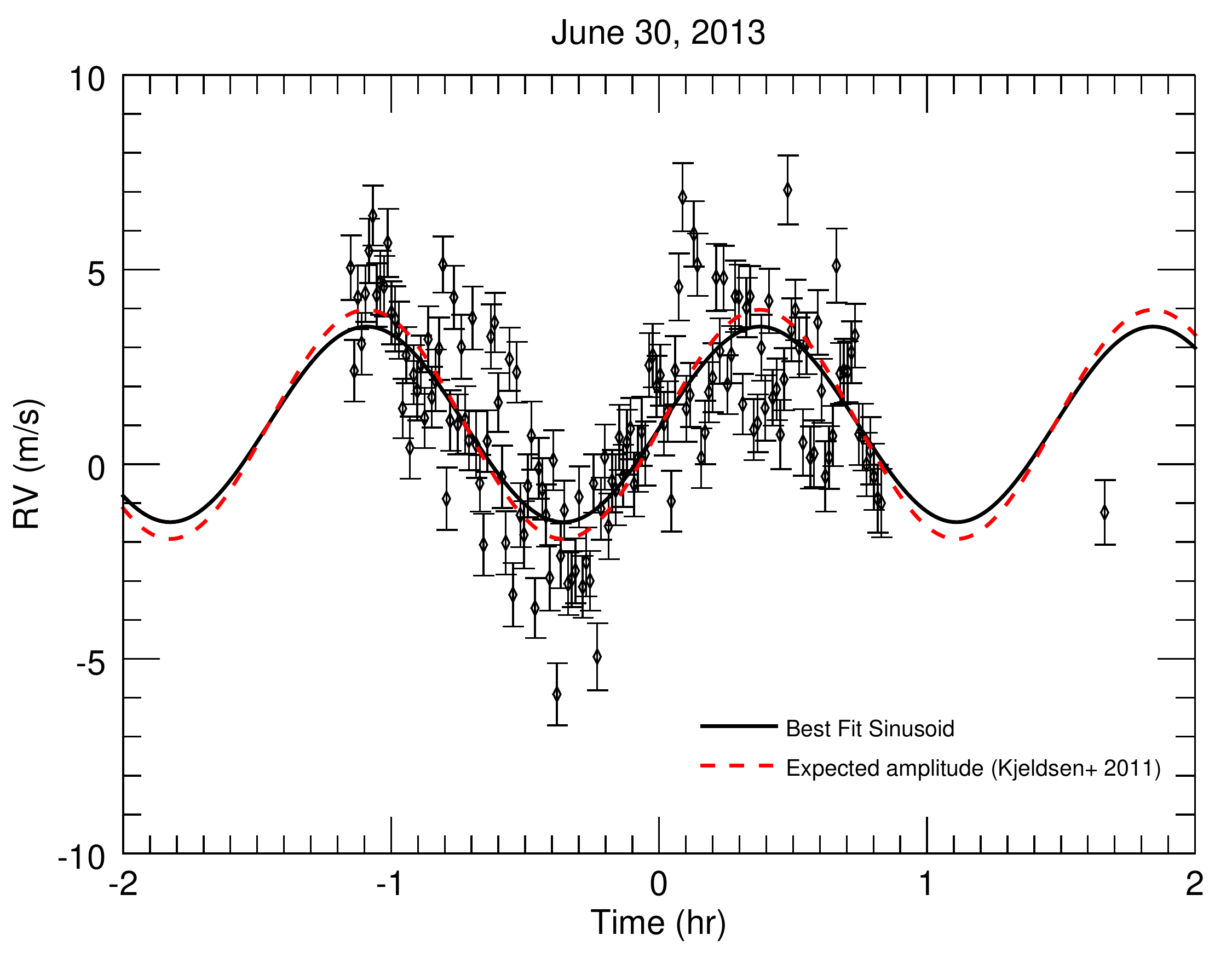}
\caption{Observations of HD 142091 targeting stellar oscillations. The black curve shows the best-fit sinusoid for the observations on June 24, 2013 (\emph{top panel}) and June 30, 2013 (\emph{bottom panel}). The red curve in each panel shows the same sine curve but with the predicted oscillation amplitude based on the scaling relation in \citet{Kjeldsen2011}, as given in \autoref{eqn:oscillation_RMS}.}
\label{fig:oscillation_comparison}
\end{figure}

To further test our scaling relation, we make use of observations of HD 142091, a 4th magnitude subgiant for which asteroseismic radial velocity observations were made specifically to target stellar p-mode oscillations. The observations were made over several hours on two separate nights in mid 2013. The first night (June 24) contains 246 observations over a 4 hour span. The second night (June 30) is separated by 6 days and contains 151 observations over a 2 hour span. Each night of observations shows a clear sinusoidal variation in the radial velocities due to the stellar p-mode oscillations, shown in \autoref{fig:oscillation_comparison}. The remaining observations of this star show evidence of a planet \citep{Baines2013}. To calculate the RV jitter for this star, we first subtract out the planet. The RV jitter of the residuals to this fit is then largely driven by the nearly 400 observations that show the stellar oscillations. Since the RMS for the two nights of observations is significantly smaller than the scatter for the remaining observations, we remove these from the jitter calculation and obtain an RV RMS of 7.6 m/s compared to the RMS during the two nights of high-frequency observations of 2.6 m/s, which provides a good check of our expected oscillation component. We show this as an `X' in the 1.5 to 1.6 M$_{\odot}$ mass bin of \autoref{fig:rms_logg2} and note that this is the only star in our sample for which we have resolved stellar oscillations. The `X' lands almost exactly where the scaling relation in \autoref{eqn:oscillation_RMS} \citep{Kjeldsen2011,Corsaro2012} predicts. We note that the blue dashed line showing the oscillation component of RV jitter shown in \autoref{fig:rms_logg2} is for a nominal 1.5 M$_{\odot}$ star. We can go one step further and compare the measured and expected oscillation amplitudes based on the actual stellar properties for this star.

Using the stellar properties of HD 142091 (M~$ =1.5~$M$_{\odot}$, $\Teff=4781$~K, [Fe/H]~$=0.25$, $\logg$~$=3.26$, R~$=4.85$~R$_{\odot}$), the expected oscillation amplitude is 3.293 m/s (2.944 m/s when using the \citet{Chaplin2009} mode lifetime). For both nights of observations we find the best-fit sinusoid to determine the amplitude of stellar p-mode oscillations. The best-fit sinusoid for the June 24th observations has an amplitude of 2.778 m/s. The amplitude for the June 30th observations is slightly smaller: 2.511 m/s. Both observations show very good agreement with the predicted amplitudes from \citet{Kjeldsen2011} and \citet{Corsaro2012}, which we show in \autoref{fig:oscillation_comparison}. The excellent agreement between the measured and expected amplitudes for a massive, evolved star is validation for the oscillation scaling relation used in this work, which has been scaled to the sun.

\subsection{Other Features of Jitter Evolution}\label{sec:jitter_evolution_features}
Here we quickly highlight a few other results and features noticed in the jitter evolution plots (Figures \ref{fig:all_rms_logg}, \ref{fig:rms_logg1}, \ref{fig:rms_logg2}, \ref{fig:rms_s1}, and \ref{fig:rms_s2}.).

\subsubsection{Jitter-Above-the-Floor Stars}\label{sec:jitter_above_floor}
Our primary concern has been to establish the ``jitter floor" and explore trends among stellar types. After establishing the jitter floor, we examined stars that appeared to be well above the floor and here we note two stars that have significantly higher RV jitter than similar stars (at $\logg \sim3.2$) and are well above our relation. These two stars, HD 207077 and HD 128095, are in fact stars that showed lower jitter in the early stages of the vetting process when the blind fitting had subtracted Keplerian signals. 

HD 207077 (M~$ =1.13~$M$_{\odot}$, $\logg=3.27$, $\sigma_{\mathrm{RV}} = 18.62$~m/s) has a strong planet candidate signal at around 600 days, and is among those listed in \citet{Luhn2019} as planet candidates. We are nearly convinced by this planet, however we have decided to not count it as a planet until more observations are made. However, the above-average jitter for this type of star provides strong evidence that this planet is real \citep[see][]{Luhn2019}. Indeed, if we subtract our best-planet model, we get an RV RMS of 6.69 m/s, where it would fall along our relation. This validates our conservative vetting procedures and supports our confidence that none of the features in our plots are the products of any subjectivity in our vetting.

HD 128095 (M~$ =1.2~$M$_{\odot}$, $\logg=3.35$, $\sigma_{\mathrm{RV}} = 25.59$~m/s) shows strong evidence of both a long period planet and a shorter period planet. However, given the small number of observations, we expect that our two-planet fit might underestimate the true jitter, as the system will be overfit. For now we have simply removed the long period planet, however, it results in a large jitter for this star. Given that it is well above our relation, we believe this provides further evidence that the two planet model is closer to correct. 

We notice several other high jitter stars in the various plots, but the above two cases show clear examples of using the jitter floor to identify stars that appear to be out of place, indicating the presence of additional RV variation that has not been accounted for, namely planets. This type of identification is useful only for large offsets typical of Jovian type planets but is still useful for prioritizing targets and optimizing the use of our resources.

\subsubsection{Active Subgiants}
We observe from \autoref{fig:all_rms_logg} a sample of stars with $ 3 < \logg < 3.5$ that are well above our jitter floor. These stars also show high levels of magnetic activity, generally unexpected for subgiant stars. However, this is the result of using a broad stellar sample, covering a wide range of masses and evolutionary stages. As explained above, these are merely intermediate-mass stars that are still spinning down, and evolving quickly before they can lose their magnetic activity. This is evidenced by the lack of any stars in this range of surface gravities that appear to have high jitter and abnormally high activity when looking at the plots separated by mass bins. We also note that a decent fraction of these active subgiants also have stellar companions, which could play a role in the observed increased activity, which was also noted in \citet{Isaacson2010}.

\subsubsection{Upturn in jitter for giant stars}
We also notice when examining \autoref{fig:all_rms_logg} an upturn in RV jitter around $\logg \sim$ 2.75. We are unable to yet confirm that this is indeed a real feature of RV jitter evolution and not merely a few high points. If it is indeed real, this could perhaps indicate that the oscillation scaling relation has a sharper scaling with $\logg$ than the theoretical relations \citep{Kjeldsen2011,Corsaro2012} suggest. Given that it is only apparent in \autoref{fig:all_rms_logg} for our entire sample and less so in the same plots broken into mass bins, it is likely not real. However, our sample includes very few stars that have evolved to $\logg \sim 2.75$, and so the lack of evidence in the plots broken up by mass could merely be due to small sample size for these types of giant stars. A future goal is to extend this further and increase our sample of giant stars for precisely this reason.

\subsubsection{Theoretical Bump in RV jitter for Intermediate Mass Stars}
As can be seen in Figures \ref{fig:rms_logg1} and \ref{fig:rms_logg2}, for stars above roughly 1.2~M$_{\odot}$, the theoretical oscillation and granulation components to RV jitter go through a period of increased RV jitter from about $\logg$ 4 to 3.75 for the 1.2~M$_{\odot}$ stars and moving to later evolutionary stages ($\logg$ 3.75 to 3.5) for the most massive stars in our sample. We find that this region corresponds to the onset of hydrogen shell burning, which has the effect of restructuring the star. This occurs differently in stars above $\sim$1.2~M$_{\odot}$ than for lower mass stars, and can be seen in the evolutionary tracks for these stars as the period where the stars move back up and left on the HR diagram as they temporarily become hotter and more luminous (see \autoref{fig:evolution}). As a relatively small effect, it is likely only noticeable in the purely theoretical tracks and is unlikely to be observable empirically due to slightly different stellar parameters ($\Teff$, M, \SHK, [Fe/H]) between stars in the sample. Even with the higher precision offered by next-generation RV instruments, we expect this bump might only be observed with a dedicated sample of nearly identical stars with similar observing strategies that specifically target granulation effects, and even then we do not expect it to be very likely.

\section{Discussion} \label{sec:discussion}
The conclusions drawn from this investigation have relied heavily on the examination of the RV jitter \emph{floor}. We remind readers that our vetting process for calculating RV jitter has undoubtedly included many stars that have additional non-stellar components which are inflating the measured RV jitter. However, by examining the \emph{floor} of RV jitter, we expect that we have primarily focused on those stars for which all dominant non-stellar components of RV jitter have been removed. 

Our results here have very serious implications for our understanding of RV jitter. We have shown the relationship between a star's evolutionary state and its radial velocity jitter. In fact, radial velocity jitter seems to track stellar evolution quite well, tracing main sequence (and post main sequence) spin down and subsequent decrease in magnetic activity until a star becomes inactive and its RV jitter follows the increase in convective power as both the characteristic size of a granular region and the p-mode oscillation amplitude grow with decreasing surface gravity. While the basic path is the same for every star, stars of different mass have subtle differences in where these transitions occur relative to the main sequence lifetimes. 

From this work, it is clear that RV jitter depends on primarily three stellar parameters: stellar mass, surface gravity, and magnetic activity. With these three parameters, we can establish an expected RV jitter. Future work will analyze this in more depth, and various methods to predict jitter. We also plan to investigate which types of stars have systematically large or small discrepancies between the observed and predicted jitter. 

Furthermore, the basis of this work is to better understand the astrophysical drivers of RV jitter in order to better identify suitable RV targets. The following section highlights the important implications of this work on RV surveys. 

\subsection{Implications for RV Surveys}
We break our implications into 3 broad categories: implications for informing target selection for RV surveys, prioritizing targets for RV follow-up, and identifying stars with abnormally high jitter.

\subsubsection{Target Selection for RV Surveys}
The RV community has long been aware of the relation between RV jitter and magnetic activity. Our results as they relate to magnetic activity align with previous results with main insight that binning by mass shows very clear relations rather than purely spectral type. We observe cleaner relations between jitter and activity due primarily to improved stellar parameters and thorough approach to calculating RV jitter. RV surveys will continue to avoid active stars, especially those that are higher mass, as they are seen to exhibit higher jitter than the low mass stars (as high as 60 m/s in 1.1 to 1.2 M$_{\odot}$ compared to less than 20 m/s in 0.8 to 0.9 M$_{\odot}$). In other words, in terms of RV jitter, more massive stars are more sensitive to activity.\footnote{We note that it is unclear why exactly this is the case. Under a purely spot-model assumption, the larger radii of the more massive stars would naively lead one to assume that RV jitter would decrease with activity due to the relatively smaller spot size compared to the stellar disk. Observing the opposite effect therefore might indicate that the spots themselves are larger on more massive stars, or perhaps have a higher spot coverage. But there are many sources of activity-induced jitter, not solely spots. Indeed, some stars have high RV jitter dominated not by spots but the bright magnetic regions. We refrain from speculating on this point as it is still unclear exactly how activity manifests in stars of different masses and in turn how that translates to the measured radial velocity. Lastly, it is worth reminding that the Calcium H\&K lines are formed in the chromosphere of stars and the velocities are measuring photospheric effects.}

In our analysis of $\logg$, we identified the jitter minimum for stars of mass $0.7 < $ M$_{\star} < 1.7$~M$_{\odot}$ (see \autoref{tbl:jitter_minima}). This will enable RV surveys to tailor their stars to those expected to be at jitter minimum, and could expand exoplanet searches into regimes that have not been previously explored, although we note that an efficient target selection requires the knowledge of the stellar mass, evolutionary status ($\logg{}$) and magnetic activity (e.g., \SHK). Our understanding of planetary system formation and evolution is hampered by our inability to detect planets, particularly in RVs in a wide variety of regimes. With this study, we can begin to strategically address this.

More generally, this study enables a more refined selection at a time when more precise and accurate stellar parameters are more readily available. It also enables surveys to better select specific focus points, for example: subgiants, main sequence stars at jitter minimum, more evolved stars with sufficiently low astrophysical jitter, etc. It also highlights areas for focused observations for further jitter studies, which may enable a refined understanding of jitter in areas where we currently suffer from selection effects, thus permitting planet searches in sparsely populated areas of the jitter-mass-$\logg$-activity space.

\subsubsection{Target Prioritization for RV Follow-up}
Similarly, we expect that this framework will be invaluable for prioritizing targets for RV-followup. While we advise prioritizing stars with low expected jitter and avoiding stars with high expected amounts of RV jitter, we acknowledge that this will not stop RV teams from giving high priority to targets that present the most scientifically interesting cases, regardless of expected jitter. However, in these cases we encourage using estimated or expected amplitude of RV jitter as well as the knowledge of the expected dominant source of RV jitter to better prepare and select proper observing strategies, in an attempt to model or remove the RV jitter. For example, \citet{Dumusque2011a} and \citet{Medina2018} give observing strategies for mitigating the effects of granulation as well as p-mode oscillations. Further, several studies have shown promising results in modeling stellar RV variability due to magnetic suppression of convective blueshift and show that proper longitudinal coverage and simultaneous photometry can significantly reduce variations \citep{Aigrain2012,Haywood2014}.

\subsubsection{Identifying Easy Planet Candidates}
As we have already seen in this work, we can use these relations to identify stars that have unusually high jitter compared to the expected value (\autoref{sec:jitter_evolution_features}). In these cases, the high RV jitter is likely due to unsubtracted companions. By comparing the observed jitter to the expected jitter, one can easily identify potential planet candidates and pick out stars that would benefit from additional observations. For example, stars in the mass range $0.9 < $ M$_{\star} < 1.1~\mathrm{M}_{\odot}$ at $\logg \sim4$ that have an RV RMS of more than 10 m/s are good candidates for potential planets.

\subsection{Limitations}
\subsubsection{Observational Biases in the CPS Sample}
In using data from the California Planet Search it is important to highlight some of the limitations imposed due to the nature of the survey. As mentioned previously, we see an instrumental noise floor of about 2.5 m/s, and so we are unable to say anything about the RV stability of stars below that level. For some stars in our sample, we have measured RV RMS less than 2.5 m/s. We have opted to report these as measured to ensure consistency, since the instrumental uncertainty indeed varies from star to star and we expect many stars to be RV stable below the 2.5 m/s level, but great care should be taken with these stars.

Furthermore, the California Planet Search did not follow any specific observing strategy for their targets. This means that the cadence and duration of observations can vary dramatically from star to star, and even for a single star throughout its observation history. Since the astrohpysical processes that drive stellar RV jitter as described in this work all operate on different timescales (minutes to years), the cadence of observations for a given star will be likely to probe some of these processes more than others, which will affect the measured RV RMS. As was shown with HD 142091, the high-cadence sampling to trace out the p-mode oscillations resulted in the majority of the observations being taken over a timespan shorter than the typical timescale for other sources of RV variation and as a result those sources were not reflected in the overall RMS. In general, the stars in this sample are observed infrequently enough that no single process should dominate the RV jitter, justifying our choice to add the theoretical granulation and oscillation components in quadrature as they will both be evenly and randomly sampled.

We also suffer from observational biases in the CPS survey, which have largely targeted inactive main sequence stars. As a result, our leverage on activity and evolution effects on RV jitter is not as strong as it could be. In both of these areas - more active and more evolved - stars in general have fewer observations. In terms of activity, this likely arises from observing a given star for a period of time, before noticing that it had high jitter and thereby removing it from the list of targets for further observations. We also find that many of the most active stars were added more recently to the sample in an effort to explore young stellar systems as part of the Formation and Evolution of Planetary Systems \emph{Spitzer} Legacy Program \citep{Meyer2006} and so have not had as long of a baseline to gather many observations. In terms of evolution, we previously noted how the stars in the ``Retired" A star survey have been added more recently and also suffer from shorter time baselines and fewer observations. Adding to this is the tendency of these stars to be slightly more massive than solar and host planets on longer periods, both of which work to reduce the expected semi-amplitude of potential planets. Thus these stars require more observations than usual to disentangle planet signals and adds to the likelihood of planetary signals being remaining in our data despite our best efforts to remove them all.

\subsubsection{Choice of Number of Observations Threshold}\label{sec:nobs}
In this work, we have chosen to measure the jitter for only those stars with 10 or more RV observations. This limit, while arbitrary, was based on having sufficient observations to see signs of a planet signal, linear trend, or other non-stellar phenomenon in the radial velocity time series. In practice, we find that the threshold for accurately calculating jitter to the 2.5 m/s level requires closer to twenty or thirty observations. However, we have a large number of stars in our sample that have between ten and twenty observations and so we have opted to make ten our threshold to include these stars, given that our goal is to identify the jitter floor. In particular, most of the stars with 10-20 observations lie in the areas where we get most of our leverage (active stars and evolved stars), which is why it was important to include them. By making the threshold 10 observations, we are able to build up a statistically significant sample to define the jitter floor for these areas where we suffer from observational bias at the expense of adding several stars with jitter above the floor.

\section{Summary and Conclusions}\label{sec:summary}
In this work we have presented a comprehensive analysis of RV jitter that builds upon previous analyses in several ways. A major improvement is the additional years of observations from which we calculate jitter. Further, our results hinge upon the precise stellar parameters provided by B17, most notable of which are the accurate measurements of $\logg$. To calculate jitter, we focus on removing any RV variations that are not due to surface features on the star itself. This primarily means removing planetary and stellar companions, but also includes removing velocities during transit for some stars that show Rossiter-McLaughlin effects, an in depth search for outliers, and other unusual circumstances which may lead to non-astrophysical RV variations. We impose a conservative subjective vetting technique that we apply on a star-by-star basis. By imposing a general method of attempting a one-planet fit to each star, we can tailor the technique to each star individually, iterating with new constraints on the fit until we favor or reject a companion fit. By not imposing a one-size-fits-all model to subtract out planets and other effects from our heterogeneous sample, we are able to verify the measurements of RV jitter for each star in our sample. 

\begin{figure}
\includegraphics[width=\columnwidth]{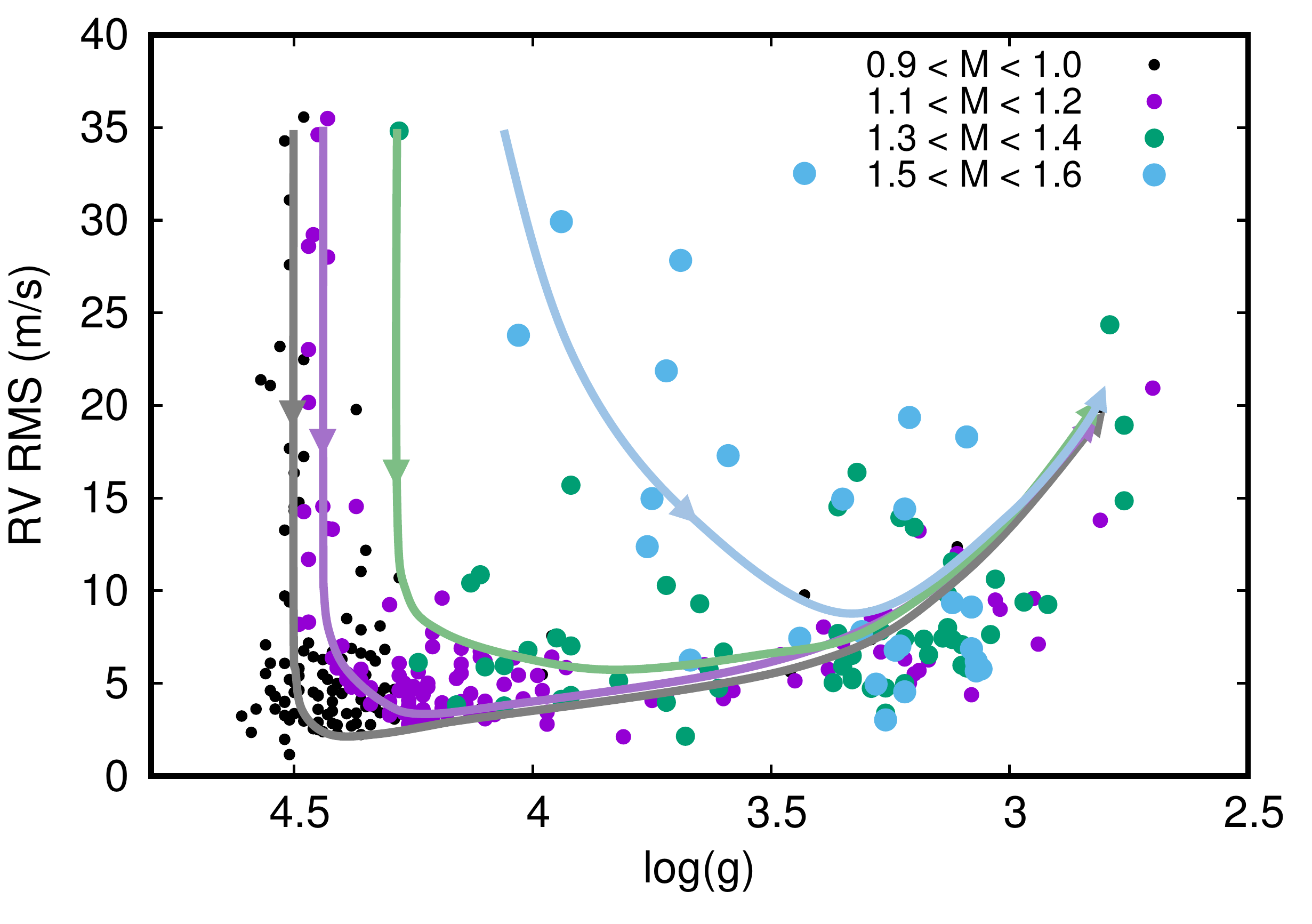}
\caption{Schematic of jitter evolution using data for stars in four mass bins. The bins are discontinuous to better highlight the effect of mass on jitter evolution. Lines have been drawn (not fit) to represent the general evolutionary path suggested by the data for each mass bin. Here it is clear how more massive stars begin to evolve before fully spinning down, such that the most massive stars move diagonally down and to the right as they transition from activity-dominated to granulation-dominated and eventually oscillation-dominated jitter.}
\label{fig:jitter_evolution}
\end{figure}

We first examined the evolutionary dependence of RV jitter by sorting our sample into mass bins of width 0.1~M$_{\odot}$. By doing so, we observe empirical evidence of two regimes of RV jitter: activity-dominated, and convection-dominated. Drawing from these observations, we conclude that RV jitter tracks stellar evolution and that most stars pass through the following stages of RV jitter:
\begin{enumerate}
\item{Zero-age main sequence stars are born with high magnetic activity that drives large RV jitter.}
\item{As a star spins down on the main sequence, it loses angular momentum through magnetic winds and RV jitter decreases with decreasing magnetic activity.}
\item{As a star becomes magnetically inactive, or ``quiet", the dominant driver of RV jitter is surface convection. The transition from magnetically-driven to convection-driven RV jitter defines the minimum level of RV jitter in a given star's evolution.}
\item{As a star moves through the subgiant and later giant regimes, the RV jitter increases as the p-mode oscillation amplitude increases and the number of granules on the star decreases due to the increased size of a characteristic granular region (which is related to the pressure scale height).}
\end{enumerate}
By investigating the different mass bins, we see subtle differences between different spectral types. As you go toward higher masses, the RV jitter minimum (and the transition from magnetically-driven to convectively-driven jitter) occurs at later evolutionary stages. This is due to both the shorter evolution timescales for higher-mass stars and the lack of convective envelopes while on the main sequence for the highest masses in our sample (above $\sim 1.3$~M$_{\odot}$), which are thereby unable to spin down due to angular momentum loss from stellar winds until they evolve off the main sequence and gain a convective envelope. These trends with mass are shown in \autoref{fig:jitter_evolution}, which shows a schematic of the jitter evolution for stars of various masses. 

We also performed a similar analysis of RV jitter by examining the RV jitter as a function of magnetic activity. Our results are consistent with many previous studies of RV jitter which indicate strong correlation between RV jitter and magnetic activity among main sequence stars. We confirm the result in \citet{Isaacson2010} that for K dwarfs, RV jitter has a very weak dependence on activity. Instead, these stars show surprisingly low levels of RV jitter for even the most magnetically active stars. 

Due to the nature of stellar evolution for more massive stars, stellar spin down for stars above the Kraft break does not occur until they have evolved off of the main sequence and have gained a convective envelope. For these more massive stars (above 1.3 M$_{\odot}$), they remain in the activity-dominated regime throughout the subgiant phase. For these stars, activity becomes the better predictor for evolved star RV jitter rather than convection.

In summary, we observe 3 different classes of stars in terms of RV jitter:
\begin{enumerate}
\item{Low mass stars ($< 0.9$~M$_{\odot}$) that are strictly activity-dominated (i.e. have not evolved enough to reach their convective phase)
\begin{itemize}
\item{Despite the wide range of S values, all of these stars seem to exhibit relatively low jitter}
\item{Not only have these stars not evolved enough to become convectively driven, but there's evidence for a lower-mass limit where stars have not yet spun down enough to reach jitter minimum.}
\end{itemize}}
\item{Solar-ish mass stars ($0.9 \leq $~M$_{\star} < 1.5$~M$_{\odot}$) that display activity and convection-dominated phases
\begin{itemize}
\item{All of these stars, including F dwarfs, display activity- and convection-dominated regimes. The two regimes meet in a ``jitter minimum," meaning that even F dwarfs, typically avoided by RV planet searches, can be amenable to RVs (see Luhn et al. in prep)}
\item{Jitter minimum for these stars occurs on the main sequence or late in its MS lifetime, near or right at the TAMS.}
\end{itemize}}
\item{More massive stars ($>1.5$~M$_{\odot}$, beyond the Kraft break) which only enter the RV surveys when they're evolved and for which activity once again becomes the key predictor of RV jitter because these stars have now developed convective envelopes (and presumably solar dynamo-like activity.)
\begin{itemize}
\item{The most evolved of these stars in our sample appear to just be reaching the expected jitter minimum and show RV stability to 3-10 m/s.}
\item{Jitter minimum for these stars occurs well into the subgiant and giant phases.}
\end{itemize}}
\end{enumerate}

Finally, we comment on the utility of these relations. From our sample of stars with well-measured stellar properties and RV jitter, we can build a model that is able to predict the expected RV jitter of a star given a variety of input stellar parameters and uncertainties. We expect that these results will be particularly beneficial for continued RV follow-up of planet candidates from \emph{K2} and \emph{TESS}. We are also able to use these relations to identify stars that have noticeably higher-than-typical RV jitter, which indicates the potential presence of unsubtracted orbital companions. We can then select these targets for further observations to fully extract the companion.

In other words, our relations have highlighted two classes of stars for which RV observations are particularly useful: those stars that are at the ``jitter minimum" of their evolution, and those stars that are already observed to have RV jitter well above the ``jitter floor". Both of these classes present good regimes in which to look for planets. In addition to identifying promising targets for RV follow-up, our results can help inform target selection for RV surveys in a similar manner by identifying which stars are the most RV stable.

\acknowledgments{
The authors thank Fabienne Bastien for her founding role in initiating the investigation and defining the project. Her insight, discussions, and advice have greatly contributed to this work.

We thank John Brewer for many useful discussions and clarifications regarding the spectroscopic stellar properties. We thank John Johnson for use of the data on HD 142091. We thank Rapha\"elle Haywood and Tim Milbourne for their discussion on solar granulation and the HARPS data. We thank Sharon Wang for helpful discussions regarding solar oscillations. 

Finally, we thank the referee, whose suggestions have helped clarify and focus the work presented here.

The data presented herein were obtained at the W. M. Keck Observatory, which is operated as a scientific partnership among the California Institute of Technology, the University of California and the National Aeronautics and Space Administration. The Observatory was made possible by the generous financial support of the W. M. Keck Foundation.

The authors wish to recognize and acknowledge the very significant cultural role and reverence that the summit of Maunakea has always had within the indigenous Hawaiian community.  We are most fortunate to have the opportunity to conduct observations from this mountain.

Keck time for this project has been awarded from many sources, primarily institutional time from the University of California, Caltech, NASA, and Yale. We thank the many observers and CPS team members who have worked over the decades to produce this invaluable data set.

This research has made use of the SIMBAD database, operated at CDS, Strasbourg, France; the Exoplanet Orbit Database and the Exoplanet Data Explorer at exoplanets.org.; and of NASA's Astrophysics Data System Bibliographic Services. This work was partially supported by funding from the Center for Exoplanets and Habitable Worlds, which is supported by the Pennsylvania State University, the Eberly College of Science, and the Pennsylvania Space Grant Consortium. This material is based upon work supported by the National Science Foundation Graduate Research Fellowship Program under Grant No. DGE1255832.}

\bibliography{\string~/Google_Drive/Research/library}

\appendix

\section{Calculation of Error on Jitter Measurements}\label{sec:jitter_error}
Since the RV RMS is already a measurement of error and represents the standard deviation of the RV observations of a given star, we seek to calculate the uncertainty in our estimate of the true standard deviation (essentially the standard deviation of the standard deviation). For simplicity of notation, we find it helpful to define the variance
\begin{equation}
v \equiv j^2,
\end{equation}
which will help in transforming between standard deviation (j) and the variance. Where the standard deviation of the data is related to the second moment of the dataset (the mean being the first moment), the variance of the sample variance is related to the fourth moment of the data. The exact statistical relation comes from \citet{Mood1974},
\begin{equation}
\sigma^2_{j^2} = \sigma^2_{v} = \frac{\mu_4}{n} - \frac{j^4 (n-3)}{n\left(n-1\right)},
\label{eqn:jitter_error}
\end{equation}
where $\mu_4$ is the fourth moment of the data, defined as
\begin{equation}
\mu_4 = \sum\left(\epsilon_{i}-\overline{\epsilon}\right)^4.
\end{equation}
However, we are interested in knowing $\sigma_{j}^2$, not $\sigma_{j^2}^2$, and so we must use error propagation               
\begin{equation}
\sigma_{j}^2 = \sigma_{v}^2 \left|\frac{\partial j}{\partial v}\right|^2
\end{equation}
which gives
\begin{equation}
\sigma_{j} = \frac{\sigma_{j^2}}{2j}
\label{eqn:uncertainty}
\end{equation}
The derivation of this formula assumes a normal distribution and does not account for individual error measurements. However our individual measurement uncertainties are largely homoskedastic, and our conclusions are not sensitive to the precision of our jitter uncertainties. Note also that this estimation of error in our calculation of RV jitter does not take into account the $\chi^2$ of any subtracted fits nor the possibility of the data containing additional companions. It is merely a measurement of the uncertainty in calculating $\sigma_{\mathrm{RV}}$ from the residuals, $\epsilon$.

\section{Notes on Individual Systems}\label{sec:notes}
Here we list many individual systems and provide notes on what alterations to the RV time series we have applied when calculating RV jitter. We provide this for completeness and reproducibility. We additionally note stars that show strong correlations between the radial velocities and activity metric \SHK, the vast majority of which have been independently extracted from the same spectra and published by \citet{Butler2017}.

\subsection{$0.7 \leq$ M$_{\star} < 0.8$~M$_{\odot}$}
\paragraph{HD 4628}
HD 4628 shows evidence of an activity cycle. However, we notice several low outliers in the activity time series that suggest rapid decreases in activity. A closer investigation reveals that these ``ramp downs" in \SHK{} occur at the beginnings and ends of the observing season. As a bright (V=5.74) target, we interpret the sudden decreases in activity as observations taken during twilight and contaminated by the solar spectrum. We do not see evidence for any correlation between the activity and the velocities and so we are satisfied with the reported jitter. We see similar effects in HD 26965, HD 69830, and HD 192310 below.

\paragraph{HD 10700}
HD 10700 (also known as Tau Ceti) has 4 reported planets all with semi-amplitudes of 0.55 m/s or less \citep{Feng2017}. As an RV standard star, it has a long baseline of observations (917 over 17 years) and there are several obvious outliers. We remove 8 points that have velocities more than 3$\sigma$ from the mean, which reduces the RV RMS from 3.23 to 2.75 m/s. After subtracting the best-fit multi-planet Keplerian, the RV jitter is essentially unchanged, 2.75 to 2.73, indicating that we are at the instrumental uncertainty for this star.

\paragraph{HD 26965}
HD 26965 has a proposed planet candidate with small semi-amplitude \citep{Diaz2018}, who also acknowledge that it could simply be stellar activity masquerading as a planet. We choose to subtract the planet given the good agreement, but we note that the RV jitter is relatively unchanged (3.6 m/s to 3.2 m/s). We also note an apparent activity cycle in this star, with low \SHK{} outliers reminiscent of the ``ramp downs" seen in HD 4628. As another bright target (V=4.43), we again attribute these outliers to solar contamination from chasing the star into twilight. We note that the lowest \SHK{} values for this star are unphysical ($< \sim 0.1$), which could indicate errors in the \SHK{} extraction for this star.

\paragraph{HD 31560}
HD 31560 shows a strong correlation between the activity and the radial velocities (Pearson coefficient 0.76). However, this correlation is based on only a few observations.

\paragraph{HD 97658}
HD 97658 shows a strong activity cycle that is not very correlated with the radial velocities (Pearson coefficient 0.31 after removing the known planet). 

\paragraph{HD 100623}
HD 100623 shows evidence of a long term linear trend, which we have subtracted out.

\paragraph{HD 116443}
HD 116443 shows evidence of an activity cycle, with no correlated radial velocities.

\paragraph{HD 170657}
HD 170657 has 3 obvious outliers that have internal errors of about 25 m/s. After removing these points, there was also a long term linear trend evident in the data, which we have subtracted.

\paragraph{HD 220339}
HD 220339 shows evidence of a possible activity cycle. More observations are necessary to confirm this signal. The activity time series show little to no correlation with the velocities.

\paragraph{HD 196124}
HD 196124 shows a correlation between the activity and the radial velocities (Pearson coefficient 0.62). It is unclear whether the star is exhibiting cycling behavior as it undergoes a large decrease in activity over several years and then was not observed until several years later when it was higher in activity. It is likely indicative of a cycle, but more data is needed.

\subsection{$0.8 \leq$ M$_{\star} < 0.9$~M$_{\odot}$}
\paragraph{HD 10476}
HD 10476 shows evidence of a possible activity cycle that is weakly correlated with the radial velocities (Pearson coefficient 0.466).

\paragraph{HD 18143}
HD 18143 has a long term linear trend, which we fit with a Keplerian (RV jitter reduced from 4.7 m/s with a linear trend to 4.3 with a Keplerian fit). This star shows evidence of an activity cycle with period near 4500 days, however the observations span at most one full period and likely only a fraction of a period so the exact cycle period is uncertain.

\paragraph{HD 20165}
HD 20165 shows evidence of an activity cycle and correlated radial velocities (Pearson coefficient 0.66).

\paragraph{HD 42250}
The velocities of HD 42250 taken prior to the Keck-HIRES upgrade appear to be systematically offset from those take post-upgrade. We include an offset of 6.6 m/s, which reduces the RV jitter from 5.4 m/s to 3.7 m/s.

\paragraph{HD 69830}
HD 69830 has 3 published planets \citep{Lovis2006}, which we are able to recover. However, similar to HD 4628 above, we notice several low outliers in the activity time suggesting rapid ``ramp downs" in \SHK{} that occur at the beginnings and ends of the observing season, more apparent in this star than in HD 4628. As a bright (V=5.95) target, we interpret the sudden decreases in activity as observations taken during twilight and contaminated by the solar spectrum. The low \SHK{} values do not affect the jitter and so a deeper dive into the cause of these low outliers is beyond the scope of this work. Nevertheless, this star remains an interesting target for investigating activity and the effects of solar contamination on measurements of stellar activity.

\paragraph{HD 114783}
HD 114783 has a single planet with a 493 day period \citep{Wittenmyer2009}. The residuals to this fit show strong periodicity near 4300 days. However, this star has a noticeable activity cycle with a period near 3000 days. Given the lack of correlation and the fact that the planetary signal and activity signal are out of phase and not on the same period, we are inclined to believe the second planet.

\paragraph{HD 124106}
HD 124106 has an obvious outlier by eye that lies about 70 m/s above the rest of the observations. The RV jitter is reduced from 15.4 m/s to 8.2 m/s when this outlier is removed. We note that this is a reasonably active star and so it is possible that this outlier was due to a flare.

\paragraph{HD 125455}
HD 125455 shows evidence of a possible activity cycle that is not correlated with the radial velocities.

\paragraph{HD 154345}
HD 154345 is a particularly interesting star. \citet{Wright2008} claimed the presence of a Jupiter twin in a 9 year orbit, while noting that the star has an activity cycle that is nearly the same period and in phase with the velocities. By looking at similar cycling G stars and not finding a correlation between the activity cycle and the RV's, the period and phase similarity of the proposed planet signal and the activity cycle was suggested to be coincidence. Later, with additional data \citet{Wright2015} reversed the claim of a planet after including several more years of additional observations that continued to show strong phase and period similarities. For this work, we again reverse the stance and decide to subtract off the planet because of the coherent RV signal. No other system in this sample with activity cycles that are correlated with the RV's shows such a clear and coherent Keplerian-like signal. Further, we note that within our own solar system the solar cycle is very similar in period to Jupiter's period. With Jupiter being the dominant radial velocity signal in the sun, a distant observer could very well have the same conundrum with our own system. Either way, this remains a very interesting system to study more closely. For now, we have subtracted off the planet, resulting in an RV jitter of 3.2 m/s, but note that the unaltered RV's yield an RV jitter of 12.7 m/s. Given that it is not an overly active star, we find that 3.2 m/s matches a little more closely with similar stars (see \autoref{fig:rms_s1}). 

\paragraph{HD 165401}
HD 165401 shows evidence of a long term linear trend, which we have subtracted out.

\paragraph{HD 189733}
The velocities for HD 189733 contain a strong Rossiter-McLaughlin signal, which contributes to the residuals and the RV jitter. Removing the points during transit results in a change in RV jitter from 15 m/s to 11 m/s.

\paragraph{HD 192310}
HD 192310 is similar to HD 4628 and HD 69830 above where we observe ``ramp downs" in the activity time series near the end of the observing season. As another bright (V=5.73) target, we believe this is another case of chasing a star into twilight, resulting in solar contamination, or errors in the \SHK{} extraction. We note that the lowest \SHK{} values for this star are unphysical ($< \sim 0.1$). This star also has a long term activity cycle with a period close to 10 years.

\paragraph{HD 219538}
HD 219538 appears to exhibit a possible decaying activity cycle, similar to HD 4915, which was identified as a possible Maunder minimum candidate in \citet{Shah2018}, and is detailed in the next mass bin below. Unlike HD 4915, there is no correlation with the velocities for HD 219538.

\paragraph{HD 131156}
HD 131156 has a long term linear trend, which we have subtracted off.

\subsection{$0.9 \leq$ M$_{\star} < 1.0$~M$_{\odot}$}

\paragraph{HD 4614}
HD 4614 shows evidence of a long term linear trend, which we have subtracted out.

\paragraph{HD 4915}
HD 4915 shows strong correlation between the radial velocities and the decaying activity cycle (Pearson coefficient 0.83). This star was recently identified as a possible Maunder minimum candidate, as the decaying activity cycle shows similar features to the the solar entrance into Maunder minimum \citep{Shah2018}.

\paragraph{HD 19467}
The directly imaged T Dwarf HD 19467 B has a period of more than 300 years and a mass of 52 $M_{\mathrm{Jup}}$ \citet{JensenClem2016}. The period of this companion is too long for our observations and exceeds the 100 year period limit imposed by RVLIN. Therefore we merely subtract the best fit linear trend from the data, since we are unable to fit this companion.

\paragraph{HD 20619}
Despite not having a strongly noticeable activity cycle, there is a strong correlation between the velocities and activity measurements for HD 20619 (Pearson coefficient 0.73).

\paragraph{HD 24496}
We have subtracted a linear trend from the velocities of HD 24496 and note the presence of an activity cycle.

\paragraph{HD 37213}
We note that the radial velocities for this star prior to the 2004 upgrade are all clustered around -5 to -10, with the post-upgrade velocities centered more closely around 0. However, given the small number of post-upgrade velocities, we take simply the full RV RMS of 5.479 m/s and note that the RV RMS of the pre-upgrade only velocities is only slightly higher: 4.588 m/s.

\paragraph{HD 40397}
HD 40397 shows evidence of a long term trend. However, removing a simple linear trend resulted in coherent periodicity remaining and so we have subtracted a full Keplerian signal. The RV RMS after removing a line was 3.424 m/s, compared to the RV RMS after removing a Keplerian of 2.557 m/s

\paragraph{HD 42618}
HD 42618 has a strong periodic signal near 4000 days, which appears by eye to be correlated with the activity cycle of this star, despite the medium correlation coefficient of 0.442 after subtracting out the planet \citep{Fulton2016}.

\paragraph{HD 46375}
HD 46375 has an outlier, which we have removed, obvious from its 20 m/s uncertainty for that observation.

\paragraph{HD 90711}
HD 90711 had an obvious outlier offset by about 200 m/s. This observation also has errors that are an order of magnitude higher than the rest of the observations and so we remove this observation. The resulting time series has a evidence of a long term linear trend, which we subtract.

\paragraph{HD 96700}
HD 96700 has been previously reported as having 2 planets \citep{Mayor?2011}. The Keck velocities show no indication of such planets and since the velocities from the initial discovery are not available, we are forced to disregard these fits. The Keck velocities also show a very strong correlation with activity (Pearson coefficient 0.91), driven in large part by one observation that has both low velocity and \SHK. When we fix the orbital solution to the \citet{Mayor?2011} orbital parameters, there is only a small decrease in RV jitter from 7.4 to 6.5.

\paragraph{HD 99491}
HD 99491 shows evidence of a strong activity cycle, which are correlated with the radial velocities (Pearson coefficient 0.74).

\paragraph{HD 105631}
HD 105631 shows a strong correlation between the radial velocities and the activity (Pearson coefficient 0.72).

\paragraph{HD 109358}
HD 109358 shows evidence of an offset between the pre- and post-upgrade velocities of about 8 m/s. We include this offset, which reduces the jitter from 4.1 m/s to 3.3 m/s.

\paragraph{HD 136352}
HD 136352 was previously reported as hosting 3 small planets \citep{Mayor?2011}. However, the velocities used in the discovery were not reported and the Keck time series is poorly sampled, leaving poor constraints on the fit. Given the number of observations for this star, a 3 planet fit to the Keck-HIRES data will be over-fit and results in a RV jitter of 1.2 m/s. In order to avoid overfitting we opt to use the raw RV RMS of 4.1 m/s. We note that \citet{Udry2019} report an RMS of 1.3 m/s for this star after fitting out all 3 planets.

\paragraph{HD 149806}
HD 149806 show evidence of an activity cycle with period near 3000 days (8 years).

\paragraph{HD 164595}
HD 164595 has been previously reported to host a planet \citep{Courcol2015}. However, the Keck velocities by themselves do not show evidence for this planet and so we do not remove it.

\paragraph{HD 181234}
HD 181234 shows a long term linear trend, which we have subtracted.

\paragraph{HD 201219}
HD 201219 shows a correlation between the activity and the radial velocities, with Pearson coefficient 0.84.

\paragraph{HD 206374}
HD 206374 shows a correlation between the activity and the radial velocities, with Pearson coefficient 0.82.

\paragraph{HD 212291}
HD 212291 shows a correlation between the activity and the radial velocities, with Pearson coefficient 0.60.

\paragraph{HD 218868}
HD 218818 shows a strong activity cycle with period near 2000 days. This activity cycle is strongly correlated with the radial velocities (Pearson coefficient 0.759).

\paragraph{HD 8389}
HD 8389 shows a slight correlation between the activity and the radial velocities, with Pearson coefficient 0.54. The activity time series also shows evidence of an activity cycle.

\paragraph{HD 45652}
HD 45652 has been previously reported to host a planet \citep{Santos2008}. However, the Keck velocities are discrepant with the SOPHIE and CORALIE and ELODIE data for this star. We remain confused by this star and have not been able to track down why the Keck data does not agree at all with the ELODIE, CORALIE, or SOPHIE data, which agree with each other over many epochs. For now we have simply not removed any planet from the Keck velocities. In fact, fixing the orbital parameters to those in \citet{Santos2008} \emph{increases} the RV RMS of the Keck velocities from 28.2 to 33.9 m/s.

\paragraph{HD 62613}
HD 62613 has slight evidence of an activity cycle.

\paragraph{HD 76445}
HD 76445 shows a long term linear trend, which we have subtracted.

\paragraph{HD 162232}
HD 162232 shows a long term linear trend, which has been subtracted. After subtracting, there is evidence for an additional companion (reducing the RV jitter from 12.2 m/s to 8.2 m/s), but not strong enough to subtract out. We only subtract the linear trend.

\subsection{$1.0 \leq$ M$_{\star} < 1.1$~M$_{\odot}$}

\paragraph{HD 1461}
There is an obvious outlier present in the time series for HD 1461. The median velocity error for all observations of this star is less than 1 m/s, but the obvious outlier has an error of 70 m/s. For this reason, we remove it from the set of observations for this star. We also note that the residuals to the known planets show strong periodicity near 4226 days. However, we find that this is the same period of the star's apparent activity cycle and appears to be in phase, so we do not attempt a third planet.

\paragraph{HD 45350}
HD 45350 is host to a known planet with published period of 963 days \citep{Endl2006}. When using the best-fit parameters from \citet{Endl2006}, we obtain a rather poor fit. In fact, our resulting best-fit is worse than when using simple blind fit for this star (reduced $\chi^2$ improves from 63 to 10 and the jitter decreases from 8 m/s to 4 m/s). Given the large number of observations after publication of the most recent best fit, we are inclined to believe that we have much better constraints on the orbital parameters now, which we report in \autoref{tbl:companions}.

\paragraph{HD 73256}
HD 73256 has been previously reported to host a short period planet \citep{Udry2003}, who see a period of 2.5 days. This is the tallest peak in the periodogram of the Keck data, however, our best fit with this period results in scatter of 63 m/s compared to the 20 m/s seen in \citet{Udry2003}. Because of the poor fit, we decide to not subtract this planet in the interest of observing the true jittter floor.

\paragraph{HD 92788}
HD 92788 has 2 confirmed planets\footnote{according to exoplanet.eu}. We find good agreement with HD 92788 b on a 325 day orbit. However, we are unable to find a believable 2 planet fit with a second planet at 162 days, as claimed in \citet{Wittenmyer2013}. Instead, we find strong evidence for a long period second planet near 11000 days. While this orbit does not have complete phase coverage, we believe it to be properly subtracting out a real companion. As we obtain more observations of this system, the long period planet's orbital parameters will be further constrained. We do not observe any signs of an activity cycle for this star, and certainly not on the timescale of the more than 25 year period. For this reason, we are confident in our subtracting this best-fit model. In addition, we have removed a single outlier which has velocity error of 15 m/s, where the median error for the remainder of the observations is 1.3 m/s.

\paragraph{HD 195019}
For this system, we have split the pre-2004 data and the post-2004 data and treated them as two separate telescopes in the fitting procedure. We chose to do this after noticing that our original best fit caused nearly every point before the Keck upgrades to have a negative velocity in the residuals and nearly every point after the upgrade to have a positive residuals. We take this as evidence that the fitting procedure is minimizing $\chi^2$ by simply straddling the two sets of data. The structure seen in the residuals also resembles the signal of a long period planet. However, neither the pre-2004 nor post-2004 data shows evidence of long period trends and so we take it to simply be an incorrect offset between the pre-2004 data and post-2004 data. By including a 16 m/s offset (found by a best-fit in RVLIN), we observe residuals that no longer appear to have suspicious structure centered around the Keck upgrade date. 

\paragraph{HD 217107}
For this system, we simply take note of 3 points that appear to be outliers when examining the residuals. The best two-planet fit after using the orbital parameters from \citet{Wright2009} as initial guesses has 3 points that are all significantly above the typical scatter of the residuals. However, the three points fall on consecutive observations, and could represent real variations on the stellar surface during those observations. We do not find evidence in the reported errors on these velocities to warrant their removal and so we have kept them in the fitting and jitter calculations. Given the large number of observations on this star (142) and the relatively low (4 m/s) jitter, we do not expect these three points to significantly affect the final jitter.

\paragraph{HD 150706}
This star has a previously published planet \citep{Boisse2012} using SOPHIE follow-up to an ELODIE planet candidate. Including the Keck data for this star, we find that it loosely agrees with the planetary parameters found by \citet{Boisse2012}. However, the Keck data alone shows strong correlation with \SHK{} (0.729 Pearson correlation coefficient), indicating that the RV jitter observed could be entirely activity-driven and not due to a planet. Indeed, the SOPHIE and ELODIE data have quite large errors compared to Keck. In addition, the long period of this planet means we do not have full phase coverage. In fact, the only points at RV maximum are the earliest ELODIE observations, which are the observations with the largest error bars. Thus we are suspicious as to the veracity of this planet. Because of the low number of Keck observations for this star, we include the planet fit for now.

\paragraph{HD 207832}
HD 207832 has two previously published planets \citep{Haghighipour2012}. With additional observations we are unable to recover convincing fits. When examining the RVs, we notice a correlation with \SHK, with a Pearson correlation coefficient of 0.6. Given that \citet{Haghighipour2012} contains no discussion on the activity of this star, we are inclined to believe that the jitter is activity-driven. Further proof of this is from the periodogram itself. With the addition of several new observations since the initial discovery, we find that neither of the two periods found for the planets (162 days and 1307 days) correspond to the peak with the highest power in the periodogram, which is near 300 days. Despite our conservative approach in general, in this case we are conservative in the opposite way: we do not quite have enough evidence to definitively claim that these planets do \emph{not} exist given the care that \citet{Haghighipour2012} put into demonstrating the planet's reality and so we use the best 2 planet model.

\paragraph{HD 1388}
This star shows signs of an obvious stellar companion, which we have fit out using RVLIN. Although our orbital parameters are highly uncertain and likely incorrect, it follows the by-eye curvature quite well and so we have kept it. The residuals show evidence of potential other companions, however this would certainly result in overfitting. For now we have excluded pre-2004 observations from our jitter calculation but not from our fit. We find that the uncertainties in the velocity measurements for this star before the Keck upgrade (4-5 m/s) are substantially higher than after the upgrade ($\sim$ 1 m/s). 

\paragraph{HD 8038}
For this star, we find a very high correlation between \SHK{} and the RV observations (Pearson correlation coefficient 0.93). We therefore take the observed RV variations as pure jitter.

\paragraph{HD 9986}
This star also shows a fairly strong activity correlation (Pearson correlation 0.5), but more importantly shows evidence of a strong activity cycle, with a periodogram peak near 1000 days, which is near the strongest peak in the RV periodogram. We expect that the observed cyclical RV variations are activity-driven.

\paragraph{HD 13931}
For this star we have removed an obvious outlier, whose velocity uncertainty of 8 m/s is significantly higher than the median uncertainty for this star (1.2 m/s). This star also has a strong planet signal, which further shows that this point is an outlier.

\paragraph{HD 18803}
HD 18803 shows evidence of a strong activity cycle, and has \SHK{} values that correlate well with the RV observations (Pearson coefficient of 0.51). We disregard any strong peaks in the RV periodogram of this star and take the RV variations for this star as jitter. 

\paragraph{HD 32923}
For this star, we notice a long term trend in the radial velocities. We therefore subtract a linear fit for this star before fitting a blind fit to the data. The best-fit planet with a line removed does not meet our criteria and so we simply remove only the linear fit from the velocities.

\paragraph{HD 39881}
HD 39881 has what appears to be either a very slight offset between the pre- and post-upgrade velocities or a very slight linear trend. The difference in RV jitter between these approaches is very small: 4.0 m/s when subtracting a linear trend and 3.8 m/s when applying an offset between the pre- and post-upgrade velocities (the raw RV RMS is 4.5 m/s). We choose to use the linear trend to better avoid over-fitting.

\paragraph{HD 52711}
HD 52711 has an outlier in the \SHK{} time series. After removing this point, there is a slight correlation between the radial velocities and the activity (Pearson coefficient 0.61), which is why we disregard the blind-fit planet.

\paragraph{HD 68168}
HD 68168 has several observations in short succession that span nearly 40 m/s. Other than these few points, the star appears to have some slight cyclical variation to it. However, we find that the \SHK{} values are well correlated (Pearson coeffecient 0.61), indicating that these observed RV variations are simply activity-driven jitter. We have also removed 2 observations that were taken during a gibbous moon through 1-3 magnitudes of clouds.

\paragraph{HD 233641}
HD 233641 shows a very strong correlation between activity metric \SHK{} and the measured velocities (Pearson correlation coefficient 0.95), again indicating activity-driven RV jitter.

\paragraph{HD 153458}
HD 153458 also shows a very strong activity-RV correlation (Pearson coefficient 0.96), indicating that RV variations are driven by activity in this star.

\paragraph{HD 157347}
HD 157347 has 3 outliers due to velocity errors more than 2.5 times the median errors for this star, although the velocity measurements themselves are consistent with the rest of the data. Nevertheless, we remove these 3 observations.

\paragraph{HD 157338}
HD 157338 has two obvious outliers in the RV time series. These are identified by eye and also by their large uncertainties (6.3 and 18 m/s) compared to the median uncertainty (1.5 m/s) for this star. It has an obvious stellar companion with a long period that cannot be well constrained. We are able to find best-fit 2 companion models with a planet and a stellar companion, however we reject these fits based on convenient lack of phase coverage near sharp RV peaks in models that find rather large eccentricities. We posit that more observations are necessary before we can definitively believe a 2 companion model. We therefore restrict ourselves to circular orbits for the obvious stellar companion, re-emphasizing that it is the \emph{jitter floor} we seek. Our stated jitter of 11.5 m/s is likely an overestimate of jitter, but we are careful not to over-subtract (our best 2 companion models yield jitter of $\sim$ 4.7 m/s). 

\paragraph{HD 85689}
HD 85689 has one extremely low point separated by more than 100 m/s, but otherwise seemingly normal. We assume that this is an anomalous point and we have removed it for that reason, which brings the jitter from 25 m/s to 6 m/s.

\paragraph{HD 92719}
The by-eye inspection of this star shows what appears to be a strong activity correlation, despite a medium Pearson coefficient (0.46). Regardless, we find no evidence of Keplerian RV variations, and so we do not perform any changes to this star.

\paragraph{HD 197076}
We see strong evidence of an activity cycle for this star, with a period of $\sim$ 2000 days. This also corresponds to the peak in the RV time series periodogram. The correlation between activity and velocity is quite strong (Pearson coefficient 0.63) and so we simply take everything as activity-induced jitter.

\paragraph{HD 12484}
HD 12484 has 1 known planet \citep{Hebrard2016} and is an active star with high jitter (43 m/s). After subtracting out the planet, we find a strong activity correlation with the residual velocities (Pearson coefficient 0.74).

\paragraph{HD 45184}
After removing an obvious outlier in both of the activity and velocity time series, we find a correlation (Pearson coefficient 0.67) between the activity index and the residual radial velocities for HD 45184 after subtracting out the best fit planet \citep{Mayor?2011}.

\subsection{$1.1 \leq$ M$_{\star} < 1.2$~M$_{\odot}$}

\paragraph{HD 13043}
The RV observations for HD 13043 show some coherence which could be evidence of an activity cycle or a planet. Further observations are necessary to fully determine what is driving the apparent coherence and so for now, we do not remove any signals.

\paragraph{HD 107148}
HD 107148 has a purported planet with period of 48 days \citep{Butler2006}. We see no evidence for a planet at 48 days and instead find a strong peak near 77 days, which RVLIN finds as the best-fit period. When we remove our best fit for this 77 day planet, a tall peak in the periodogram shows up at $\sim$20 days. Our best fit to a 2 planet model is not convincing enough to accept. However we note that this star is worth further observations to reliably reject or accept the presence of a second planet. For this work, we merely take the RMS of the residuals to a one-planet fit as the RV jitter. 

\paragraph{HD 141004}
HD 141004 has an obvious outlier that we have removed. The error on this measurement was more than 2.5 times the typical error for this star, and the measured velocity is more than 5 times the mean jitter for this star. In this case, the resulting difference in jitter is not extreme: 5.126 m/s when the outlier is included and 5.007 m/s with it removed. However, the point remains a clear outlier and so we remove it.

\paragraph{HD 28005}
HD 28005 also contains a very obvious outlier. The error is significantly more than 2.5 times the typical error for this star. Removing the outlier, our measurement of jitter reduces substantially from 13.14 m/s to 7.24 m/s. 

\paragraph{HD 159222}
HD 159222 shows very strong correlation between RV observations and \SHK, with a Pearson correlation coefficient of 0.69. We therefore take the RV observations as activity-induced jitter.

\paragraph{HD 182572}
We have removed an outlier from the observations of HD 182572. The outlier is obvious by its larger-than-normal errors. However, the velocity of this observation is not largely displaced from the mean RV measurement and so the RV jitter is largely unaffected by the presence of this outlier.

\paragraph{HD 195564}
HD 195564 shows evidence of a clear stellar companion, which we remove. The residuals to our best fit stellar companion show coherence and what appears to be a somewhat sinusoidal signal. However, we are unable to arrive at a good two-body model and so we simply subtract the loosely fit stellar companion and treat the residuals as jitter.

\paragraph{HD 1293}
HD 1293 has 12 Keck-HIRES observations spanning 5 years (mid 2007 to mid 2012). The radial velocities rise close to linearly over the course of these observations and so we subtract a linear trend for this star. There are strong deviations from this line, which could be evidence of an additional companion that cannot yet be subtracted out due to the limited number of points. This target presents a good case for future observations to correctly fit the long term linear trend and to determine the nature of the additional perturbations to that line, whether it be pure RV jitter or an additional companion.

\paragraph{HD 77818}
HD 77818 shows evidence of a small stellar companion. There are enough observations to observe a slight curvature in the long period trend, however we resort to subtracting a linear trend until more observations are obtained. 

\paragraph{HD 128095}
HD 128095 also shows evidence of a long period stellar companion. For this star, the curvature is enough to produce a believable fit to the time series. However, the residuals to this fit show variation of $\pm$ 30 m/s. When we attempt a two-planet fit, we find a surprisingly good fit to the time-series with a roughly circular 472 day planet, and a fairly eccentric ($e=0.63$) stellar companion. However, given that there are only 14 observations for this star, we expect that our 2-planet fit has overfit, given that we are attempting to fit 14 data points with 12 free parameters. We therefore simply remove the long period trend, and encourage follow up of this star to investigate the legitimacy of a second planet (in \autoref{sec:jitter_above_floor}, we find further evidence that points toward the second companion being real).

\paragraph{HD 207077}
The periodogram of observerations of HD 207077 exhibit a strong peak near 600 days. The resulting best-fit planet model presents a convincing case for a 606 day planet with eccentricity 0.20. Despite the good fit and improved jitter from 18.66 m/s to 6.69 m/s, we decide to disfavor this planet for now and for our jitter analysis. In doing so, we can investigate how the raw jitter for this star matches with similar star, and use this as a case to favor of disfavor the planet. HD 207077 is discussed further in \autoref{sec:jitter_above_floor}.

\paragraph{HD 10212}
The RV timeseries for HD 10212 shows a clear stellar companion that passes through periapse. However, observations are sparse and only contain 2 points with negative radial velocities near pericenter. This caused RVLIN a few problems in obtaining a good fit. We noticed that no model could perform a good fit for both of these two points, with residuals for these points separated by more than 100 m/s (with every other residual near or less than 100 m/s). At first inspection, neither of these two points seemed to be a clear outlier. However upon further investigation, we find that the median $\chi$ value for one of these observations to be abnormally high, indicating that the radial velocity was poorly measured due to a poor fit to the absorption in the spectrum itself. When we remove this point, our fit improves dramatically, from a jitter of 28 m/s (and reduced $\chi^2$ of 193) to 7 m/s (reduced $\chi^2$ 23).

\subsection{$1.2 \leq$ M$_{\star} < 1.3$~M$_{\odot}$}

\paragraph{HD 8907}
HD 8907 is a fairly active star, and shows good correlation between the activity and RV time series (Pearson coefficient 0.61). Given this and the lack of a best-fit model, we expect all RV variations are driven by the activity of the star.

\paragraph{HD 28237}
We notice a strong correlation between RV's and activity in this fairly active star (Pearson coefficient 0.58), suggesting that RV jitter in this star is activity-driven.

\paragraph{HD 52265}
We find little evidence of the 59 day second planet for this star published in \citet{Wittenmyer2013}, and are unable to get a fit without invoking high eccentricity. Regardless, the best two-planet fit using Keck data results in an RV jitter of 4.1 m/s, compared to 4.6 m/s in the one-planet fit. By selecting the one-planet fit, we have not changed the RV jitter enough to affect any results.

\paragraph{HD 74156}
For this star, we find that the residuals to the 2 planet fit are all below 0 m/s for data before the Keck upgrade (with a mean of -8.7 m/s compared to a mean of 3 m/s post-upgrade), but not by more than the average scatter for the post-upgrade residuals. In fact, the lowest velocity residuals are all post-upgrade. For this reason, we have not split the pre- and post-upgrade velocities in the fit but have taken only the post-upgrade velocities when calculating jitter.

\paragraph{HD 114613}
This is a star with a known planet \citep{Wittenmyer2014} on a 3624 day period. However, the Keck data show an activity correlation (Pearson coefficient 0.695) and do not show evidence of this planet. We therefore consider the velocities to be purely activity-induced RV jitter.

\paragraph{HD 179949}
After the discovery of a companion to HD 179949 in \citet{Butler2006}, this star had additional observations starting again in 2009. After removing the best-fit companion, we notice that the pre-upgrade residuals cluster around 10 m/s, while the post-upgrade observations cluster around -10 m/s. When including an offset in the fit, the RV jitter is reduced from 15 m/s to 10 m/s. 

\paragraph{HD 188512}
HD 188512 shows evidence of a long term linear trend, which we have subtracted out.

\paragraph{HD 198802}
We find a strong activity correlation for HD 198802 (Pearson coefficient 0.78), which leads us to disregard the best fit single-planet model and leave the RV variations as activity-driven jitter.

\paragraph{HD 11970}
After subtracting the stellar companion to HD 11970, we note a correlation between the radial velocities and the activity (Pearson coefficient 0.73).

\paragraph{HD 25311}
HD 25311 shows evidence of RV variation beyond simply activity-driven variations (given the lack of correlation between activity and velocities). We find that with a linear trend and a period of 70 days, we can reduce the RV jitter to nearly 10 m/s. However, this fit relies on a convenient lack of phase coverage during the RV maximum of the orbit and so we are hesitant to accept this best-fit planet. We instead simply subtract a linear trend. 

\paragraph{HD 25457}
We remove an obvious outlier from this star. The outlier is not as obvious when comparing simply the uncertainties in the velocities, although it is much larger than the typical uncertainty on the velocity for this star (13 m/s vs. 5 m/s). However, it is also noticeable by eye as an outlier in that it is below -175 m/s when nearly every other velocity for this star is within $\pm$30 m/s.

\paragraph{HD 38801}
After subtracting the best-fit companion to HD 38801, we note a correlation between the radial velocities and the activity (Pearson coefficient 0.64).

\paragraph{HD 88134}
HD 88134 shows evidence of a long term linear trend, which we have subtracted out.

\paragraph{HD 117497}
This star has the minimum number of observations for our jitter calculation (10) but over a very short baseline (close to 2 months). For this star, our jitter estimate is therefore relatively uncertain. 

\paragraph{HD 144363}
HD 144363 has clear evidence of a stellar companion (RV variations with semi-amplitude 6 km/s). However, the small number of points makes an exact fit difficult, and the residuals to the best fit remain of order 500 m/s. This could suggest spectral contamination from the binary companion, leading to spurious RV measurements, but for now it is unclear whether the poor fit is due to measurement error or the lack of complete coverage.

\paragraph{HD 154144}
After subtracting a linear trend, we find a strong correlation between the radial velocities and the activity (Pearson coefficient 0.68).

\paragraph{HD 156342}
HD 156342 shows evidence of a long term linear trend, which we have subtracted out.

\subsection{$1.3 \leq$ M$_{\star} < 1.4$~M$_{\odot}$}

\paragraph{HD 21847}
HD 21847 is another active star that shows strong correlation between activity and radial velocities (Pearson coefficient 0.78). The RV jitter for this star is relatively high (34 m/s) but appears to be purely activity-induced. 

\paragraph{HD 39828}
HD 39828 showed evidence of a long period stellar companion due to a strong linear trend, decreasing by about 400 m/s over the 8 and a half years of observations. Since there is no indication of curvature, we have subtracted a linear trend.

\paragraph{HD 95088}
HD 95088 had an obvious outlier that we have removed. This outlier was identified by eye as it had a velocity measurement that was 100 m/s removed from the remaining points. The RV jitter is decreased from 23 m/s to 9 m/s once the outlier is removed.

\paragraph{HD 106270}
HD 106270 shows a clear planetary signal \citep{Johnson2011}. However, it also shows a strong activity correlation (Pearson coefficient 0.645). Despite the strong correlation, the two signals do not seem to be entirely in phase and the planetary signal has sufficient phase coverage to indicate that the planetary signal is indeed correct.

\subsection{$1.4 \leq$ M$_{\star} < 1.5$~M$_{\odot}$}

\paragraph{HD 38529}
HD 38529 shows evidence of a strong activity cycle, however, this cycle does not appear to induce correlated radial velocities. The planets found in \citet{Wright2009} are not in phase and show little signs of being activity-induced RV variations.

\paragraph{HD 63754} 
Despite obtaining a planet fit for HD 63754, we find that we are quite likely overfitting in this case. Instead we simply subtract off a linear trend. Additional observations would confirm if the curvature suggested by the planet fit is indeed present.

\paragraph{HD 33142}
HD 33142 has 1 previously known planet \citep{Johnson2011}. After subtracting out the best fit planet, we find evidence for an additional planet near 860 days. We compare the 1-planet, 1-planet plus trend, 2-planet, and 2-planet plus trend fits and find that there is not strong enough evidence to accept either of the 2-planet fits. We therefore take the 1-planet fit with a trend as it resulted in a lower $\chi^2$ than the 1-planet fit alone. This system is worth further scrutiny to determine whether there is indeed a second planet in this system. If it does, our current jitter measurement will be slightly inflated: 8.7 m/s instead of 6.0 m/s.

\paragraph{HD 99706}
HD 99706 was originally published as a one planet system with a linear trend \citep{Johnson2011}. Later, \citep{Bryan2016} used additional observations to refine the fit and published a 2-planet solution. With the addition of a few more observations, we find that the current dataset is inconsistent with the \citet{Bryan2016} two planet solution. We use simply a 1 planet plus trend solution.

\paragraph{HD 140025}
We note that the radial velocity observations of HD 140025 have a strong correlation with activity (Pearson coefficient 0.785). 

\paragraph{HD 142091}
HD 142091 was discussed in detail in \autoref{sec:oscillation_validation}. The majority of the observations of this star were obtained on two separate nights in 2013 to capture stellar p-mode oscillations. These observations therefore skew the RMS toward lower values expected from purely stellar oscillations. We therefore remove these two nights and calculate the RV RMS of the remaining observations. The oscillation observations are used separately to validate the theoretical scaling relations used in this work.

\paragraph{HD 155413}
HD 155413 shows evidence of a stellar mass companion, and so we have simply subtracted a linear trend.

\paragraph{HD 219553} 
HD 219553 has a potential planet based on a 1-planet fit. However, the periodogram does not show a clear, sharp peak at the best-fit period. Given the few observations and poor phase-coverage of the best-fit planet fit, we opt to disregard the best-fit one planet solution and treat it as pure RV jitter. If there is indeed a planet, the RV RMS drops from 18.6 to 5.8. As such, this is a system where we may be able to use its height above the expected jitter floor to provide additional evidence for or against the one planet solution.

\paragraph{HD 224032}
This system shows evidence of a long term linear trend, which we have subtracted off.

\subsection{$1.5 \leq$ M$_{\star} < 1.6$~M$_{\odot}$}

\paragraph{HD 67767}
The velocities of HD 67767 pre-upgrade show much larger variation due to several high points than the post-upgrade velocities (14 m/s vs. 6 m/s). However, the post-upgrade velocities show evidence of an activity correlation. If the velocities are indeed correlated with activity, then the high velocity measurements pre-upgrade could simply be activity-induced. Therefore, we find it necessary to include all velocities in the calculation of RV jitter (11 m/s when combined).

\paragraph{HD 18645}
HD 18645 shows a slight correlation between activity and radial velocity (Pearson coefficient 0.538).

\paragraph{HD 22682}
HD 22682 has unusually large errors in the measured velocities. It is not a faint star, nor is it rapidly rotating. Rather, it is bright (V$=6.67$) and slowly rotating ($v\sin{i} = 0.8$ km/s.). 

\paragraph{HD 31423}
The velocities for HD 31423 show a rising trend in velocities spanning several hundred m/s and crossing 0 m/s from mid 2011 to mid 2012. A single point removed from the rest by more than a year (late 2013) has a measured velocity near 3000 m/s. As the latest observation, we are unable to constrain a fit with this point included. We do not expect this to be an errant point, but in order to accurately measure the jitter, we remove this point from the time series. The resulting rising trend we subtract as a simple linear trend.

\paragraph{HD 72440}
HD 72440 presents a very interesting case. We observe what appears to be a strong downward linear trend spanning 400 m/s. However, when examining the \SHK{} time series for this star, we notice they  \emph{very} strongly correlate with the radial velocities with a Pearson coefficient of 0.984. While the RV trend is almost certainly indicative of a stellar companion, the corresponding very large drop in S is hard to explain and could be interpreted as an extraordinarily large RV-activity correlation. We remain puzzled by this system.

\paragraph{HD 85440}
We obtain a reasonable planet fit for HD 85440. However, the RV measurements are highly correlated with the activity measurements (Pearson coefficient 0.929), and so we disfavor the planet, simply counting all variation as activity-induced jitter.

\paragraph{HD 100337} 
HD 100337 also shows a strong correlation between the activity and RV measurements (Pearson coefficient 0.789).

\paragraph{HD 193391}
HD 193391 shows a strong correlation between the activity and RV measurements (Pearson coefficient 0.860).

\subsection{$1.6 \leq$ M$_{\star} < 1.7$~M$_{\odot}$}

\paragraph{HD 12137}
HD 12137 appears to be a binary system. After subtracting a linear trend, we find evidence of a 5000 day period planet. However, we are reluctant to subtract it due to the few number of observations. We simply subtract a line for now.

\paragraph{HD 45210}
HD 45210 also shows a long term linear trend, which we have subtracted out.

\subsection{M$_{\star} > 1.7$~M$_{\odot}$}

\paragraph{HD 225021}
HD 225021 shows evidence of a linear trend, which we have removed, resulting in an RV jitter of 8 m/s instead of 27 m/s. 

\paragraph{HD 31543}
HD 31543 shows clear evidence of a binary companion. However, the final two observations have very large errors (10-20 m/s) and are the only two observations after the velocities have turned over and begin to increase again. Because of the large uncertainties for those measurements, the orbit is not well-constrained and other solutions exist.

\paragraph{HD 44506}
HD 45506 shows a coherent downward trend in the velocities, spanning a range of 100 m/s. We have therefore subtracted a linear trend which results in a jitter of 8 m/s (26 m/s otherwise).

\paragraph{HD 64730}
HD 64730 shows evidence of a stellar companion, however it has only completed one velocity minimum. In addition, there are strong velocity variations that result in high eccentricity fits for this system. Due to the few observations, we cannot fully constrain the orbit and can merely report an RV jitter about a poor fit. The resulting RV jitter is 49 m/s, beyond the range of the plots in Figures \ref{fig:rms_logg1} \& \ref{fig:rms_logg2}.

\paragraph{HD 96683}
HD 96683 shows evidence of a long term trend but with some slight curvature. We have opted to include the full Keplerian fit, which results in an RV RMS of 7.4 m/s, but note that this is likely an overfit. The RV RMS about a linear trend is 40 m/s, and about a parabola is 7.7 m/s, indicating that the curvature is indeed significant.

\subsection{Stars Removed from the Sample}\label{sec:removed}
Here we list the handful of stars that were removed from the sample along with the reasons for their removal.

\paragraph{HD 92855}
The cadence of observations for HD 92855 is not high enough to provide a Keplerian fit to the 30 km/s variations in the RVs. We expect that such a high amplitude variation likely arises from a stellar companion of nearly equal mass. We are therefore suspicious of contamination in the RV's from the companion that also contributes to the poor fit.

\paragraph{HD 4741}
HD 4741 shows evidence of a potentially long-period stellar-mass companion (about 8 km/s variation). However, we have exactly 10 observations of this star over a span of 10 years, but only 5 have any significant leverage, as the first 4 and final 3 observations are clustered in a period of about a week. This star was observed too infrequently in between those sets of observations that a Keplerian fit would be overfitting. Indeed, while we are able to fit a Keplerian that results in 7 m/s of jitter, the resulting eccentricity is likely unphysical ($e=0.99$) and indicative of the fitter ``chasing" a single point. We expect this best fit to be an overfit.

\paragraph{HD 202568}
HD 202568 has very large internal error bars for every radial velocity measurement (8 m/s). In addition, the observations span only one and a half years, giving only a very small time baseline on which to anchor the jitter.

\paragraph{HD 34957}
HD 34957 has evidence of spectral contamination in that all spectral lines have blueward asymmetries. We remove this star as a double-lined binary.

\paragraph{HD 75732}
HD 75732 (also known as 55 Cnc) has 5 planets in the system, which are interacting gravitationally at a scale that is discernible in the radial velocities \citep{Nelson2014}. For that reason, a multi-planet Keplerian fit is not sufficient for this system, and it requires an N-body dynamical model, as in \citet{Nelson2014}, which is beyond this analysis. We therefore remove it from our sample, as we know we cannot recreate the best fit for this system. For reference, \citet{Nelson2014} measure a jitter from Keck to be 3 to 3.5 m/s depending on whether the high cadence observations were treated as independent or perfectly correlated.

\paragraph{HD 16297}
Despite meeting our threshold of having 10 or more observations, HD 16297 has large uncertainties on the velocity measurements prior to the Keck-HIRES upgrade. When these points are removed, the remaining number of good observations is only 6, and so we remove it from our sample.

\paragraph{HD 69076}
HD 69076 has large amplitude variations (20 m/s) that indicate the presence of a stellar companion on what appears to be a short orbit. However, we are unable to obtain a believable best fit Keplerian for this system and so we are suspicious of spectral contamination that may be affecting the measured RV's.

\paragraph{HD 88656}
Like HD16297 above, HD 88656 does not meet our threshold for needing 10 or more observations once we remove the highly-uncertain and clearly inconsistent velocities prior to the Keck upgrade.

\end{document}